\documentclass[useAMS,usenatbib]{mn2e}
\usepackage{graphicx}
\usepackage{subfigure}
\usepackage{comment} 
\usepackage{placeins}

\newcommand{\kms}{km~s\ensuremath{^{-1}}}

\newcommand{\ms}{m~s\ensuremath{^{-1}}}

\title[]{Observational constraints on tidal effects using orbital eccentricities.\thanks{Based on observations made at the 1.93-m telescopes at Observatoire de Haute-Provence (CNRS), France with the SOPHIE spectrograph.}}

\begin{document}
\bibliographystyle{mn2e}

\author[N. Husnoo et al.]{Nawal Husnoo$^1$, Fr\'ed\'eric Pont$^1$, Tsevi Mazeh$^2$, Daniel Fabrycky$^{3}$, Guillaume H\'ebrard$^{4,5}$,\newauthor
Fran\c cois Bouchy$^{4,5}$, Avi Shporer$^{6}$\\
$^1$ School of Physics, University of Exeter, Exeter, EX4 4QL, UK \\
$^2$ School of Physics and Astronomy, Tel Aviv University, Tel Aviv 69978, Israel\\
$^3$ Harvard-Smithsonian Centre for Astrophysics, Garden Street, Cambridge, MA\\
$^4$ Institut d'Astrophysique de Paris, UMR7095 CNRS, Universit\'e Pierre \& Marie Curie, 98bis boulevard Arago, 75014 Paris, France\\
$^5$ Observatoire de Haute-Provence, CNRS/OAMP, 04870 Saint-Michel-l'Observatoire, France\\
$^{6}$ Las Cumbres Observatory Global Telescope network, 6740 Cortona Drive, suite 102, Goleta, CA 93117,
USA\\
}

\maketitle

\label{firstpage}

\begin{abstract}
We have analysed radial velocity measurements for known transiting exoplanets to study the empirical signature of tidal orbital evolution for close-in planets. Compared to standard eccentricity determination, our approach is modified to focus on the rejection of the null hypothesis of a circular orbit. We are using a MCMC analysis of radial velocity measurements and photometric constraints, including a component of correlated noise, as well as  Bayesian model selection to check if the data justifies the additional complexity of an eccentric orbit. We find that among planets with non-zero eccentricity values quoted in the literature, there is no evidence for an eccentricity detection for the $7$ planets CoRoT-5b, WASP-5b, WASP-6b, WASP-10b, WASP-12b, WASP-17b, and WASP-18b. In contrast, we confirm the eccentricity of HAT-P-16b, $e=0.034\pm0.003$, the smallest eccentricity that is reliably measured so far for an exoplanet as well as that of WASP-14b, which is the planet at the shortest period ($P=2.24$ d), with a confirmed eccentricity, $e= 0.088\pm0.003$. As part of the study, we present new radial velocity data using the HARPS spectrograph for CoRoT-1, CoRoT-3, WASP-2, WASP-4, WASP-5 and WASP-7 as well as the SOPHIE spectrograph for HAT-P-4, HAT-P-7, TrES-2 and XO-2.

We show that the dissipative effect of tides raised in the planet by the star and vice-versa explain all the eccentricity and spin-orbit alignment measurements available for transiting planets. We revisit the mass-period relation \citep{Mazeh2005, Pont2011} and consider its relation to the stopping mechanism of orbital migration for hot Jupiters. In addition to CoRoT-2 and HD 189733 \citep{Pont2009b}, we find evidence for excess rotation of the star in the systems CoRoT-18, HAT-P-20, WASP-19 and WASP-43.
\end{abstract}


\section{Introduction}
Most of the information we have about the formation, evolution and structure of exoplanets have come from the study of transiting planets. This is possible because the combination of radial velocity measurements with transit photometry can provide powerful constraints on the physical and orbital parameters of an exoplanet, such as the planetary mass, radius, orbital eccentricity, etc.

A selection effect due to geometry means that most transiting planets with radial velocity confirmation are found on very short period orbits with $P\sim1-20$ days. The close-in planets with periods of a few days are expected to experience strong tidal effects \citep[e.g.][]{Rasio1996}, which should increase sharply with decreasing period and these orbits are thus expected to circularise on a timescale much smaller than the system age. A higher tendency for such circular orbits is indeed observed in the sample of transiting planets, as compared to those from radial velocity surveys. This has been interpreted as a signature for tidal circularization. The transition from eccentric orbits to circular orbits at short period has also been seen in binary star systems, e.g. \citet{Mathieu1988} and \citet{Mazeh2008a}. Over the last few years, we have carried out a monitoring programme to obtain several radial velocity measurements of known transiting planetary systems with the intention of refining the orbital properties such as orbital eccentricity and spin-orbit alignment angle. We have used the SOPHIE spectrograph in the Northern hemisphere and the HARPS spectrograph in the Southern hemisphere \citep[e.g.][ESO Prog. 0812.C-0312]{Loeillet2008,Hebrard2008, Husnoo2011, Pont2011}.

One issue is the difficulty of measuring the orbital eccentricity of exoplanets for faint stars, especially for low-mass planets. While it is impossible to prove that an orbit is circular, with $e=0$ exactly, we can place an upper limit on the eccentricity of a given orbit (e.g. reject $e>0.1$ at the 95\% confidence level). In fact, a number of eccentric orbits have been detected at short period, but follow-up observations using photometry or additional radial velocity measurements led to the conclusion that some of these eccentricities had originally been overestimated. For example, the WASP-10 system \citep{Christian2009} was revisited by \citet{Maciejewski2011}, who showed that the initially reported eccentricity ($e=0.059^{+0.014}_{-0.004}$) had been overestimated and was in fact compatible with zero. The orbital eccentricity of WASP-12b \citep[$e=0.049\pm0.015$]{Hebb2009} is similarly compatible with zero \citep{Husnoo2011}, and the original detection was possibly due to systematic effects (weather conditions,  instrumental drifts, stellar spots or scattered sunlight).

The eccentricity distribution at short period has a crucial importance for any  theory of planetary formation and orbital evolution. Planets on orbits that are consistent with circular gather in a well-defined region of the mass-period plane, close to the minimum period for any given mass \citep{Pont2011}. We now show that there are no exceptions to this pattern, and revisit some apparent exceptions as reported in the literature. As an ensemble, the totality of transiting planets considered in this study are in agreement with classical tide theory, with orbital circularisation due to tides raised on the planet by the star and tides on the star raised by the planet, to varying degree depending on the position of the planet-star system in the mass-period plane.

In this study, we consider new radial velocity measurements made with HARPS and SOPHIE, as well as measurements present in the literature. We use photometric constraints in the form of the orbital period $P$ and mid-transit time $T_{tr}$, both of which can be measured accurately using transit photometry, and we also consider constraints from the secondary eclipse where available. In fact, if we define the orbital phase $\phi=(t-T_0)/P$ to be zero at mid-transit time $T_0=T_{tr}$, a planet on a circular orbit would have a mid-occultation phase of $\phi=0.5$ (by symmetry). A planet that is on an eccentric orbit will, however, have a mid-occultation phase different from 0.5 (unless the orbital apsides are aligned along the line of sight). This allows us to place a constraint on the $e\cos\omega$ projection of the eccentricity, as given by \citet{Winn2005a} (slightly modified):
\begin{equation}
e\cos\omega\simeq\frac{\pi}{2}\left(\phi_{\rm occ}-0.5\right),
\end{equation}
\noindent
to first order in $e$, where we now define $\phi_{\rm occ}$ to be the phase difference between the mid-transit time and the mid-occultation time, i.e. $\phi_{\rm occ} = ({T_{sec}-T_{tr}})/P$, where $T_{sec}$ is the time of the secondary eclipse following the transit time $T_{tr}$. The component $e\sin\omega$ is dependent on a ratio involving the durations of the occultation and transit \citep{Winn2005a},
\begin{equation}
e\sin\omega\simeq\frac{T_{\rm tra}-T_{\rm occ}}{T_{\rm tra}+T_{\rm occ}},
\end{equation}
to first order in $e$, where $T_{\rm tra}$ and $T_{\rm occ}$ are the transit and occultation durations respectively, although this constraint is weaker than the one on $e\cos\omega$.

In addition to the reanalysis of radial velocity measurements with photometric constraints, we also introduce two modifications to the Markov Chain Monte Carlo (MCMC) process commonly used by teams analysing radial velocity data to work out the orbital parameters of transiting exoplanets. This involves a new treatment of the correlated noise present in most radial velocity datasets, as well as analysing the data in model selection mode to check if an eccentric orbit is indeed justified, given the additional complexity of the eccentric version of a Keplerian orbit. 

The present time is significant in the study of exoplanets, because a number of high quality measurements are now available for the three main observable effects of tides: circularisation, synchronisation and spin-orbit alignment. In Section~\ref{sec:observations}, we describe our new radial velocity measurements obtained with SOPHIE and HARPS for 10 objects, as well as the measurements we collected from the literature for this study. We then describe the analysis we performed, in Section~\ref{sec:analysis}. In Section~\ref{sec:results}, we describe the objects in the classes ``eccentric'', ``compatible with circular, $e<0.1$'', and ``poorly constrained'' and present the updated eccentricities, as shown in Table~\ref{tab:results}. In Section~\ref{sec:discussion}, we consider our orbital eccentricity estimates in the light of tidal effects inside the planet due to the star and vice-versa. We find that our results are compatible with classical tidal theory, removing the need for perturbing stellar or planetary companions to excite non-negligible eccentricities in short period orbits.

\citet{Winn2010b} presented a discussion of the available measurements of the projected spin-orbit alignment angles, and found that hot planet-hosting stars ($T_{\textrm{eff}}>6250$ K) had random obliquities whereas cooler stars ($T_{\textrm{eff}}<6250$ K) tended to have aligned rotations. These authors suggested that this dichotomy can be explained if all these stars harbouring a planetary system start off with a random obliquity following some dynamical interaction, but only cool stars with a significant convective layer are able to undergo tidal effects leading to alignment. We verify in Section~\ref{sec:discussion} that the strong exceptions WASP-8 and HD 80606 are indeed systems with weak tidal interactions, and that the observation that these two are misaligned, is not incompatible with tidal theory.

In a number of cases, such as HD 189733 \citep{Henry2008}, WASP-19 \citep{Hebb2010} and CoRoT-2 \citep{Lanza2009a}, the rotational period of the star is known from photometric monitoring. Assuming the results of \citet{Winn2010b} are correct in the sense that  the convective layer in G dwarfs would cause tidal dissipation that aligns the stellar equator with the planetary orbit, the negligible value of the projected spin-orbit angle $\lambda$ means that the obliquity is indeed zero, i.e. the stellar equators are aligned with the orbital planes. In this case, a measurement of the projected equatorial rotational velocity of the star ($v\sin i$) through Doppler broadening yields the rotational period of the star. This means that for G dwarfs at least, we have enough information to observe the effect of tidal interactions on the stellar rotation. We show in Section~\ref{sec:discussion} that in addition to CoRoT-2 and HD 189733 \citep{Pont2009b}, we find evidence for excess rotation of the star in the systems CoRoT-18, HAT-P-20, WASP-19 and WASP-43.

The preliminary results from this study were published in \citet{Pont2011}, and the exact numerical values of the eccentricities have been updated in this paper to reflect our new choice of radial velocity measurement correlation timescale $\tau=1.5$ d (see Section~\ref{sec:tau}), as opposed to $\tau=0.1$ d in \citet{Pont2011}. The overall results, i.e. the clear separation between orbits that are consistent with circular and eccentric orbits in the mass-period plane, does not change in this paper. The mass-period relation of \citet{Mazeh2005} is still clearly present, with low-mass hot Jupiters on orbits that are consistent with circular clumping in a definite region of the mass-period plane, with heavier objects moving closer in, to shorter periods. This strongly suggests that tidal effects are involved in the stopping mechanism of these objects. A similar stopping mechanism can be seen at higher planetary masses, but destruction of the planet is not excluded in many cases. Other effects, such as spin-orbit alignment and stellar spin-up also point strongly towards the scenario of \citet{Rasio1996} where the short period orbits of hot Jupiters are formed by dynamical scattering, which produces eccentric and misaligned orbits. This is followed by tidal dissipation which leads to circularisation at short period, spin-orbit alignment and synchronisation of the rotation of the host star.

\section{Observations}
\label{sec:observations}
We include 73 measurements for 6 objects with the HARPS spectrograph (Tables~\ref{tab:rv_corot1}, \ref{tab:rv_corot3}, \ref{tab:rv_wasp2}, \ref{tab:rv_wasp4}, \ref{tab:rv_wasp5} and \ref{tab:rv_wasp7} ) and 45 measurements for 4 objects with the SOPHIE  spectrograph (Tables~\ref{tab:rv_hatp4}, \ref{tab:rv_hatp7}, \ref{tab:rv_tres2} and \ref{tab:rv_xo2}). Both are bench-mounted, fibre-fed spectrograph built on the same design principles and their thermal environments are carefully controlled, to achieve precise radial-velocity measurements. The two instruments have participated in the detection and characterisation of numerous transiting exoplanets, notably from the WASP and CoRoT transit searches. The wavelength calibrated high-resolution spectra from the instruments are analysed using a cross correlation technique which compares them with a mask consisting of theoretical positions and widths of the stellar absorption lines at zero velocity \citep{Pepe2002}. 

We carried out a literature survey and collected radial velocity measurements for 54 transiting planets, as well as other relevant data such as the orbital periods and the time of mid-transit. For the cases of CoRoT-1, CoRoT-2 and GJ-436, we also used the secondary eclipse constraint on the eccentricity component $e\cos\omega$ from \citet{Alonso2009}, \citet{Alonso2009a} and \citet{Deming2007}, respectively. 

Given the rapid rate of announcement of new transiting exoplanets, we had to stop the clock somewhere, and we picked the 1st of July 2010. We selected only objects that had been reported in peer-reviewed journals or on  the online preprint archive ArXiV.org. Moreover, we selected systems with well measured parameters (planetary radius $R_p$ and mass $M_p$ to within 10\%) and excluded faint objects ($V>15$). At that time, 64 such systems were known. We reanalyse the existing radial velocity data for 54 transiting systems, providing additional radial velocity measurements for 10 systems described above, and include 10 systems without further reanalysis of orbital ephemeris. These systems are listed in Table~\ref{tab:results}. In Section~\ref{add_systems}, we include a further 16 systems, most of which had been discovered in the mean time. The planets involved in this study are listed on the webpage http://www.inscience.ch/transits/, where we also include the parameters $v\sin i$ (the projected rotation velocity of the host star), $P_{\rm rot}$ (the orbital period of the host star) and the projected spin-orbit angle $\lambda$ where available.

\begin{table}
\centering
\begin{tabular}{ l r r }
\hline
Time  & RV   & $\sigma_{\rm RV}$ \\
\ [BJD-2450000] &  [\kms] &[\kms]  \\
\hline
4385.86631 & 23.4168 & 0.0190\\
4386.83809 & 23.6726 & 0.0139\\
4387.80863 & 23.3155 & 0.0149\\
4419.81749 & 23.5811 & 0.0123\\
4420.80300 & 23.3290 & 0.0118\\
4421.81461 & 23.6586 & 0.0113\\
4446.77797 & 23.3936 & 0.0145\\
4447.75517 & 23.4562 & 0.0130\\
4448.77217 & 23.6982 & 0.0126\\
4479.67146 & 23.3161 & 0.0123\\
4480.65370 & 23.6836 & 0.0140\\
4481.63818 & 23.5129 & 0.0173\\
4525.59523 & 23.6451 & 0.0116\\
4529.56406 & 23.3324 & 0.0127\\
4530.58002 & 23.5743 & 0.0105\\
4549.58179 & 23.5793 & 0.0280\\
4553.49391 & 23.3652 & 0.0124\\
4554.57636 & 23.6696 & 0.0157\\
4768.77120 & 23.7041 & 0.0092\\
4769.76601 & 23.4802 & 0.0104\\
4770.80872 & 23.3613 & 0.0108\\
4771.76514 & 23.6955 & 0.0102\\
4772.76824 & 23.4379 & 0.0109\\
4773.76896 & 23.3980 & 0.0095\\
\hline
\end{tabular}
\caption{HARPS radial velocity measurements for CoRoT-1 (errors include random component only).}
\label{tab:rv_corot1}
\end{table}

\begin{table}
\centering
\begin{tabular}{ l r r }
\hline
Time  & RV   & $\sigma_{\rm RV}$ \\
\ [BJD-2450000] &  [\kms] &[\kms]  \\
\hline
4768.52895 & -56.5039  & 0.0036\\
4769.51404 & -58.1970  & 0.0038\\
4770.51329 & -56.1017    & 0.0046\\
4772.52655 & -55.9171  & 0.0047\\
4773.52957 & -58.2227 & 0.0042\\
\hline
\end{tabular}
\caption{HARPS radial velocity measurements for CoRoT-3 (errors include random component only).}
\label{tab:rv_corot3}
\end{table}

\begin{table}
\centering
\begin{tabular}{ l r r }
\hline
Time  & RV   & $\sigma_{\rm RV}$ \\
\ [BJD-2450000] &  [\kms] &[\kms]  \\
\hline
4766.56990 &-27.8402 &0.0033\\
4767.52666 &-27.6797 &0.0023\\
4768.56373 &-27.7842 &0.0018\\
4769.54823 &-27.7343 &0.0017\\
4770.54665 &-27.7131 &0.0026\\
4771.54501 &-27.8099 &0.0031\\
4772.56012 &-27.6489 &0.0027\\
4773.56432 &-27.8568 &0.0019\\ \hline
\end{tabular}
\caption{HARPS radial velocity measurements for WASP-2 (errors include random component only).}
\label{tab:rv_wasp2}
\end{table}

\begin{table}
\centering
\begin{tabular}{ l r r }
\hline
Time  & RV   & $\sigma_{\rm RV}$ \\
\ [BJD-2450000] &  [\kms] &[\kms]  \\
\hline
4762.60256 &57.6687 &0.0028\\
4763.62220 &57.5637 &0.0022\\
4764.58386 &57.9085 &0.0035\\
4765.59031 &57.9871 &0.0038\\
4768.60378 &57.9109 &0.0022\\
4769.58081 &57.9784 &0.0023\\
4769.71186 &58.0331 &0.0017\\
4770.58784 &57.6591 &0.0024\\
4770.72474 &57.7930 &0.0023\\
4771.57892 &57.6311 &0.0021\\
4771.68481 &57.5752 &0.0019\\
4772.59125 &57.9518 &0.0025\\
4773.59429 &57.9811 &0.0018\\
4773.70377 &58.0346 &0.0024\\ \hline
\end{tabular}
\caption{HARPS radial velocity measurements for WASP-4 (errors include random component only).}
\label{tab:rv_wasp4}
\end{table}

\begin{table}
\centering
\begin{tabular}{ l r r }
\hline
Time  & RV   & $\sigma_{\rm RV}$ \\
\ [BJD-2450000] &  [\kms] &[\kms]  \\
\hline
4768.63152 & 19.7967 &0.0022\\
4768.73169 & 19.8696 &0.0018\\
4769.62838 & 20.1047 &0.0023\\
4770.62473 &  20.1231 &0.0022\\
4770.76117 &  20.2255 &0.0031\\
4771.60846 & 19.7737 &0.0017\\
4771.71520 & 19.7446 &0.0021\\
4772.63762 &  20.2588 &0.0023\\
4772.73505 &  20.2071 &0.0022\\
4773.62311 & 19.8540 &0.0021\\
4773.73277 & 19.9582 &0.0025\\ \hline
\end{tabular}
\caption{HARPS radial velocity measurements for WASP-5 (errors include random component only).}
\label{tab:rv_wasp5}
\end{table}

\begin{table}
\centering
\begin{tabular}{ l r r }
\hline
Time  & RV   & $\sigma_{\rm RV}$ \\
\ [BJD-2450000] &  [\kms] &[\kms]  \\
\hline
4762.53711 &-29.4388 & 0.0024\\
4763.57798 &-29.4994 & 0.0023\\
4764.50127 &-29.5469 & 0.0032\\
4765.54456 &-29.3948 & 0.0032\\
4767.54077 &-29.3485 & 0.0031\\
4768.57924 &-29.5636 & 0.0021\\
4769.64528 &-29.5332 & 0.0018\\
4770.64161 &-29.4468 & 0.0022\\
4771.62297 &-29.3421 & 0.0019\\
4772.65474 &-29.4212 & 0.0022\\
4773.63924 &-29.5829 & 0.0020\\ \hline
\end{tabular}
\caption{HARPS radial velocity measurements for WASP-7 (errors include random component only).}
\label{tab:rv_wasp7}
\end{table}

\begin{table}
\centering
\begin{tabular}{ l r r }
\hline
Time  & RV   & $\sigma_{\rm RV}$ \\
\ [BJD-2450000] &  [\kms] &[\kms]  \\
\hline
5003.41234 & -1.3253 & 0.0206\\
5005.47636 & -1.3669 & 0.0122\\
5006.50185 & -1.3244 & 0.0123\\
5007.42327 & -1.4822 & 0.0129\\
5008.39084 & -1.4143 & 0.0131\\
5009.38832 & -1.3465 & 0.0132\\
5010.39331 & -1.4711 & 0.0131\\
5011.42884 & -1.4181 & 0.0129\\
5012.46735 & -1.3504 & 0.0131\\
5013.45923 & -1.4487 & 0.0128\\
5014.43881 & -1.4106 & 0.0123\\
5015.48148 & -1.3212 & 0.0124\\
5016.41444 & -1.4666 & 0.0119\\ \hline
\end{tabular}
\caption{SOPHIE Radial velocity measurements for HAT-P-4 (uncertainties include random component only).}
\label{tab:rv_hatp4}
\end{table}

\begin{table}
\centering
\begin{tabular}{ l r r }
\hline
Time  & RV   & $\sigma_{\rm RV}$ \\
\ [BJD-2450000] &  [\kms] &[\kms]  \\
\hline
5002.48517 & -10.2995 & 0.0100\\
5003.52118 & -10.6910 & 0.0103\\
5004.59910 & -10.2681 & 0.0137\\
5005.49926 & -10.6377 & 0.0101\\
5006.55335 & -10.2564 & 0.0101\\
5007.53107 & -10.5975 & 0.0101\\
5008.47624 & -10.4027 & 0.0106\\
5010.43095 & -10.5681 & 0.0102\\
5011.52259 & -10.3835 & 0.0102\\
5013.60648 & -10.3090 & 0.0093\\
5014.57426 & -10.6862 & 0.0101\\
5015.58518 & -10.2680 & 0.0103\\
5016.54123 & -10.6808 & 0.0084\\ \hline
\end{tabular}
\caption{SOPHIE radial velocity measurements for HAT-P-7 (uncertainties include random component only).}
\label{tab:rv_hatp7}
\end{table}

\begin{table}
\centering
\begin{tabular}{ l r r }
\hline
Time  & RV   & $\sigma_{\rm RV}$ \\
\ [BJD-2450000] &  [\kms] &[\kms]  \\
\hline
5005.57207 & -0.4489 & 0.0107\\
5006.57091 & -0.2231 & 0.0100\\
5007.57482 & -0.2619 & 0.0105\\
5008.44948 & -0.4742 & 0.0106\\
5010.44234 & -0.4505 & 0.0107\\
5011.51233 & -0.2499 & 0.0110\\
5013.59448 & -0.4207 & 0.0114\\
5014.56301 & -0.1604 & 0.0108\\
5015.57356 & -0.5090 & 0.0107\\
5016.55442 & -0.2197 & 0.0110\\ \hline
\end{tabular}
\caption{SOPHIE radial velocity measurements for TrES-2 (uncertainties include random component only).}
\label{tab:rv_tres2}
\end{table}

\begin{table}
\centering
\begin{tabular}{ l r r }
\hline
Time  & RV   & $\sigma_{\rm RV}$ \\
\ [BJD-2450000] &  [\kms] &[\kms]  \\
\hline
4878.41245 & 46.7905 & 0.0091\\
4879.38681 & 46.9667 & 0.0084\\
4886.39349 & 46.7748 & 0.0095\\
4887.44867 & 46.9583 & 0.0084\\
4888.47514 & 46.7722 & 0.0085\\
4889.40965 & 46.8778 & 0.0086\\
4890.46546 & 46.8994 & 0.0085\\
4893.41643 & 46.8202 & 0.0087\\
4894.44335 & 46.8073 & 0.0121\\ \hline
\end{tabular}
\caption{SOPHIE radial velocity measurements for XO-2 (uncertainties include random component only).}
\label{tab:rv_xo2}
\end{table}

\clearpage
\section{Analysis}
\label{sec:analysis}
We used the radial velocity data , as well as the constraints on the orbital period $P$ and mid-transit time $T_{tr}$ (and $e\cos\omega$ where available) from photometry as described in Section \ref{sec:observations}. To calculate the median values of the derived parameters and their corresponding uncertainties, we marginalise over their joint probability distribution using a Markov Chain Monte Carlo analysis with the Metropolis-Hastings algorithm. This has been described in the past by \citet{holman2006} and our implementation is described in \cite{Pont2009b}. We model the radial velocity using a Keplerian orbit and run the MCMC for 500,000 steps, the first 50,000 of which are then dropped to allow the MCMC to lose memory of the initial parameters. We verify that the autocorrelation length of each chain is much shorter than the chain length to that ensure the relevant region of parameter space is properly explored.

Although this procedure is common practice in the community, we bring two changes. The first is a modification to the merit function that is used to work out the likelihood of a set of parameters given the data, to include the effects of correlated noise. This is described in Section~\ref{sec:sigmar}. The second modification we bring is that we consider not only the case of an orbital model with a free eccentricity $e$, but we also work out the likelihood for a circular orbit (i.e. with $e$ fixed at zero). We then compare the two models, by including a penalty for the additional complexity in the eccentric one (i.e. two additional degrees of freedom). We do this by using the Bayesian Information Criterion, as described in Section~\ref{sec:modelselection}. 

We report the median value in the chain for each parameter, as well as the central 68.4\% confidence interval on the parameter. A circular orbit model for an orbit that is in fact eccentric would artificially make the uncertainties in the derived parameters smaller, so in the case of the systemic velocity $V_0$ and the semi-amplitude $K$, we report the median values from a circular orbit model, yet we include the confidence intervals derived from the eccentric model.

\subsection{The treatment of correlated noise}
\label{sec:sigmar}
Correlated noise can be important in the analysis of transit light curves \citep{Pont2006}, and we included this in the analysis of radial velocity measurements in the case of WASP-12 \citep[see][]{Husnoo2011}. If we assume uncorrelated Gaussian noise when analysing data that is affected by correlated noise, we run the risk of overestimating the importance of a series of measurements that were obtained in quick succession, and this can have implications for example in estimating the orbital eccentricity.

From \citet{Sivia2006}, the likelihood function for some data, given a model, is given by:
\begin{equation}
{\rm P}({\rm\bf D}|{\bf \theta},I)=\frac{{\rm exp}\left[-\frac{1}{2}({\bf F}-{\bf D})^{\rm T}{\rm\bf C}^{-1}({\bf F}-{\bf D})\right]}{\sqrt{(2\pi)^N{\rm det}({\rm\bf C})}},
\label{eqn:likelihood}
\end{equation}
where $\bf D$ is the radial velocity time series data expressed as a vector, $\bf \theta$ is the vector of model parameters, $\bf F$ is the predicted values from the Keplerian model. In this case, $\chi^2$ is defined by 
\begin{equation}
\chi^2=({\bf F}-{\bf D})^{\rm T}{\rm\bf C}^{-1}({\bf F}-{\bf D}),
\end{equation}

\ \\
where $\bf C$ is the covariance matrix, which remains constant throughout the MCMC analysis for each system. 
In the case of independent measurements, the components of the covariance matrix $\bf C$ would be obtained using

\begin{equation}
C_{k,k'}= \left\{ 
\begin{array}{l l}
  \sigma_k^2 & \quad \mbox{for $k=k'$}\\
  0 & \quad \mbox{otherwise,}\\ \end{array} \right.
\end{equation}
whereas in the presence of some correlated noise, we modify this to include a squared exponential covariance kernel so that

\begin{equation}
C_{k,k'}=\delta_{k,k'}\sigma_k^2+\sum\limits^M_{i=1}\sigma_i^2\exp-\frac{(t_k-t_{k'})^2}{2\tau_i^2}
\label{covariancesum}
\end{equation}
where $\sigma_k$ is the formal uncertainty on each measurement $k$ as obtained from the data reduction for that measurement, and the sum over $M$ terms having the form $\sigma_i^2\exp-\frac{(t_k-t_{k'})^2}{2\tau_i^2}$ allows us to include a number of stationary covariance functions to account for correlations in the noise, occuring over the timescales of hours to days. 

In practise, it can be tricky to estimate the values of $\tau_i$ and $\sigma_i$ for radial velocity datasets, especially where the number of measurements is few or the phase-coverage is incomplete. Given the small datasets, we elect to use a single time-invariant correlation term, setting $M=1$. The timescale is now called $\tau$, and  the corresponding value of $\sigma_i$ is now called  $\sigma_r$, where the subscript $r$ indicates ``red noise'' \citep{Pont2006}. A fully Bayesian analysis would require that we assign priors to these two parameters and then marginalise over them.  In practise, the sparse sampling and small datasets for radial velocity observations mean that it is very difficult to perform Bayesian marginalisation over these two parameters, and the results would depend on the prior space chosen (eg: $\tau$, $\log \tau$, etc). We therefore use a single pair of parameters for $\tau$ and $\sigma_r$. Using a single term in the sum in Equation~\ref{covariancesum} makes the expression less flexible, but prior experience shows that correlations over the $\sim 1$ day timescale are particularly important for radial velocity datasets, especially for measurements taken in the same night \citep[see for example,][ for the case of WASP-12]{Husnoo2011}. For datasets where the reduced $\chi^2$ for a given model (circular or eccentric) was larger than unity, we estimated $\sigma_r$ by repeating the MCMC analysis with different values of $\sigma_r$ until the the best-fit orbit resulted in a reduced $\chi^2$ of unity for some optimal value of $\sigma_r$. We discuss the estimation of $\tau$ in the next sub-section.

\subsection{Estimation of $\tau$}
\label{sec:tau}
There is a degeneracy between $\sigma_r$ and $\tau$ for the time sampling typical of our RV data: if we assume a long timescale compared to the interval of time between the measurements, we are asserting that we have a reason to believe that several measurements may have been systematically offset in the same direction. A measurement that occurs within that timescale but is offset to a very different extent from nearby measurements (e.g. if the correlation timescale $\tau$ has been overestimated) will require a larger value of $\sigma_r$ for the dataset as a whole to yield a reduced $\chi^2$ of unity.

\begin{figure*}
\resizebox{8cm}{!}{\includegraphics{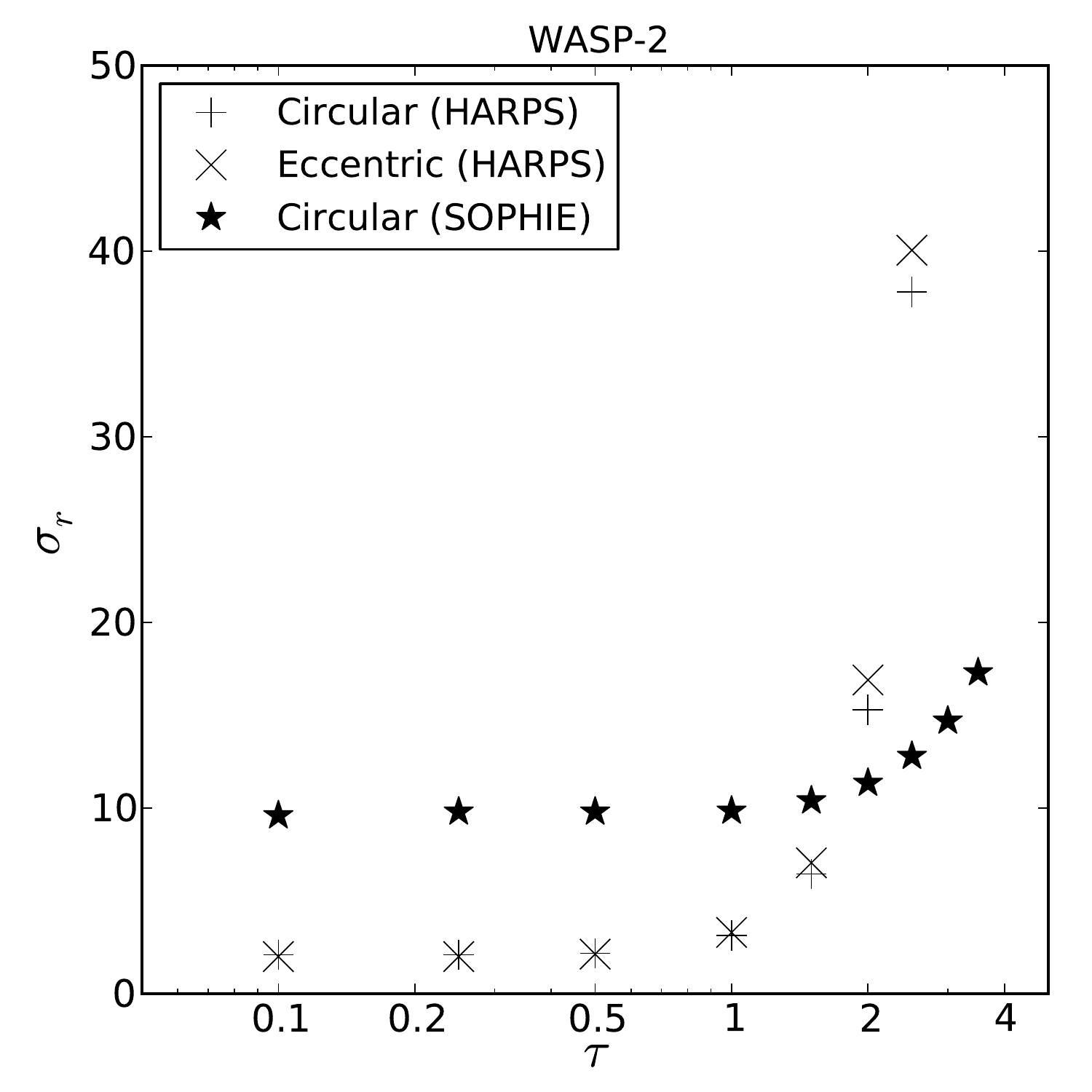}}
\resizebox{8cm}{!}{\includegraphics{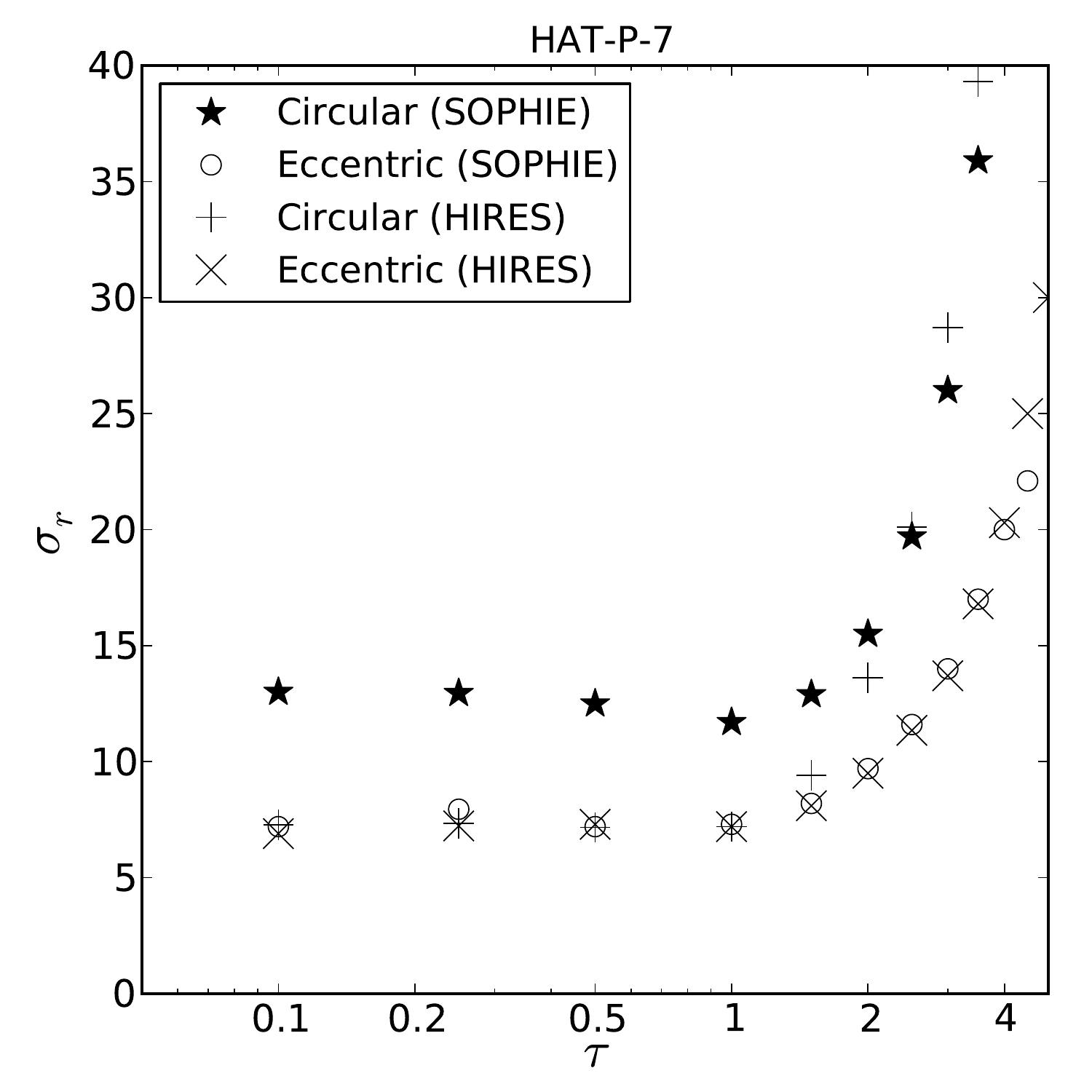}}
\resizebox{8cm}{!}{\includegraphics{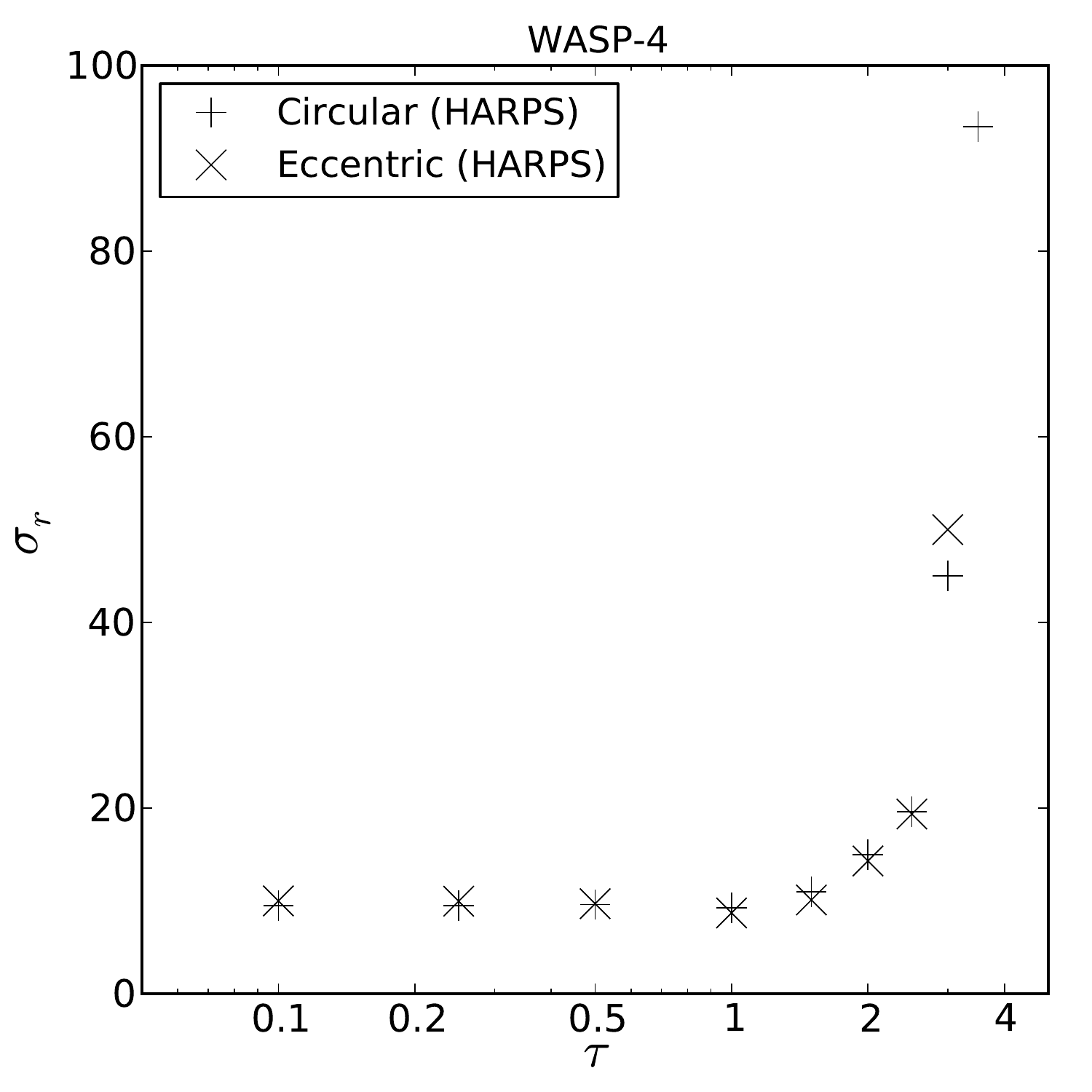}}
\resizebox{8cm}{!}{\includegraphics{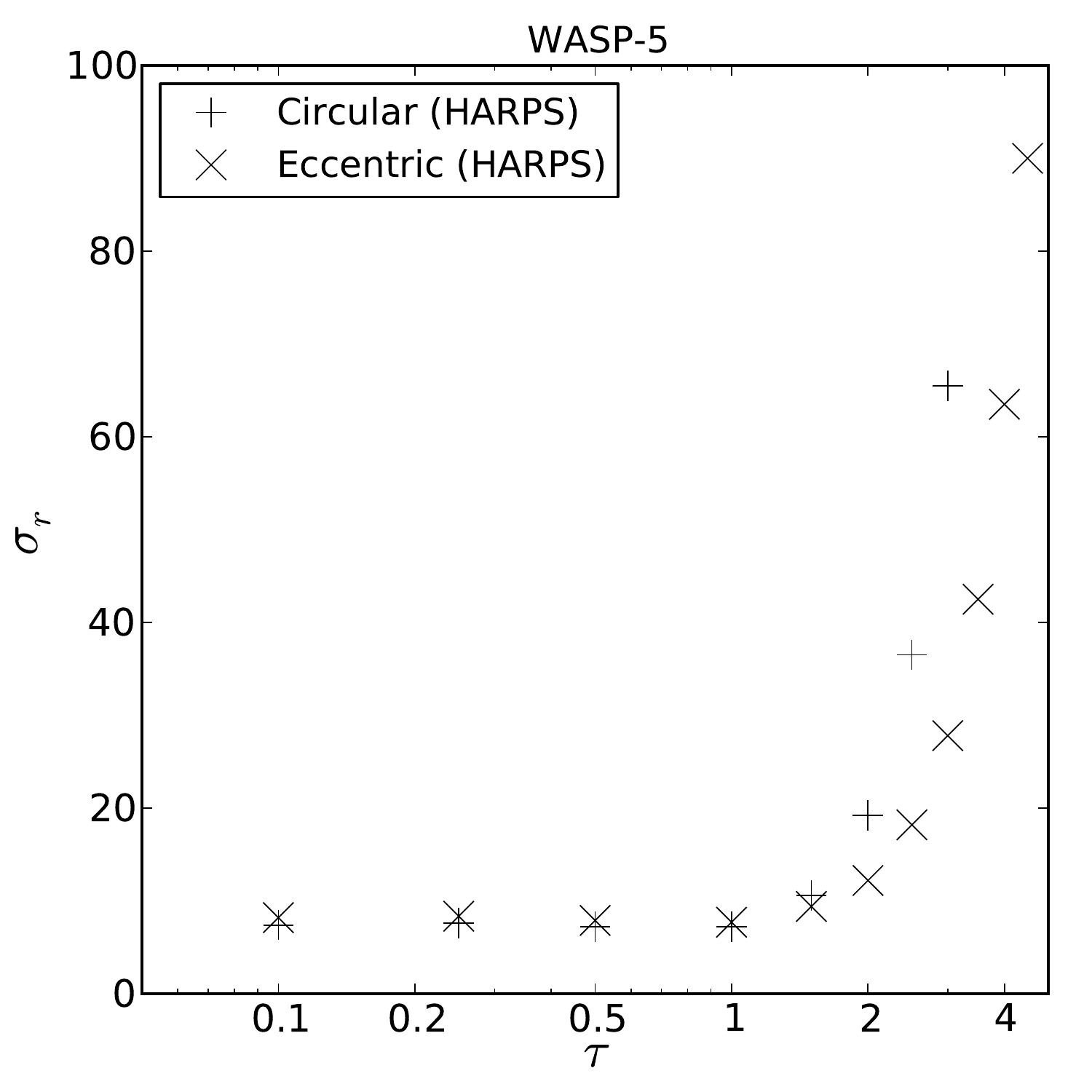}}
\caption{Plot showing the degeneracy between $\sigma_r$ and $\tau$ for objects where the orbital parameters are well constrained (see Sections~\ref{sec:sigmar} and \ref{sec:tau}) because a sufficient number of radial velocity measurements is available and provides sufficient phase coverage. The object being studied is shown in the title for each panel, and the instrument used for the measurements are shown in parenthesis. As can be seen on each plot, the optimal $\sigma_r$ that gives a reduced $\chi^2$ of unity for each dataset increases slowly with $\tau$ for $\tau<1.5$ d, but increases faster after 1.5 d. This hints that the systematic effects occur on a timescale of 1.5 d, and could be related to the weather. Note the SOPHIE data for WASP-2 did not provide full phase coverage --- the solution did not converge for an eccentric model and the knee at $\tau\sim1.5$ d is less pronounced.}
\label{fig:tausigma}
\end{figure*}

To estimate $\tau$, we looked at several datasets for each of the instruments HARPS, HIRES and SOPHIE. We repeated the analysis in Section~\ref{sec:sigmar} using values of $\tau$ in the range 0.1--5~d, to check for weather-related correlations. To see the effects of choosing between an eccentric orbit or a circular orbit on our estimation of $\tau$, we carried each analysis twice, by adjusting $\sigma_r$ (see Section~\ref{sec:sigmar}) to obtain a reduced $\chi^2$ of unity (within 0.5\%) for each orbital model (circular and eccentric). We plotted the optimal values of $\sigma_r$ against $\tau$ for several objects using data obtained from different instruments separately, as shown in Figure~\ref{fig:tausigma}. For WASP-2, we used our new HARPS measurements (Table~\ref{tab:rv_wasp2}) as well as SOPHIE measurements from \citet{Cameron2007}.  For WASP-4 and WASP-5, we used our new HARPS measurements (Tables~\ref{tab:rv_wasp4} and \ref{tab:rv_wasp5}), and for HAT-P-7, we used our new SOPHIE measurements (Table~\ref{tab:rv_hatp7}) as well as HIRES measurements from \citet{Winn2009b}. We found that for those datasets and objects where the orbital elements were well-constrained the plot showed a gentle increase in $\sigma_r$ with $\tau$, for $\tau\leq 1.5$d, then increased much faster for these datasets at a timescale of $\tau> 1.5$ d. For objects that have been observed with multiple instruments, this characteristic timescale is independent, both of the instrument used or the assumption about the eccentricity (i.e. free eccentricity or $e$ fixed at zero), suggesting that the correlated noise is probably related to weather conditions. We therefore assumed a correlation timescale of $1.5$ d in the rest of this study, unless otherwise noted. This means that we are accounting for the red noise in the same-night measurements, and for measurements that are taken further apart in time, this procedure reduces to the more familiar ``jitter'' term. The value of $\sigma_r$ inferred at $\tau=1.5$ d in some cases varies by a few percent depending on the model chosen, i.e. eccentric or circular, and varies across datasets, as discussed later.

\begin{figure*}
\resizebox{16cm}{!}{\includegraphics{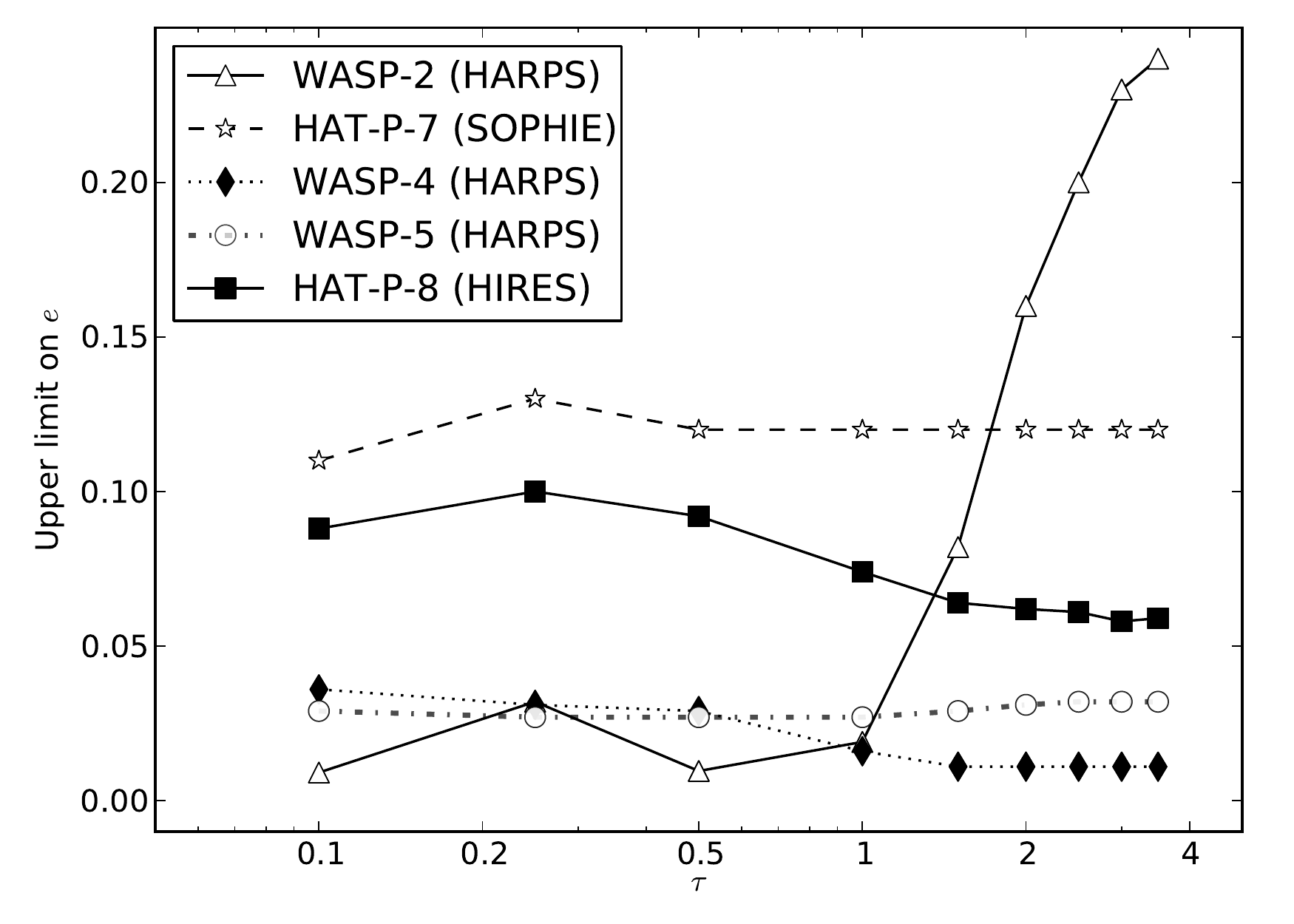}}
\caption{Plot showing the effect of varying $\tau$ on the 95\% upper limit on the derived eccentricity for WASP-2, HAT-P-7, WASP-4 and WASP-5. Except for WASP-2, where the phase coverage of the HARPS data is incomplete (see Figure~\ref{fig:wasp2ph}), varying $\tau$ has no effect on the 95\% upper limit of the derived eccentricity for timescales of a few days, $\tau>1.5$ d. Note: each line on this plot is made a single dataset.}
\label{fig:tau_ecc}
\end{figure*}

We also investigated the effects of varying $\tau$ on our final results. For the same systems discussed above, we plotted the 95\% upper limit on the eccentricity as obtained from each dataset separately. The results are shown in Figure~\ref{fig:tau_ecc}, where it is clear that the choice of $\tau$ has no effect on the final result for $\tau\geq1.5$ d. The only exception is WASP-2 (HARPS), where we only have 8 measurements and the phase coverage is not as complete as for the other objects (see Figure~\ref{fig:wasp2ph}). Similarly, the derived parameters $V_0$ and $K$ did not vary appreciably with $\tau$.

\subsection{Model selection}
\label{sec:modelselection}

Determining whether an orbit is consistent with circular (Model 1) or eccentric (Model 2), is an exercise in model selection. If we assume the prior probability of the circular and eccentric models are the same, we can use the Bayesian Information Criterion (BIC) \citep{Liddle2007} to decide between the two models. This is equivalent to working out the Bayes factor $P({\rm data}|{\rm Model_1})/P({\rm data}|{\rm Model_2})$, subject to the assumptions described below. The Bayes factor is the ratio of marginal likelihoods for each model, each of which is given from
\begin{equation}
P({\rm data}|{\rm Model_j}) = \displaystyle\int_{\Theta_j} L(\Theta_j|{\rm data}) \times P(\Theta_j|{\rm Model_j}) d\Theta_j
\end{equation}
where $\Theta_j$ represent the vector of parameters for each model $j$, $L(\Theta_j|{\rm data})$ is the likelihood and $P(\Theta_j|{\rm Model_j})$ is the joint posterior distribution of the parameters.

As described in Section \ref{sec:analysis} above and in \cite{Pont2009b}, the MCMC process produces the joint posterior distribution for the parameters, and we also obtain a maximum likelihood $L_{\rm max}$, corresponding to the smallest value of $\chi^2$ (as given in Section~\ref{sec:sigmar}) for each model. We then use the Bayesian Information Criterion \citep{Liddle2007} as given by,

\begin{equation}
{\rm BIC}=-2\ln L_{\rm max} +k \ln N,
\label{eqn:bic}
\end{equation}
where $N$ is the number of measurements, $k$ is the number of parameters in the model used. This simplifies the expression for the marginal likelihood by performing the integration using Laplace's method and assumes a flat prior. If we replace $L_{\rm max}$ with the expression given by ${\rm P}({\rm\bf D}|{\bf \theta},I)$ in equation \ref{eqn:likelihood} above,

\begin{equation}
{\rm BIC}=\chi_{\rm min}^2 +k \ln N + \ln \left((2\pi)^N|{\bf C}|\right),
\label{eqn:bic2}
\end{equation}

where $\chi_{\rm min}^2$ is the minimum value of $\chi^2$ achieved by the model, $N$ is the number of measurements, $k$ is the number of parameters in the model, and $|{\bf C}|$ is the determinant of the correlation matrix given in Section~\ref{sec:sigmar} above.

The radial velocity  data for a Keplerian orbit involves 6 free parameters: the period $P$, a reference time such as the mid-transit time $T_{\rm tr}$, a semi-amplitude $K$, a mean velocity offset $V_0$, the argument of periastron $\omega$ and the eccentricity $e$. Following \citet{Ford2006}, we use the two projected compoments $e\cos\omega$ and $e\sin\omega$ instead of $e$ and $\omega$, to improve the efficiency of the MCMC exploration. In this study, we use the period $P$, mid-transit time $T_{\rm tr}$ and their corresponding uncertainties as {\it a priori} information. We thus count them as two additional measurements in the calculation of the BIC, while the number of free parameters in each model (circular or eccentric) is now decreased by two. In this case, a circular model would have 2 free parameters ($V_0$ and $K$), while an eccentric model would have 4 free parameters ($V_0$, $K$, $e$, and $\omega$).

The term $k\ln N$ thus penalises a model with a larger number of parameters (for example, an eccentric orbit), and we seek the model with smallest BIC. For each object, we repeated the MCMC analysis using the optimal value for $\sigma_r$ for a circular orbit and an eccentric orbit separately, at $\tau=1.5$ d, unless otherwise noted. We call these two families. For each family, we performed a fit with a circular model and an eccentric model. In most cases, the two families agreed on a circular model (indicated by ``C'' in Table~\ref{tab:results}) or an eccentric model (indicated by ``E'' if $e>0.1$ in Table~\ref{tab:results}), indicating this with a smaller BIC$_c$ or a smaller BIC$_e$ respectively. If the two families favoured a circular (or eccentric) orbit, we give the parameters from the family using an optimal value of $\sigma_r$ for the circular (or eccentric) orbit. In a number of such cases, however, the upper limits on the orbital eccentricity were larger than $e=0.1$. We labelled these eccentricities as ``poorly constrained'' (indicated by ``P'' in Table~\ref{tab:results}). In a few cases, the small number of measurements or the quality of measurements (e.g. for faint targets, or low mass planets) meant the two families disagreed: the family using the optimal value of $\sigma_r$ for a circular orbit gave a smaller value of BIC$_c$, favouring the circular orbit and the family using the optimal value of $\sigma_r$ for an eccentric orbit gave a smaller value of BIC$_e$, favouring the eccentric orbit. We labelled these cases ``poorly constrained'' as well.


\section{Results}
\label{sec:results}
The results of this study are shown in Table~\ref{tab:results}. We place constraints on the eccentricities of transiting planets for which enough data is available. We analysed radial velocity data for 54 systems. For 8 systems, we used our new radial velocity data (described in Section~\ref{sec:observations}) in addition to existing RVs from the literature. For the other 46 systems, we reanalysed existing RVs from the literature. 

In Section~\ref{sec:noteccentric}, we describe the planets for which we do not consider the evidence for an orbital eccentricity compelling, despite previous evidence of a departure from circularity ($e>1\sigma$ from zero), followed by Section~\ref{sec:circular}, where we describe the planets for which we consider the orbital eccentricity to be either so small as to be undetectable or compatible with zero. In Section~\ref{sec:eccentric} we describe planets that can be safely considered to be on eccentric orbits and finally, in Section~\ref{sec:unknown}, we describe the planets for which we consider the orbital eccentricity to be poorly constrained (as described in Section~\ref{sec:modelselection}). In the following Sections we also include a discussion of the evidence for eccentricity for 26 other systems from the literature.

\subsection{Planets with orbits that no longer qualify as eccentric according to this study}
\label{sec:noteccentric}
In a number of cases in the past, the derived eccentricity from an MCMC analysis deviated from zero by more than 1$\sigma$, for example CoRoT-5b, GJ436b, WASP-5b, WASP-6b, WASP-10b, WASP-12b, WASP-14b, WASP-17b and WASP-18b. In this Section, we discuss the cases of 7 planets, CoRoT-5b, WASP-6b, WASP-10b, WASP-12b, WASP-17b, WASP-18b and WASP-5b, that are shown to have orbital eccentricities that are compatible with zero.

\ \\
{\bf CoRoT-5}\\
CoRoT-5b is a $0.46$ M$_j$ planet on a $4.03$ day orbit around a F9 star (V=14.0), first reported by \cite{Rauer2009}. Using $6$ SOPHIE measurements (one of which is during the spectroscopic transit, which we ignore in this study) and $13$ HARPS measurements, the authors derived a value of eccentricity $e=0.09^{+0.09}_{-0.04}$. In our study, we used the formal uncertainties quoted with the data without any additional noise treatment, since they resulted in a reduced $\chi^2$ less than unity for both an eccentric and a circular orbit. We imposed the prior information from photometry $P=4.0378962(19)$ and $T_{tr}=2454400.19885(2)$ from the \cite{Rauer2009} and obtained a value of $\chi^2_c=15.97$ for the circular orbit and a value of $\chi^2_e=13.50$ for the eccentric orbit ($e=0.086^{+0.086}_{-0.054}$, $e<0.26$). Using $N=20$, $k=3$ and $k=5$ for the circular (two datasets, each with one $V_0$ and a single $K$) and eccentric orbits respectively, we obtained BIC$_{c}=151.05$ and BIC$_{e}=154.57$. A smaller value of BIC$_{c}$ means the circular orbit cannot be excluded.

\ \\
{\bf WASP-6}\\
WASP-6b is a $0.50$ M$_j$ planet on a $3.36$ day orbit around a G8 star (V=11.9), first reported by \cite{Gillon2009b}. Using $35$ CORALIE measurements and $44$ HARPS measurements (38 of which occur near or during a spectroscopic transit, which we ignore in this study), the authors derived a value of eccentricity $e=0.054^{+0.018}_{-0.015}$. In our study, we used the 35 CORALIE measurements and the $6$ HARPS measurements that were not taken in the single night where the spectroscopic transit was observed. We used $\sigma_r=0$ \ms\  for CORALIE (the data produces a reduced $\chi^2=0.89$ when fitted with a circular orbit, indicating overfitting) but for HARPS we used $\tau=1.5$ d and $\sigma_r=4.15$ \ms\ to obtain a reduced $\chi^2$ of unity for the circular orbit. We obtained a value of $\chi^2_c=38.09$ for the circular orbit and a value of $\chi^2_e=33.58$ for the eccentric orbit ($e=0.041\pm0.019$, $e<0.075$). Using $N=43$ (41 RVs and two constraints from photometry), $k=3$ and $k=5$ for the circular (two datasets, each with one $V_0$ and a single $K$) and eccentric orbits respectively, we obtained BIC$_{c}=333.25$ and BIC$_{e}=336.27$. We repeated the calculations, using $\sigma_r=0$ for CORALIE (the data produces a reduced $\chi^2=0.85$ when fitted with an eccentric orbit, indicating overfitting) but for HARPS we used $\tau=1.5$ d and $\sigma_r=3.59$ \ms\ to obtain a reduced $\chi^2$ of unity for the eccentric orbit. We obtained a value of $\chi^2_c=39.20$ for the circular orbit and a value of $\chi^2_e=34.47$ for the eccentric orbit ($e=0.043\pm0.019$, $e<0.075$). Using $N=43$, $k=3$ and $k=5$ for the circular and eccentric orbits respectively, we obtained BIC$_{c}=333.60$ and BIC$_{e}=336.39$. We therefore find that the circular orbital solution cannot be excluded, but the possibility that $e>0.1$ is rejected.

\ \\
{\bf WASP-10}\\
WASP-10b is a $2.96$ M$_j$ planet on a $3.09$ day orbit around a K5 star (V=12.7), first reported by \cite{Christian2009}. Using $7$ SOPHIE measurements and $7$ FIES measurements, the authors derived a value of eccentricity $e=0.059^{+0.014}_{-0.004}$. The FIES data yielded a reduced $\chi^2$ less than unity with both eccentric and circular orbits, indicating overfitting, so we set $\sigma_r=0$ \ms.

For the SOPHIE data, used $\tau=1.5$ d, $\sigma_r=54.5$ \ms\ to  obtain a reduced $\chi^2$ of unity for the circular orbit. We reanalysed all the radial velocity measurements, and applied the prior from photometry $P=3.0927636(200)$ and $T_{tr}=2454357.8581(4)$ from \cite{Christian2009}. We obtained a value of $\chi^2_c=13.49$ for the circular orbit and a value of $\chi^2_e=7.47$ for the eccentric orbit ($e=0.049\pm 0.022$, less significant than the original claim). Using $14$ measurements and two priors from photometry ($N=16$), $k=3$ and $k=5$ for the circular (two datasets, each with one $V_0$ and a single $K$) and eccentric orbits respectively, we obtained BIC$_{c}=151.50$ and BIC$_{e}=151.01$. This now appears to show only a marginal support for an eccentric orbit.

\begin{figure*}
\resizebox{15cm}{!}{\includegraphics{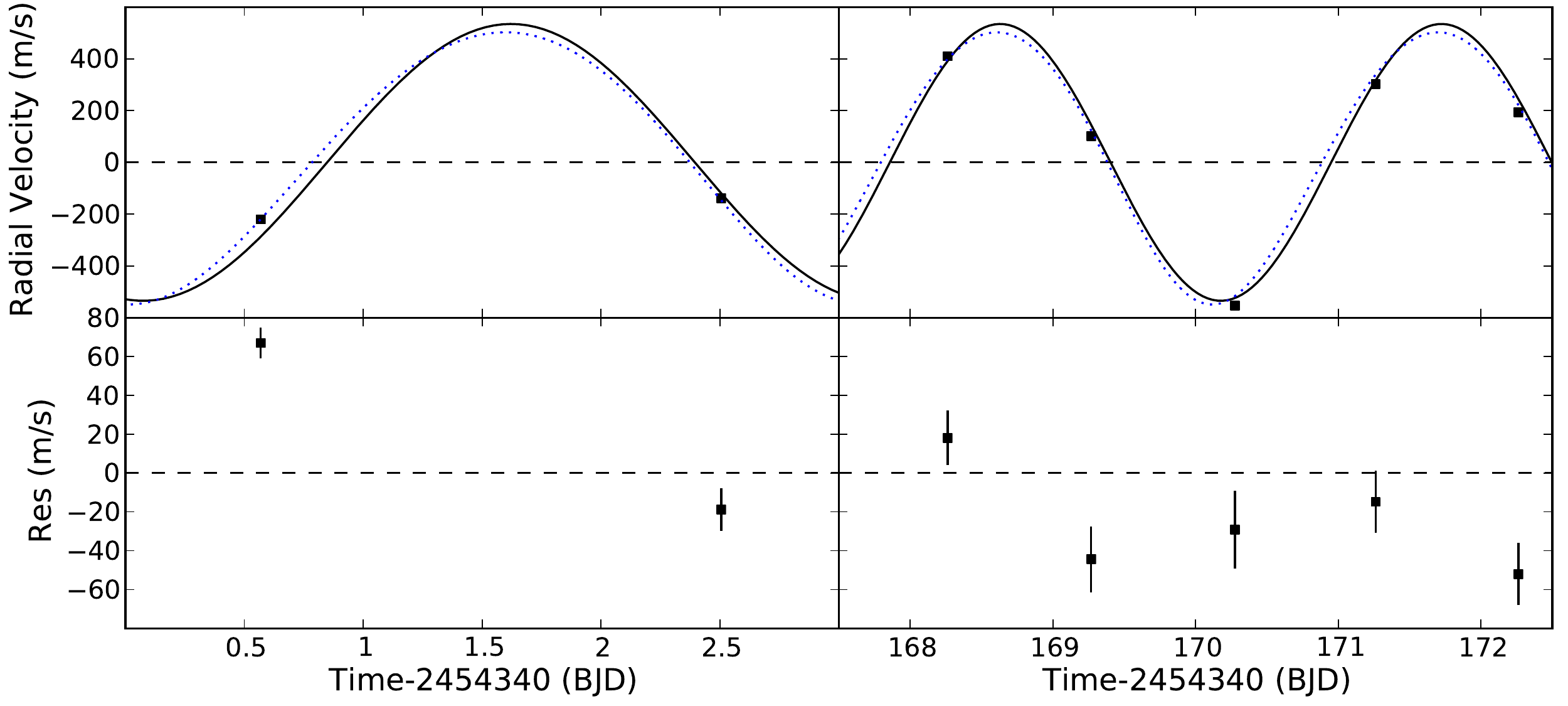}}
\caption{Plot showing SOPHIE radial velocity data from \citet{Christian2009} for WASP-10, plotted against time. The plot has been split along the time axis into two panels (left and right) to remove the ~160 days without measurements, for clarity. A circular orbit (solid line) and an orbit with the best-fit eccentricity ($e=0.048$) are overplotted. The residuals relative to the circular orbit are shown in the bottom panel.}
\label{fig:wasp10}
\end{figure*}

We plotted the SOPHIE radial velocity data against time, as shown in Figure~\ref{fig:wasp10} and overplotted a circular orbit as well as an eccentric orbit. Due to the long time between the first two measurements and the last five, we plot them in separate panels, shown on the left and right respectively. It is clear that the first measurement is pulling the eccentricity upwards, and we suspect from experience that the long term drifts in the SOPHIE zero point in HE mode for faint targets could have affected the first two measurements. We therefore repeated our calculations using only the last five measurements from the SOPHIE dataset and the whole FIES dataset, and set $\sigma_r=45.5$ \ms\ for SOPHIE. This time, we obtained a value of $\chi^2_c=11.81$ for the circular orbit and a value of $\chi^2_e=7.64$ for the eccentric orbit ($e=0.043\pm 0.035$). Using $12$ measurements and two priors from photometry ($N=14$), $k=3$ and $k=5$ for the circular (two datasets, each with one $V_0$ and a single $K$) and eccentric orbits respectively, we obtained BIC$_{c}=128.71$ and BIC$_{e}=129.83$, this time favouring the circular orbit. We repeated this calculation, and set $\sigma_r=0$ \ms\ for both SOPHIE and FIES, as each dataset gave a reduced $\chi^2$ of less than unity for the eccentric orbit (SOPHIE reduced $\chi^2=0.64$, FIES reduced $\chi^2=0.45$). This time, we obtained a value of $\chi^2_c=19.30$ for the circular orbit and a value of $\chi^2_e=10.65$ for the eccentric orbit ($e=0.080\pm 0.055$). Using $12$ measurements and two priors from photometry ($N=14$), $k=3$ and $k=5$ for the circular (two datasets, each with one $V_0$ and a single $K$) and eccentric orbits respectively, we obtained BIC$_{c}=128.33$ and BIC$_{e}=124.97$, this time favouring the eccentric orbit once again. It is therefore unclear to us whether or not the orbital eccentricity is non-zero as claimed in \cite{Christian2009}.

\citet{Maciejewski2011b}, used transit timing variation analysis and reanalysed the radial velocity data, to obtain an eccentricity that is indistinguishable from zero ($e=0.013\pm0.063$). They argued instead that the original detection of an eccentricity had been influenced by starspots. The difference between our value of eccentricity and that derived by \citet{Maciejewski2011b} is probably due to the fact that the latter used a two planet model, which can reduce the derived eccentricity further --- sparse sampling of the radial velocity from a two planet system can lead to an overestimated eccentricity.

\ \\
{\bf WASP- 12}\\
WASP- 12b is a $1.41$ M$_j$ planet on a $1.09$ day orbit around a F9 star (V=11.7), first reported by \cite{Hebb2009}. Using SOPHIE measurements, the original authors derived a value of eccentricity $e=0.049\pm0.015$. \citet{Husnoo2011} used new SOPHIE radial velocity measurements, as well as the original transit photometry from \citet{Hebb2009} and the secondary eclipse photometry from \citet{Campo2011} to suggest that  the eccentricity was in fact compatible with zero ($e= 0.017^{+0.015}_{-0.010}$).

\ \\
{\bf WASP-17}\\
WASP-17b is a $0.50$ M$_j$ planet on a $3.74$ day orbit around a F6 star (V=11.6), first reported by \cite{Anderson2010}. Using $41$ CORALIE measurements (three of which are during the spectroscopic transit, which we ignore in this study) and $3$ HARPS measurements, the authors considered three cases: first imposing a prior on the mass $M_*$ of the host star, secondly imposing a main-sequence prior on the stellar parameters and thirdly with a circular orbit. They derived values of eccentricity $e=0.129^{+0.106}_{-0.068}$ and $e=0.237^{+0.068}_{-0.069}$ for the first two cases respectively. We set $\sigma_r=0$ for both HARPS and CORALIE since we obtained a reduced $\chi^2$ of slightly less than unity for both eccentric and circular orbits for either dataset alone, indicating overfitting. We obtained a value of $\chi^2_c=37.98$ for the circular orbit and a value of $\chi^2_e= 35.94$ for the eccentric orbit. Using $41$ measurements and two priors from photometry ($N=43$), $k=3$ and $k=5$ for the circular (two datasets, each with one $V_0$ and a single $K$) and eccentric orbits respectively, we obtained BIC$_{c}=399.31$ and BIC$_{e}=404.80$.  We thus find that the circular orbit cannot be excluded, agreeing with the third case ($e=0$, fixed) considered in \cite{Anderson2010} and rejecting the two derived values of eccentricity in that paper.

\ \\
{\bf WASP-18}\\
WASP-18b is a $10.3$ M$_j$ planet on a $0.94$ day orbit around a F6 star (V=9.3), first reported by \cite{Hellier2009a}. Using $9$ CORALIE measurements (we drop the third measurement in our final analysis, since it produces a 5-$\sigma$ residual that is not improved by an eccentric orbit, suggesting that it is a genuine outlier), the authors derived a value of eccentricity $e=0.0092\pm0.0028$. In our study, we set $\tau=1.5$ d and $\sigma_r=20.15$ \ms\ to obtain a reduced $\chi^2$ of unity for the circular orbit. We obtained a value of $\chi^2_c=8.17$ for the circular orbit and a value of $\chi^2_e=6.64$ for the eccentric orbit ($e=0.007\pm0.005$, $e<0.018$). Using $N=10$, $k=2$ and $k=4$ for the circular (one dataset, with one $V_0$ and a single $K$) and eccentric orbits respectively, we obtained BIC$_{c}=75.34$ and BIC$_{e}=78.41$. We repeated the calculations using $\sigma_r=22.5$ \ms\ to obtain a reduced $\chi^2$ of unity for the eccentric orbit. We obtained a value of $\chi^2_c=7.14$ for the circular orbit and a value of $\chi^2_e=6.00$ for the eccentric orbit ($e=0.008\pm0.005$, $e<0.019$). Using $N=10$, $k=2$ and $k=4$ for the circular (one dataset, with one $V_0$ and a single $K$) and eccentric orbits respectively, we obtained BIC$_{c}=75.50$ and BIC$_{e}=78.97$.  We thus find that the circular orbit cannot be excluded, in contrast to \cite{Hellier2009a}. The possibility that $e>0.1$ is excluded.


\ \\
{\bf WASP-5 (new HARPS data) \label{sec:wasp5}}\\
WASP-5b is a 1.6 M$_j$ planet on a $1.63$ day orbit around a G4 star (V=12.3), first reported by \citet{Anderson2008}. \citet{Gillon2009} used z-band transit photometry from the VLT to refine the eccentricity to $e=0.038^{+0.026}_{-0.018}$, and the authors made a tentative claim for the detection of a small eccentricity. We analysed our 11 new HARPS measurements for WASP-5 and the 11 CORALIE RVs from \citet{Anderson2008} using the photometric constraints on the orbital period $P=1.6284246(13)$ and mid-transit time $T_{tr}=2454375.624956(24)$ from \citet[]{Southworth2009b}.

\begin{figure*}
	\resizebox{8cm}{!}{\includegraphics{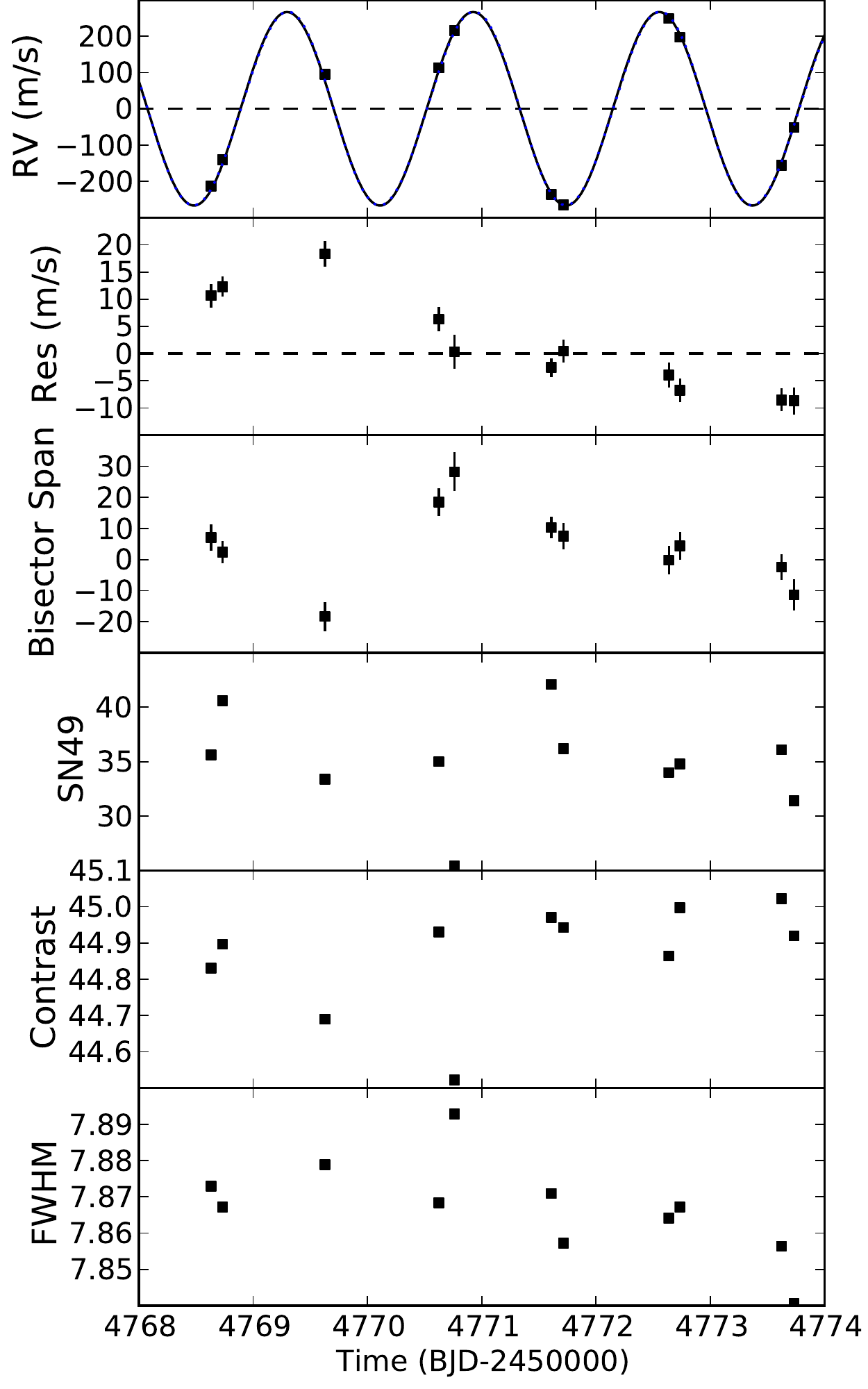}} 
	\resizebox{8cm}{!}{\includegraphics{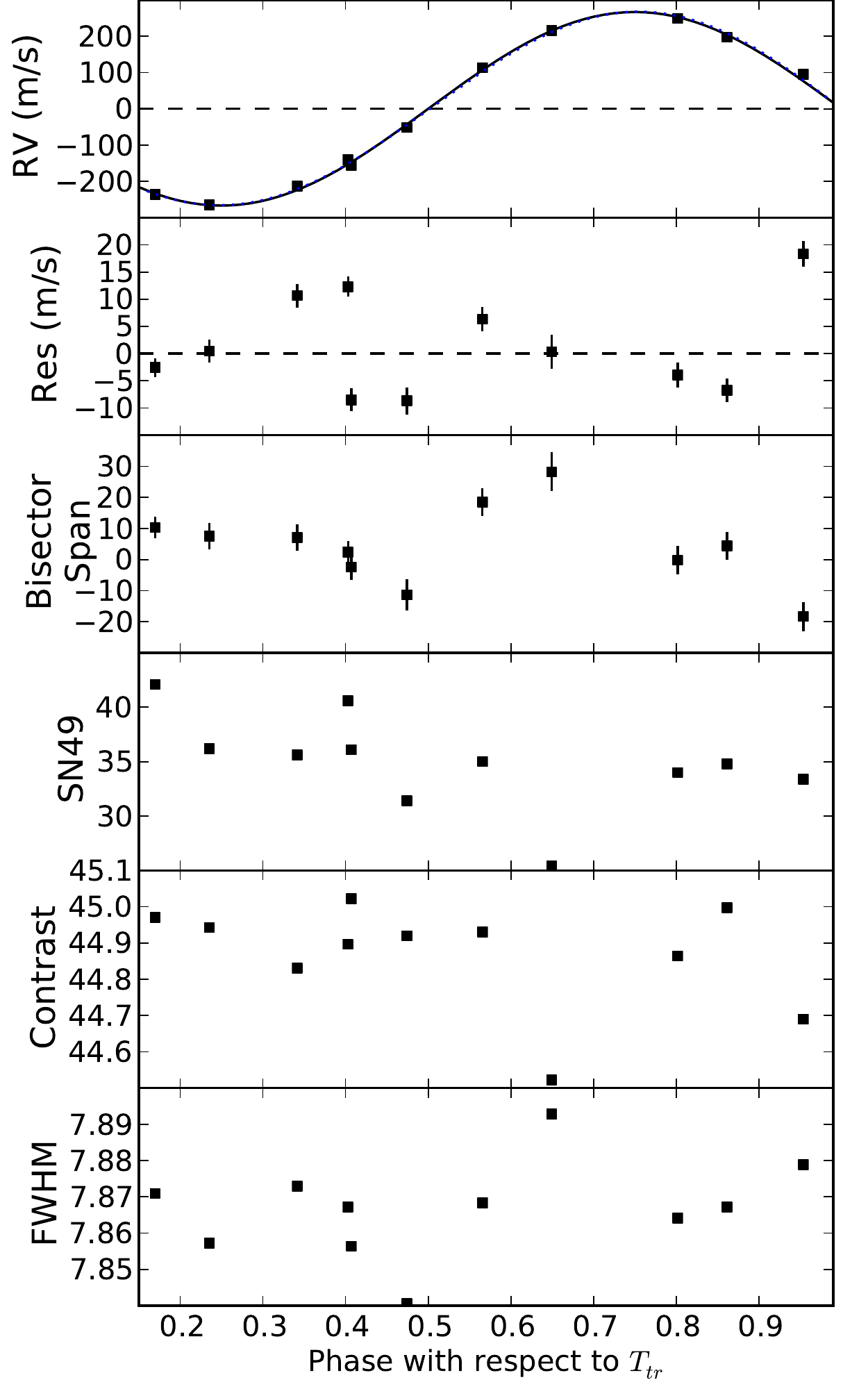}} 
	\caption{HARPS measurements of WASP-5 plotted against time ({\bf left}) and phase with respect to the mid-transit time $T_{tr}$ ({\bf right}). In each case, a solid line is overplotted to represent a circular orbit and the residuals are plotted for this circular orbit. It is clear that a signal is present in the residuals (see text). An eccentric orbit with the best-fit value of $e=0.012$ is overplotted in both panels with a dotted line, but it is indistinguishable from the circular solution at this scale. Note the trend that is apparent in the residuals (second panel from the top on both the time and phase plots). We correct for this using a linear acceleration term in our model (see Figure~\ref{fig:wasp5_circ_line}).}
	\label{fig:wasp5_circ_noline}
\end{figure*}

We use $\tau=1.5$ d, $\sigma_r=10.6$ \ms\ for HARPS and $\sigma_r=4.3$ \ms\ for CORALIE to obtain a value of reduced $\chi^2$ of unity for the circular orbit for each dataset separately. We ran the MCMC twice: the first time fitting for the systemic velocity $v_0$ and semi-amplitude $K$, and the second time adding two parameters $e\cos\omega$ and $e\sin\omega$ to allow for an eccentric orbit. The best fit result is shown in Figure~\ref{fig:wasp5_circ_noline}. The residuals for a circular orbit are plotted, and a signal is clearly present in the residuals. The value of $\chi^2$ for the circular orbit is $24.36$ and that for an eccentric orbit is $20.57$. This results in a value of BIC$_c=169.40$ for the circular orbit and BIC$_e=171.97$ for the eccentric orbit, given 22 measurements, 2 constraints from photometry and 3 and 5 free parameters respectively for each model.

We repeated the above analysis using $\tau=1.5$ d, $\sigma_r=9.4$ \ms\ for the HARPS dataset to obtain a value of reduced $\chi^2$ of unity for the eccentric orbit and $\sigma_r=0$ \ms\ for CORALIE (which resulted in a reduced $\chi^2$ of 0.58). This time, we obtained a value of $\chi^2$ for the circular orbit is $27.35$ and that for an eccentric orbit is $23.00$. This leads to a value of BIC$_c=170.37$ for the circular orbit and BIC$_e=172.38$ for the eccentric orbit. Once again, the circular orbit is favoured.

A keplerian model, circular or eccentric ($e=0.012\pm0.007$) does not account for the scatter in the data the HARPS dataset as shown in Figure~\ref{fig:wasp5_circ_noline}. We have therefore plotted the radial velocity measurements, the bisector span, the signal to noise at order 49, the contrast and full width at half maximum for the cross-correlation function against the same time axis. The trend in radial velocity residuals can be seen to be correlated with both the bisector span and the full width at half maximum of the cross correlation function. This suggests a line shape change that's related to either weather effects or instrumental systematics. The timescale of this variation is compatible with both scenarios. The bisector inverse span is generally directly correlated with the residuals, which weighs against a scenario involving stellar activity, but this is not so clear for the first three measurements --- the drift could be due to stellar activity or an additional planetary or stellar companion.

\begin{figure*}
	\resizebox{8cm}{!}{\includegraphics{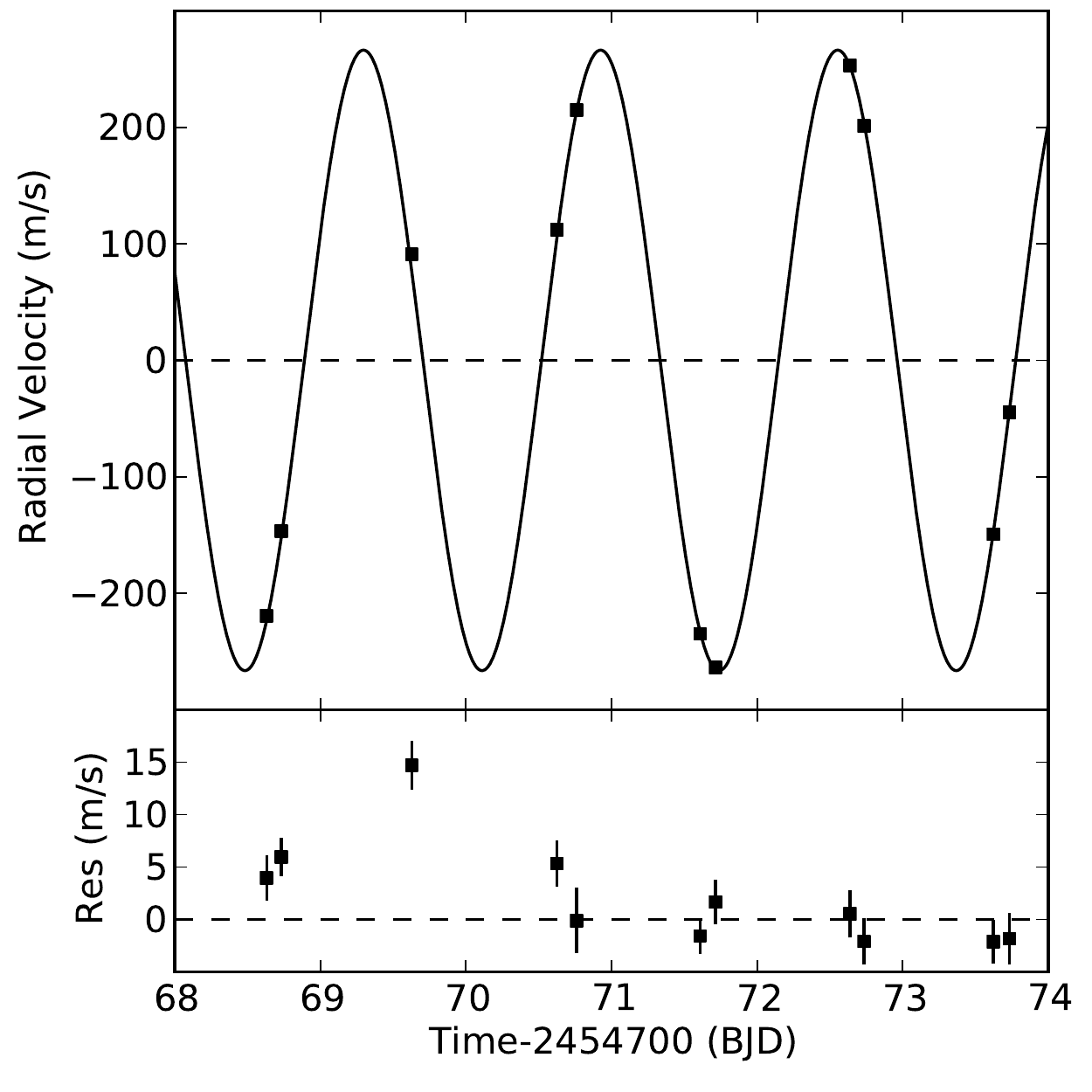}} 
	\resizebox{8cm}{!}{\includegraphics{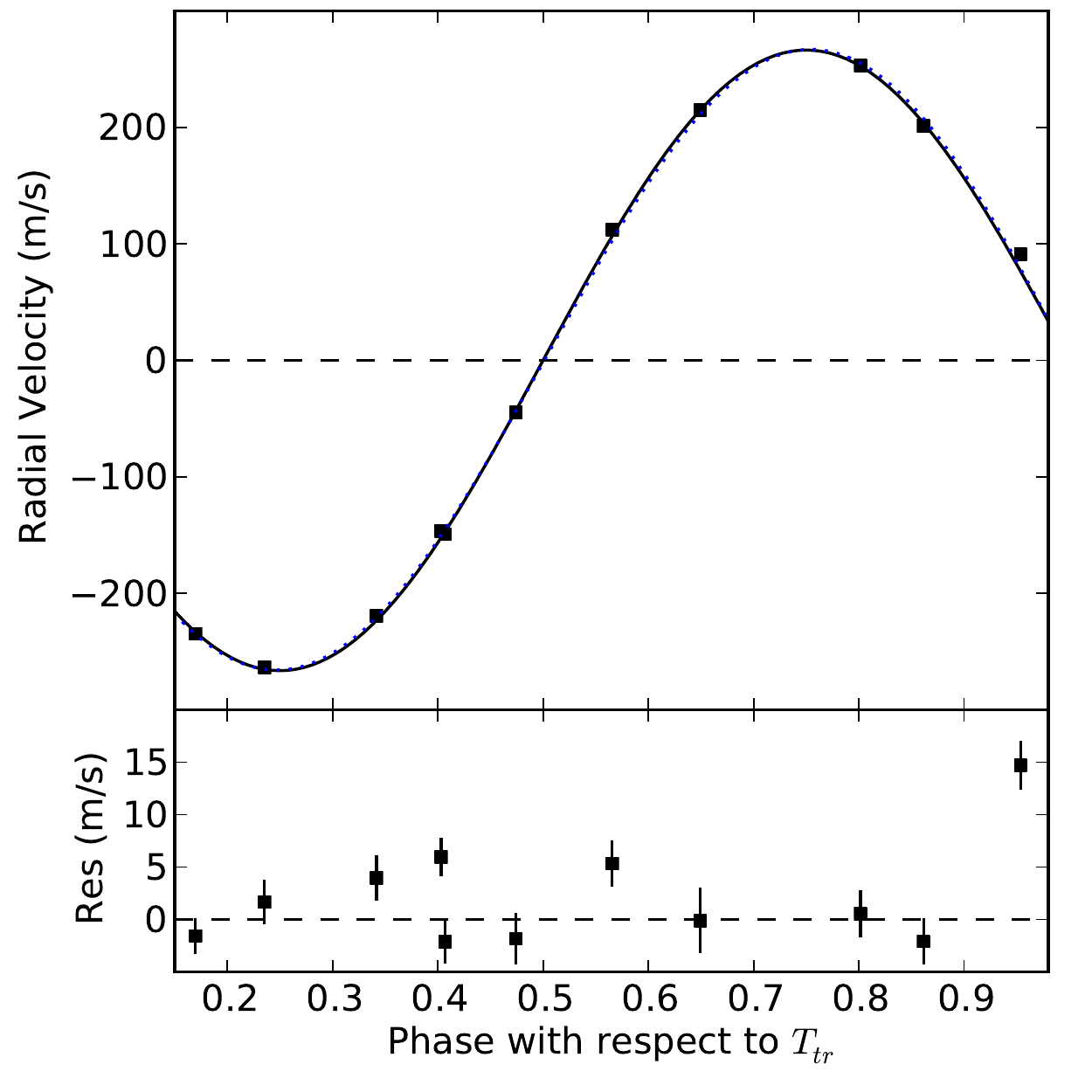}} 
	\caption{HARPS measurements of WASP-5 plotted against time ({\bf left}) and phase with respect to the mid-transit time $T_{tr}$ ({\bf right}). In each case, a solid line is overplotted to represent a circular orbit and the residuals are plotted for this circular orbit. It is clear that a signal is present in the residuals (see text). A model for an eccentric orbit with the best-fit value of $e=0.013$ is overplotted in both panels with a dotted line, but it is indistinguishable from the circular solution at this scale. Both include the linear trend (see Section~\ref{sec:wasp4} and Figure~\ref{fig:wasp5_circ_noline}).}
	\label{fig:wasp5_circ_line}
\end{figure*}

We extended the model with a linear acceleration of the form
 \begin{equation}
v(t) = v_{\rm keplerian}(t) + \dot{\gamma}(t-t_0),
\label{eqn:lintrend}
\end{equation}

and fitted the HARPS data alone using $t_0=2454768$ (to allow the MCMC to explore values of $\dot{\gamma}$ more efficiently) and reran the MCMC twice: once for a circular orbit and once for an eccentric orbit. Firstly, we used $\sigma_r=10.6$ \ms\ for the HARPS dataset, and the linear trend for a circular orbit resulted in $\dot{\gamma}=-2.6\pm2.9$ \ms\ yr$^{-1}$ and that for an eccentric orbit is $\dot{\gamma}=-2.0\pm2.9$ \ms\ yr$^{-1}$. The best fit result is shown in Figure~\ref{fig:wasp5_circ_line} and the residuals for a circular orbit are plotted in the bottom panel. The value of $\chi^2$ for the circular orbit is $10.24$ and that for an eccentric orbit is $7.70$. This results in a value of BIC$_{c, lin}=71.91$ for the circular orbit and BIC$_{e, lin}=74.49$ for the eccentric orbit, given 11 (N=13) measurements, 2 constraints from photometry and 3 and 5 free parameters respectively for each model. We repeated these calculations using $\sigma_r=9.4$ \ms\ for the HARPS dataset, and the linear trend for a circular orbit resulted in $\dot{\gamma}=-3.7\pm1.3$ \ms\ yr$^{-1}$ and that for an eccentric orbit is $\dot{\gamma}=-3.3\pm1.3$ \ms\ yr$^{-1}$. The value of $\chi^2$ for the circular orbit is $15.46$ and that for an eccentric orbit is $13.50$. This leads to a value of BIC$_{c, lin}=69.72$ for the circular orbit and BIC$_{e, lin}=72.89$ for the eccentric orbit. The circular orbit is not excluded, and the possibility that $e>0.1$ is excluded. The results for both models, one including the linear trend but excluding the CORALIE data, and one including the CORALIE data but excluding the linear trend are shown in Table~\ref{tab:wasp5}. In both cases, we give results for the case where $\sigma_r$ is chosen to yield a reduced $\chi^2$ of unity for the circular orbit. We attempted to repeat this using both the CORALIE and HARPS datasets, but we were unable to obtain a fit with the MCMC, because of the long time scale between the two datasets.

\begin{table*}
\centering
\begin{tabular}{l c c  c }
\hline
Parameter	& \citet{Anderson2008}					&	HARPS only, {\it this work}  &	HARPS \& CORALIE, {\it this work}	\\
			& 									&	(with linear trend)		 &  (no linear trend) \\ \hline
Centre-of-mass velocity $V_0$ [\ms] 	& 20010.5$\pm$3.4 	& 20018$\pm$12  		 & 20009.9$\pm$7.4 (HARPS)  	\\
Orbital eccentricity $e$ 				& 0 (adopted)			& $0.013\pm0.008$ ($<0.029$)& $0.012\pm 0.007$ ($<0.026$)\\
Argument of periastron $\omega$ [$^o$]& 0 (unconstrained)	& 0 (unconstrained)			 & 0 (unconstrained)			\\ 
$e\cos\omega$	& 				--					&  0.002$\pm$0.003 		&  0.003$\pm$0.003 \\
$e\sin\omega$		&  				--					& 0.012$\pm$0.010		& 0.011$\pm$0.009		\\
Velocity semi-amplitude K [m$\,$s$^{-1}$]		& 277.8$\pm$7.8	& 266.4$\pm$1.3		& 266.9$\pm$1.3			\\
\hline
\end{tabular}
\caption{System parameters for WASP-5. Left: \citet{Anderson2008}. Right: Results from our HARPS radial velocity data alone, and results from using both our HARPS data and the original CORALIE data in \citet{Anderson2008}. Median values for $V_0$ and $K$ are quoted for the circular orbits, as well as 68.3\% confidence limits obtained from the eccentric solution (see section Analysis).}
\label{tab:wasp5}
\end{table*}


\subsection{Planets on circular orbits}
\label{sec:circular}
We establish that 20 planets have orbital eccentricities compatible with zero and the 95\% upper limits are smaller than $e_{95}=0.1$. In this Section, we describe the planets WASP-4b, HAT-P-7b, TrES-2 and WASP-2b, for which we introduce new RVs. We also establish that the 95\% upper limits on the eccentricities of WASP-5b, WASP-12b and WASP-18b, which have been described in Section~\ref{sec:noteccentric} above. In addition, we give the 95\% upper limits on the eccentricities of CoRoT-1b, CoRoT-3b, HAT-P-8b, WASP-3b, WASP-16b, WASP-19b, WASP-22b, WASP-26b and XO-5b in Table~\ref{tab:results}. We discuss the evidence for circular orbits for HAT-P-13b, HD189733b, HD209458b and Kepler-5b at the end of this section.

\ \\
{\bf WASP-4 (new HARPS data) \label{sec:wasp4}}\\
WASP-4b is a 1.2 M$_j$ planet on a $1.34$ day orbit around a G7 star (V=12.5), first reported by \citet{Wilson2008}. We analysed our 14 new HARPS measurements and the 14 CORALIE measurements from \citet{Wilson2008} for WASP-4 and used the photometric constraints on the orbital period $P=1.33823214(71)$ and mid-transit time $T_{tr}=2454697.797562(43)$ from \citet{Winn2009a}.

\begin{figure*}
	\resizebox{8cm}{!}{\includegraphics{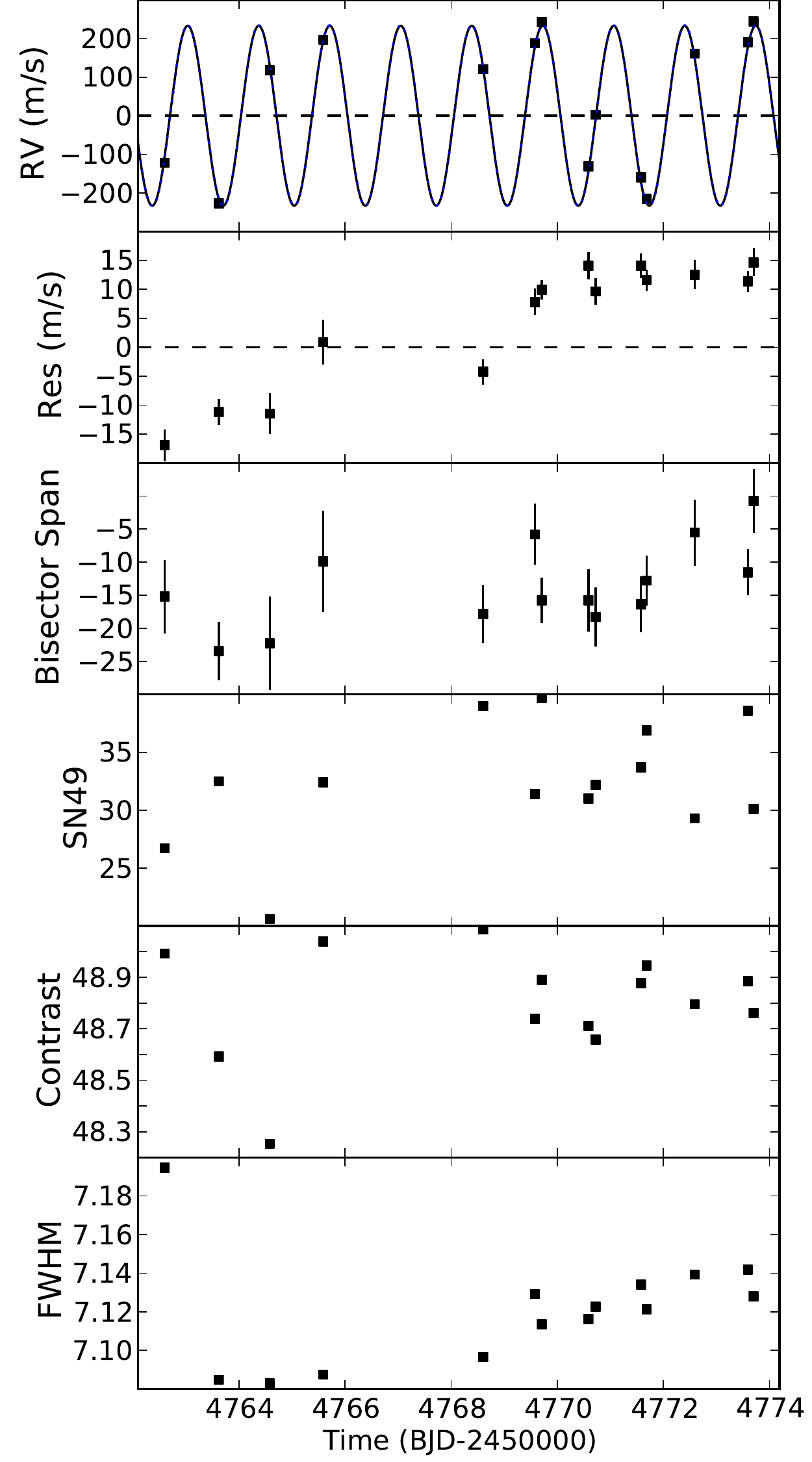}}
	\resizebox{8cm}{!}{\includegraphics{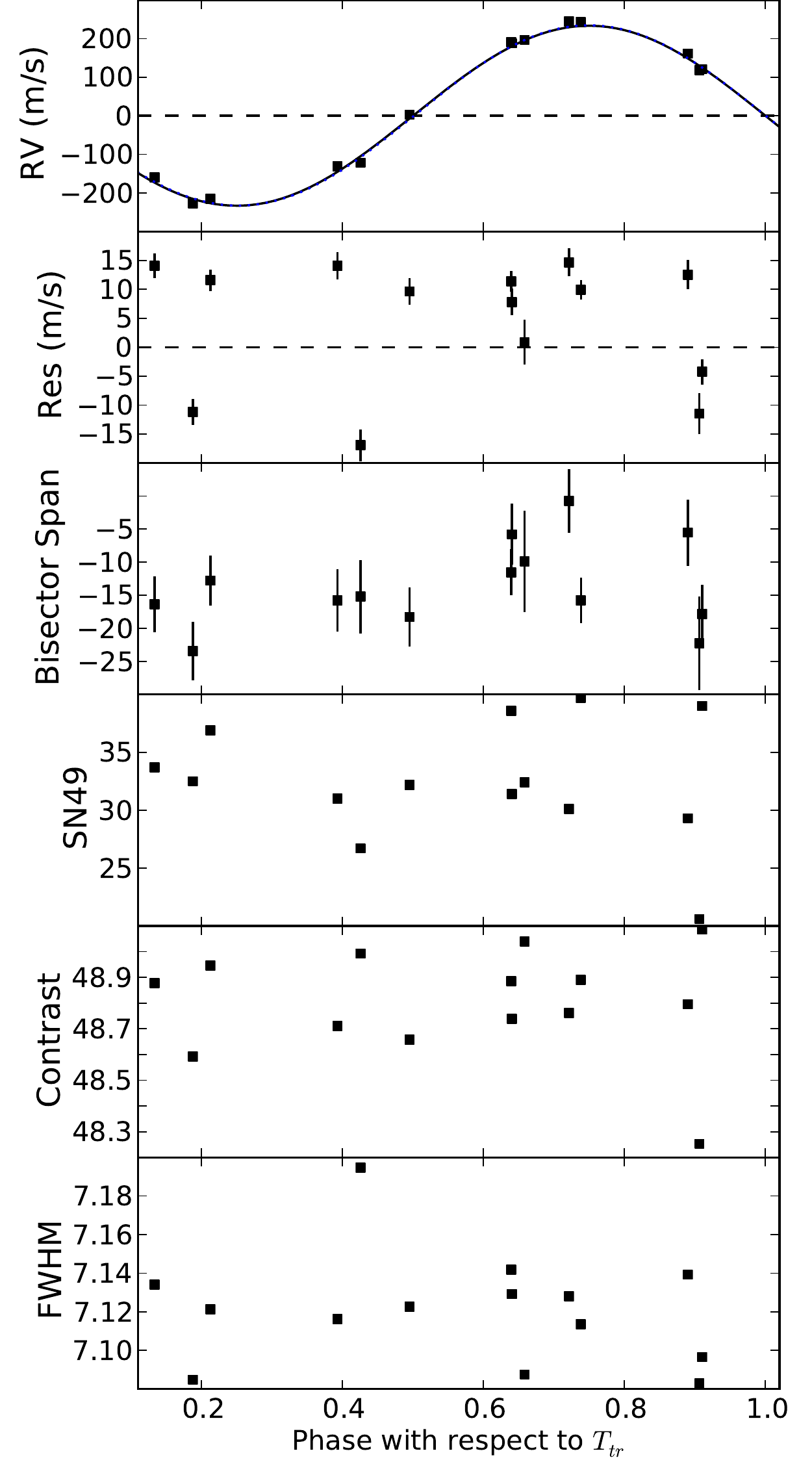}}
	\caption{HARPS measurements of WASP-4 plotted against time ({\bf left}) and phase with respect to the mid-transit time $T_{tr}$ ({\bf right}). In each case, a solid line is overplotted to represent a circular orbit and the residuals are plotted for this circular orbit. It is clear that a signal is present in the residuals (see text). An eccentric orbit with the best-fit value of $e=0.005$ is overplotted in both panels with a dotted line, but it is indistinguishable from the circular solution at this scale. Note the trend that is apparent in the residuals (second panel from the top on both images). We attempt to correct for this by repeating our calculations with a linear acceleration term in the model (see Section~\ref{sec:wasp4} and Figure~\ref{fig:wasp4_circ_line}).}
\label{fig:wasp4_circ_noline}
\end{figure*}

We estimate $\tau=1.5$ d, $\sigma_r=11$ \ms\ for the HARPS dataset and $\sigma_r=4.5$ \ms\ for the CORALIE dataset to obtain a reduced $\chi^2$ of unity for a circular orbit for each dataset separately. We ran the MCMC twice: the first time fitting for the systemic velocity $v_0$ and semi-amplitude $K$ only, ie. a circular orbit ($k=2$), and the second time adding two parameters $e\cos\omega$ and $e\sin\omega$ to allow for an eccentric orbit ($k=4$). The best fit result is shown in Figure~\ref{fig:wasp4_circ_noline}. The residuals for a circular orbit are plotted, and a signal is clearly present in the residuals. The value of $\chi^2$ for the circular orbit is $27.13$ and that for an eccentric orbit is $24.32$. This leads to a value of BIC$_c=208.62$ for the circular orbit and BIC$_e=212.55$ for the eccentric orbit, given 14 measurements, 2 constraints from photometry and 2 and 4 free parameters respectively for each model. 

We repeated the calculations, estimating $\tau=1.5$ d, $\sigma_r=10.1$ \ms\ for the HARPS dataset and $\sigma_r=7.1$ \ms\ for the CORALIE dataset to obtain a reduced $\chi^2$ of unity for an eccentric orbit for each dataset separately. The value of $\chi^2$ for the circular orbit is $27.29$ and that for an eccentric orbit is $24.33$. This leads to a value of BIC$_c=208.72$ for the circular orbit and BIC$_e=212.51$.

Note the trend that is apparent in the residuals in Figure~\ref{fig:wasp4_circ_noline}. We have therefore plotted the radial velocity measurements, the bisector span, the signal to noise at order 49, the contrast and full width at half maximum for the cross-correlation function against the same time axis. For most measurements, the trend in radial velocity residuals can be seen to be correlated with both the bisector span and the full width at half maximum of the cross correlation function. This suggests a line shape change that's related to either stellar activity, weather effects or instrumental systematics. The timescale of this variation is compatible with all three scenarios.

We repeated the calculations for the HARPS dataset alone, and added a linear component to the radial velocity model in the same way we did for WASP-5 in Section~\ref{sec:noteccentric} and we set $t_0=2454762$ (to allow the MCMC to explore values of $\dot{\gamma}$ more efficiently) and reran the MCMC twice: once for a circular orbit and once for an eccentric orbit. We set $\tau=1.5$ d and $\sigma_r=11$ \ms\ for the HARPS dataset.

\begin{figure*}
	\resizebox{8cm}{!}{\includegraphics{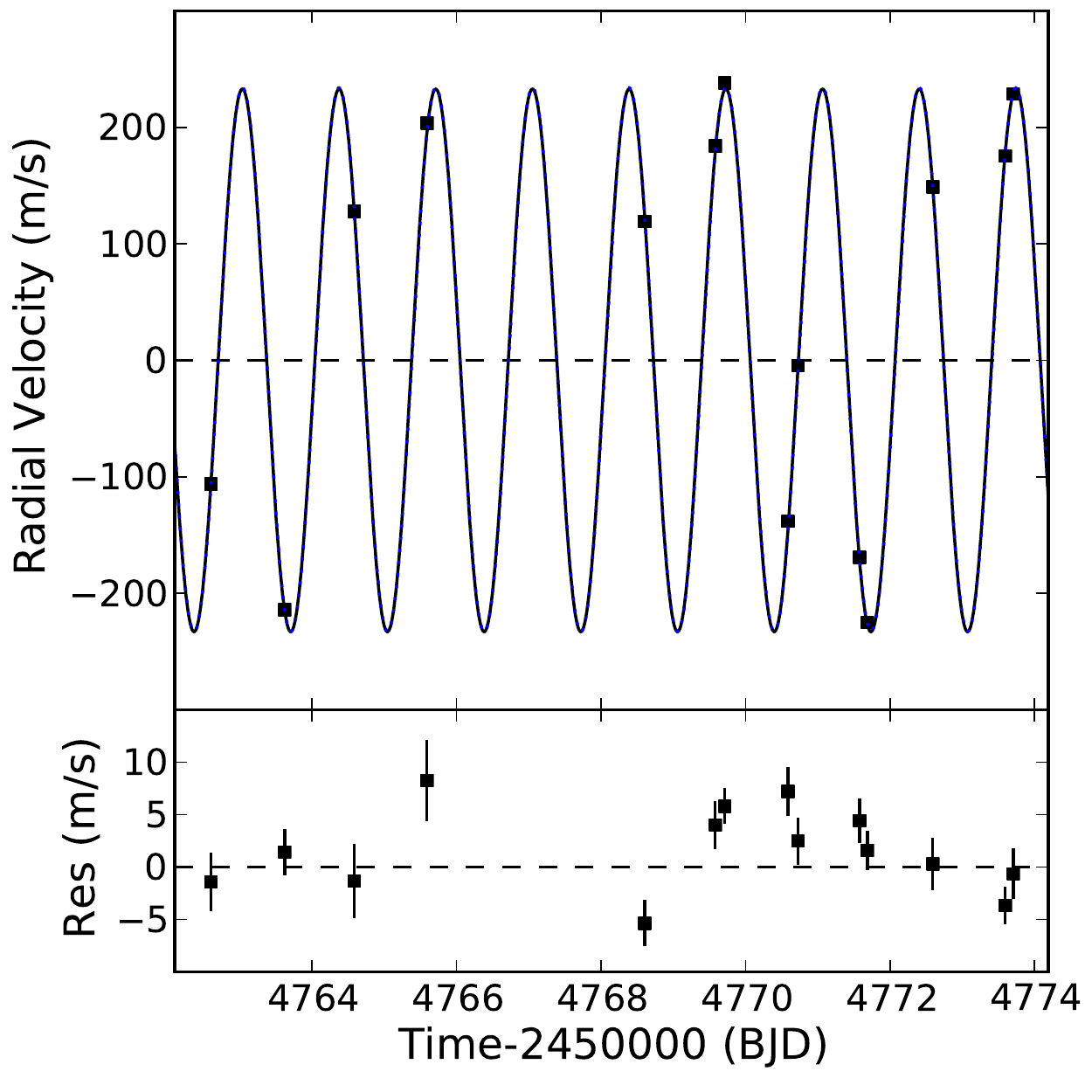}}
	\resizebox{8cm}{!}{\includegraphics{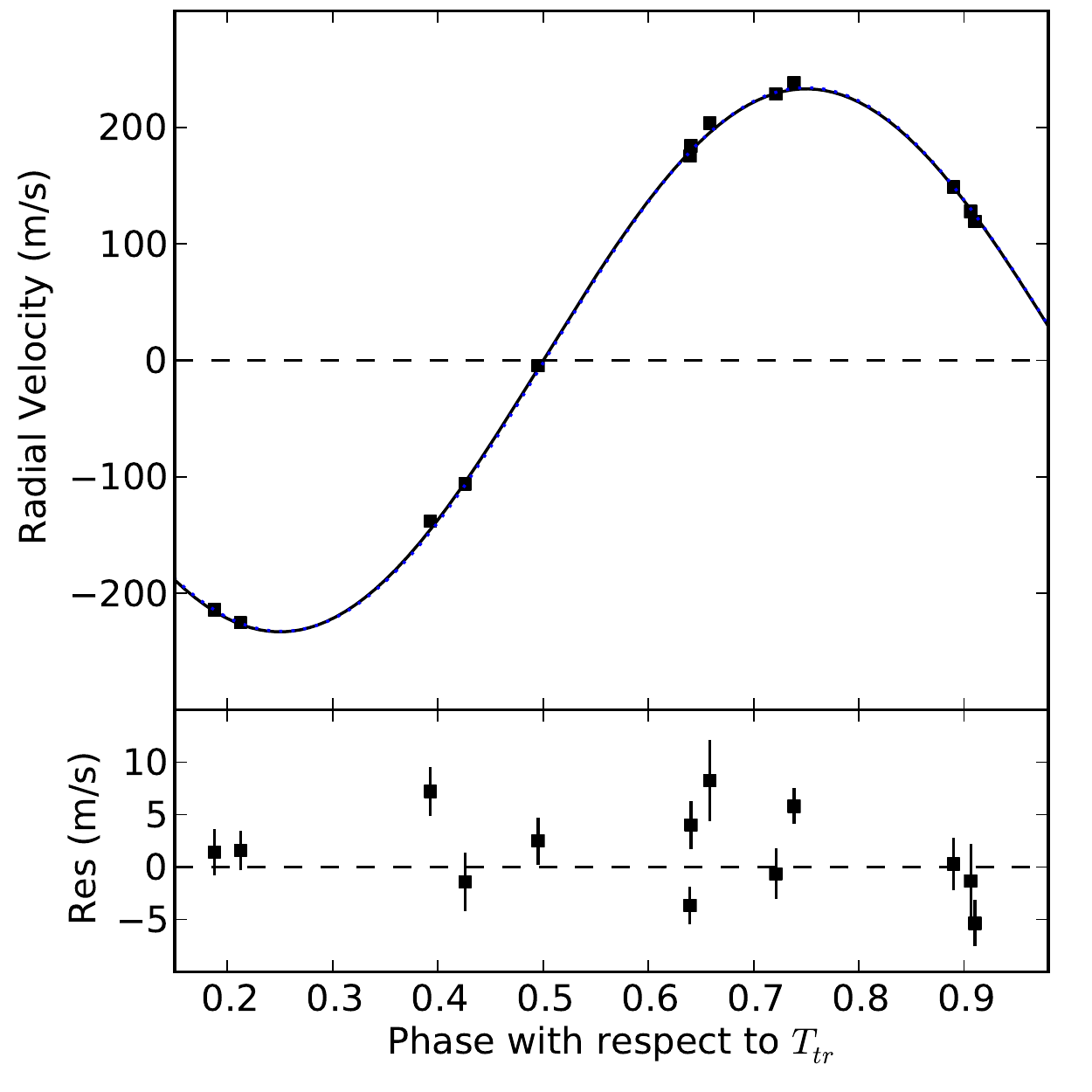}}
	\caption{HARPS measurements of WASP-4 plotted against time ({\bf left}) and phase with respect to the mid-transit time $T_{tr}$ ({\bf right}). In each case, a solid line is overplotted to represent a circular orbit and the residuals are plotted for this circular orbit. It is clear that a signal is present in the residuals (see text). An eccentric orbit with the best-fit value of $e=0.004$ is overplotted in both panels with a dotted line, but it is indistinguishable from the circular solution at this scale. The linear trend from Figure~\ref{fig:wasp4_circ_noline} has now been included in the model.}
\label{fig:wasp4_circ_line}
\end{figure*}

The best fit result is shown in Figure~\ref{fig:wasp4_circ_line}. The residuals for a circular orbit are plotted, and a signal is clearly present in the residuals. The linear trend for a circular orbit results in $\dot{\gamma}=1023\pm490$ \ms\ yr$^{-1}$ and that for an eccentric orbit is $\dot{\gamma}=919\pm500$ \ms\ yr$^{-1}$.

The value of $\chi^2$ for the circular orbit is $9.83$ and that for an eccentric orbit is $7.51$. This leads to a value of BIC$_c=92.02$ for the circular orbit and BIC$_e=95.26$ for the eccentric orbit, given 14 measurements, 2 constraints from photometry and 3 and 5 free parameters respectively for each model.

We repeated the calculations, setting $\tau=1.5$ d, $\sigma_r=10.05$ \ms\ for the HARPS dataset. The value of $\chi^2$ for the circular orbit is $10.30$ and that for an eccentric orbit is $8.07$. This leads to a value of BIC$_c=91.18$ for the circular orbit and BIC$_e=94.49$.  In all cases, the circular orbit is not excluded. The results for both models, one including the linear trend but excluding the CORALIE data, and one including the CORALIE data but excluding the linear trend are shown in Table~\ref{tab:wasp4}. In both cases, we give results for the case where $\sigma_r$ is chosen to yield a reduced $\chi^2$ of unity for the circular orbit. We reject the possibility that $e>0.1$.

\begin{figure*}
\resizebox{8cm}{!}{\includegraphics{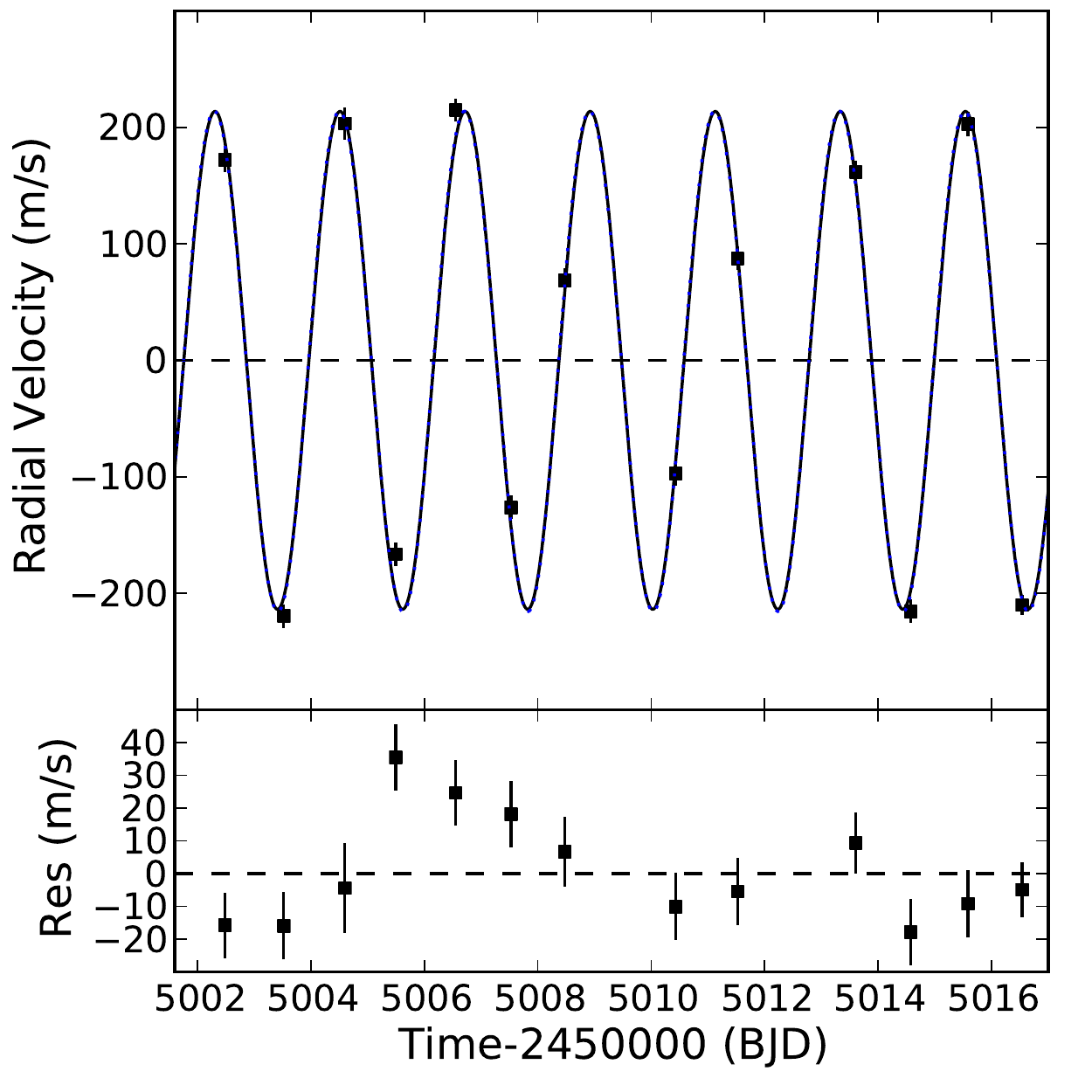}}
\resizebox{8cm}{!}{\includegraphics{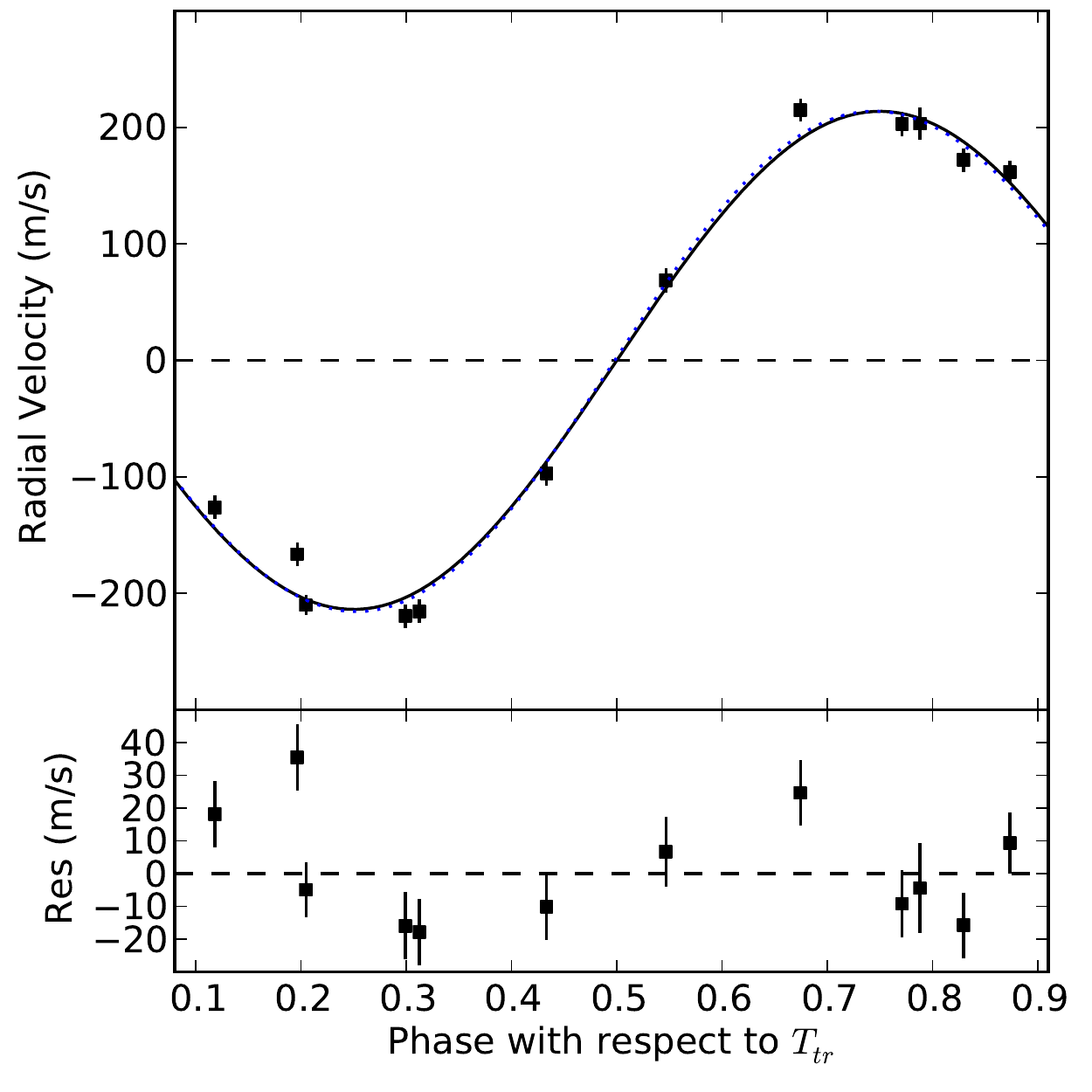}}
\caption{Plot showing our new SOPHIE radial velocity data for HAT-P-7, plotted against time (left), and orbital phase (right) with respect to $T_{tr}$. A circular orbit (solid line) and an orbit with the best-fit eccentricity (dotted line, but almost undistinguishable from the circular solution since $e=0.014$) are overplotted. The residuals relative to the circular orbit are shown in the bottom panels.}
\label{fig:hatp7sophie}
\end{figure*}

\begin{table*}
\centering
\begin{tabular}{l c  c c }
\hline
Parameter	& \citet{Wilson2008}					&	HARPS only, {\it this work}  &	HARPS \& CORALIE, {\it this work}	\\
			& 									&	(with linear trend)				&  (no linear trend) \\ \hline

Centre-of-mass velocity $V_0$ [\ms] 	& 57733$\pm$2 		& 57773$\pm$10 				& 57790.8$\pm$5.7 	\\
Orbital eccentricity $e$ 				& 0 (adopted)			&  $0.004\pm0.003$ ($<$0.011)	&  $0.005\pm0.003$ ($<$0.011)\\
Argument of periastron $\omega$ [$^o$]& 0 (unconstrained)	& 0 (unconstrained)				& 0 (unconstrained) \\ 
$e\cos\omega$	& 				--					& 0.004$\pm$0.003				&  0.003$\pm$0.003	\\
$e\sin\omega$		&  				--				& $-$0.002$\pm$0.004			& $-$0.004$\pm$0.004 \\
Velocity semi-amplitude K [m$\,$s$^{-1}$]		& 240$\pm$10	& 233.1$\pm$2.1				& 233.7$\pm$2.0	\\
\hline
\end{tabular}
\caption{System parameters for WASP-4. Left: \citet{Wilson2008}. Right: Results from our HARPS radial velocity data alone, and results from using both our HARPS data and the original CORALIE data in \citet{Wilson2008}. Median values for $V_0$ and $K$ are quoted for the circular orbits, as well as 68.3\% confidence limits obtained from the eccentric solution (see section Analysis).}
\label{tab:wasp4}
\end{table*}



\ \\
{\bf HAT-P-7 (new SOPHIE data)}\\
HAT-P-7b is a $1.8$ M$_j$ planet on a $2.20$ day orbit around an F6 star (V=10.5), first reported by \cite{Pal2008}. We use 13 new SOPHIE radial velocity measurements and 16 out of the 17 HIRES measurements in  \citet{Winn2009b} (we drop one in-transit measurement) to work out the orbital parameters of HAT-P-7b. We impose the period $P=2.204733(10)$ d as given from photometry in \citet{Welsh2010} and mid-transit time $T_{tr}=2454731.67929(43)$ BJD as given from photometry in \citet{Winn2009b}. We set $\tau=1.5$ d, $\sigma_r=9.41$ \ms\ for HIRES 
and $\sigma_r=12.9$ \ms\ for SOPHIE to obtain a reduced $\chi^2$ of unity for the best-fit circular orbit for each dataset separately. We used $29$ measurements in all, and count the two constraints from photometry as two additional data points to obtain $N=31$, and used $k=4$ for the circular orbit (two $V_0$, one for each dataset, the semi-amplitude $K$ and a constant drift term $\dot{\gamma}$, since \citet{Winn2009b} found evidence for a distant companion in the system and we set $t_0=2454342$). We repeated this analysis with an eccentric orbit $k=6$ (4 degrees of freedom for the circular orbit with two datasets and a linear acceleration, and 2 additional degrees of freedom for the eccentricity, $e\cos\omega$ and $e\sin\omega$). The orbital parameters are given in Table \ref{tab:hatp7}, and the radial velocity dataset is plotted in Figure \ref{fig:hatp7sophie}, with residuals shown for a circular orbit. The Figure also shows models of a circular and an eccentric orbit (with $e=0.014$), but they are almost undistinguishable. For the circular orbit, we obtained $\chi^2=26.94$, and a value of BIC$_c=222.81$ and for the eccentric orbit, we obtained $\chi^2=23.98$ and a value of BIC$_e=226.72$. We repeated the calculations and set $\tau=1.5$ d, $\sigma_r=8.2$ \ms\ for HIRES and $\sigma_r=8.2$ \ms\ for SOPHIE to obtain a reduced $\chi^2$ of unity for the best-fit eccentric orbit. For the circular orbit, we obtained $\chi^2=35.65$ and a value of BIC$_c=224.14$ and for the eccentric orbit, we obtained $\chi^2=31.89$ and a value of BIC$_e=227.25$. We therefore find that the circular orbit cannot be excluded for HAT-P-7b. Further, we exclude the possibility that $e>0.1$.

\begin{table*}
\centering
\begin{tabular}{l c  c }
\hline
Parameter	& HIRES, \citet{Winn2009b}					&	HIRES+SOPHIE, {\it this work}	\\ \hline
Centre-of-mass velocity $V_0$ [\ms] & $-$51.2$\pm$3.6		& $-$49.96$\pm$6.0 (HIRES) and $-$10510$\pm$10 (SOPHIE)  	\\
Orbital eccentricity $e$ 				& $e_{99\%}<$0.039		&  0.014$\pm0.010$ ($e<0.038$)\\
Argument of periastron $\omega$ [$^o$]& --					& 0 (unconstrained)			\\ 
$e\cos\omega$	& 			$-$0.0019$\pm$0.0077		& $-$0.007$\pm$0.004	\\
$e\sin\omega$		&  		0.0037$\pm$0.0124		& $-$0.011$\pm$0.015	\\
Velocity semi-amplitude $K$ [m$\,$s$^{-1}$]	& 211.8$\pm$2.6	& 213.8$\pm$1.2	\\
Constant radial acceleration $\dot{\gamma}$ [m$\,$s$^{-1}$yr$^{-1}$]	& 21.5$\pm$2.6 & 21.1$\pm$4.2\\
\hline
\end{tabular}
\caption{System parameters for HAT-P-7. Left: \citet{Winn2009b}. Right: Results from our SOPHIE radial velocity data. Median values for $V_0$ and $K$ are quoted for the circular orbits, as well as 68.3\% confidence limits obtained from the eccentric solution. The upper 95\% limit is also given for the eccentricity from our analysis.}
\label{tab:hatp7}
\end{table*}

\ \\
{\bf TrES-2 (new SOPHIE data)}\\
TrES-2b is a $1.3$ M$_j$ planet on a $2.47$ day orbit around a G0 star (V=11.4), first reported by \cite{ODonovan2006}. We use 10 new SOPHIE radial velocity measurements and the 11 HIRES measurements in \cite{ODonovan2006} to work out the orbital parameters of TrES-2b. We impose the period $P=2.470614(1)$ d and mid-transit time $T_{tr}=2453957.63492(13)$ BJD as given from photometry in \citet{Raetz2009}.

\begin{figure*}
\resizebox{8cm}{!}{\includegraphics{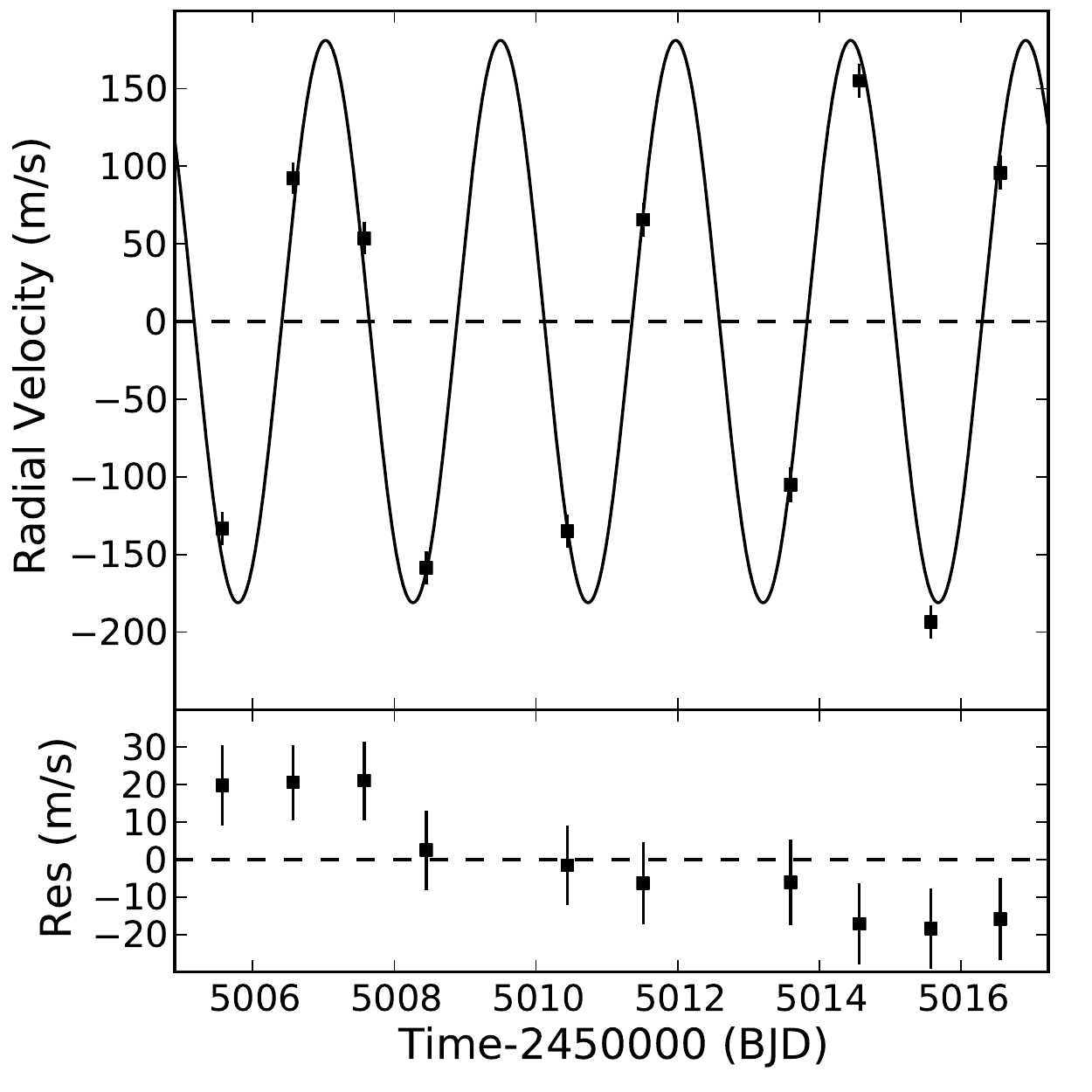}}
\resizebox{8cm}{!}{\includegraphics{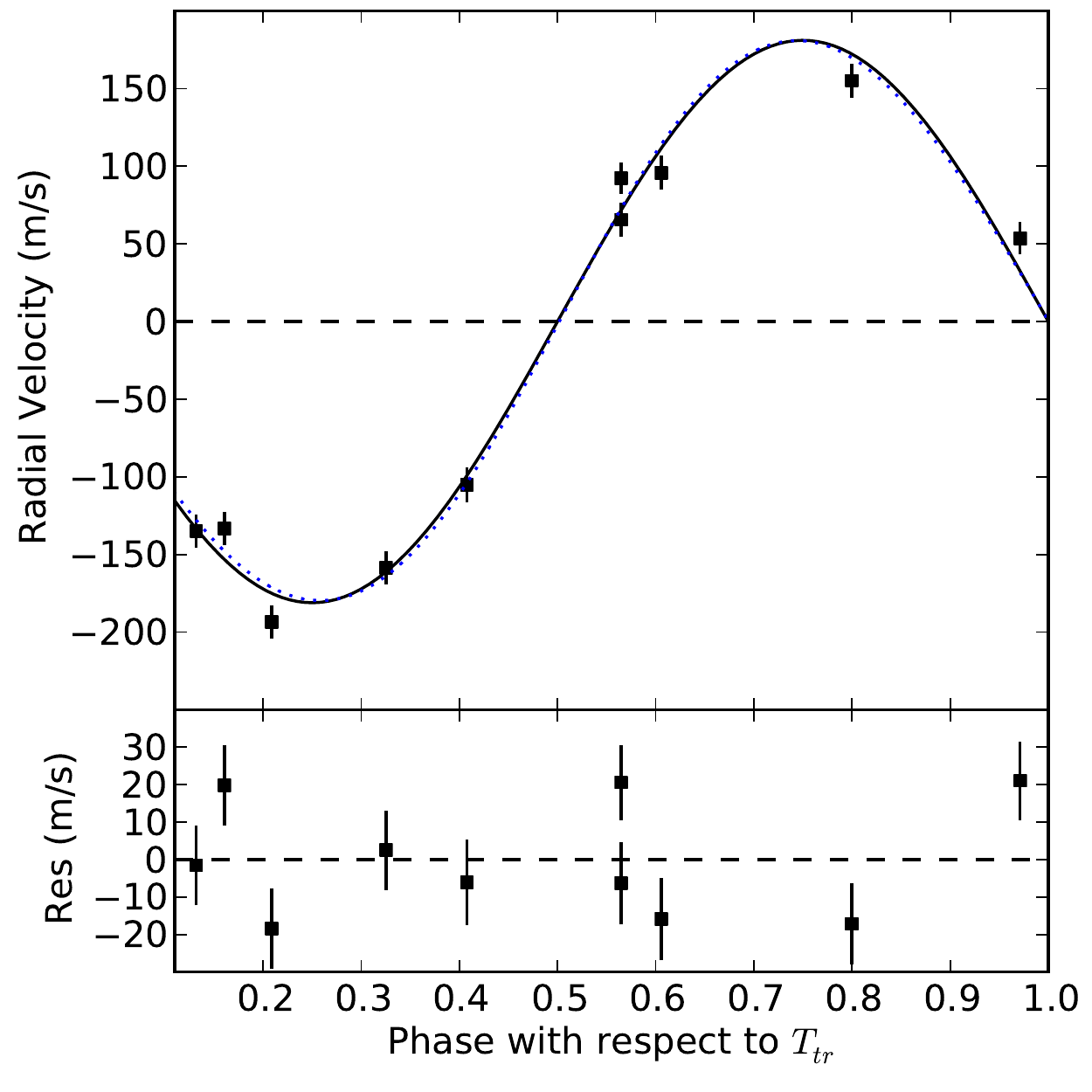}}
\caption{Plot showing our new SOPHIE radial velocity data for TrES-2, plotted against time (left) and orbital phase with respect to $T_{tr}$ (right). A circular orbit (solid line) and an orbit with the best-fit eccentricity (dotted line, but almost undistinguishable from the circular solution since $e=0.023$) are overplotted. The residuals relative to the circular orbit are shown in the bottom panel.}
\label{fig:tres2}
\end{figure*}

We set $\tau=1.5$ d and $\sigma_r=6.8$ \ms\ for SOPHIE to obtain a reduced $\chi^2$ of unity for the best-fit circular orbit (using the SOPHIE data alone), and set $\sigma_r=0$\ms\ for the HIRES data since a circular orbit for that dataset alone yields a reduced $\chi^2$ of 0.72, indicating over-fitting. We used $21$ measurements in all, and count the two constraints from photometry as two additional datapoints to obtain $N=23$, and used $k=3$ for the circular orbit (two $V_0$, one for each dataset, and the semi-amplitude $K$). We repeated this analysis with an eccentric orbit $k=5$ (three degrees of freedom for the circular orbit, and two additional degrees of freedom for the eccentricity, $e\cos\omega$ and $e\sin\omega$). The orbital parameters are given in Table \ref{tab:tres2}, and the radial velocity dataset is plotted in Figure \ref{fig:tres2}, with residuals shown for a circular orbit. The Figure also shows models of a circular and an eccentric orbit (with $e=0.023$), but they are almost undistinguishable. For the circular orbit, we obtained $\chi^2=18.00$, yielding a value of BIC$_c=160.30$ and for the eccentric orbit, we obtained $\chi^2=15.91$ and a value of BIC$_e=164.48$. We repeated the calculations and set $\sigma_r=8.45$ \ms\ for SOPHIE to obtain a reduced $\chi^2$ of unity for the best-fit circular orbit (using the SOPHIE data alone), while we set $\sigma_r=0$\ms\ for the HIRES data since an eccentric orbit for that dataset alone yields a reduced $\chi^2$ of 0.56, indicating over-fitting. For a circular orbit, we obtained $\chi^2=15.97$, resulting in a value of BIC$_c=159.38$ and for an eccentric orbit, we obtained $\chi^2=13.88$ and a value of BIC$_e=163.56$. We therefore find that the circular orbit cannot be excluded for TrES-2b. Furthermore, we exclude the possibility that $e>0.1$.

\begin{table*}
\centering
\begin{tabular}{l c  c }
\hline
Parameter	& HIRES, \citet{ODonovan2006}					&	HIRES, SOPHIE, {\it this work}	\\ \hline
Centre-of-mass velocity $V_0$ [\ms] 	& -- 						& $-$29.8$\pm$2.4 (HIRES), $-$315.5$\pm$5.0 (SOPHIE) 	\\
Orbital eccentricity $e$ 				& 0 (adopted)				&  0.023$\pm$0.014, $e<0.051$\\
Argument of periastron $\omega$ [$^o$]& 0 (unconstrained)		& 0 (unconstrained)			\\ 
$e\cos\omega$	& 				--						& 0.002$\pm$0.009	\\
$e\sin\omega$		&  				--					& $-$0.022$\pm$0.016	\\
Velocity semi-amplitude K [m$\,$s$^{-1}$]		& 181.3$\pm$2.6	& 181.1$\pm$2.5			\\
\hline
\end{tabular}
\caption{System parameters for TrES-2. Left: \citet{ODonovan2006}. Right: Results from our HARPS radial velocity data. Median values for $V_0$ and $K$ are quoted for the circular orbits, as well as 68.3\% confidence limits obtained from the eccentric solution. The 95\% limit on the eccentricity is also given.}
\label{tab:tres2}
\end{table*}

\ \\
{\bf WASP-2 (new HARPS data)}\\
WASP-2b is a $0.85$ M$_j$ planet on a $2.15$ day orbit around a K1 star (V=12), first reported by \cite{Cameron2007}. We use 8 new HARPS radial velocity measurements and 7 of the original 9 SOPHIE measurements (we drop the first measurement, which has an uncertainty about 15 times larger than the rest, and the fifth, which shows a 3-$\sigma$  deviation at a phase close to the transit) in \cite{Cameron2007} to work out the orbital parameters of WASP-2b. We impose the period $P=2.15222144(39)$ d and mid-transit time $T_{tr}=2453991.51455(17)$ BJD as given from photometry in \citet{Southworth2010}. We used $15$ measurements in all, and count the two constraints from photometry as two additional datapoints ($N=17$) and used $k=3$ for the circular orbit (two $V_0$, one for each dataset, and the semi-amplitude $K$).

\begin{figure*}
\resizebox{8cm}{!}{\includegraphics{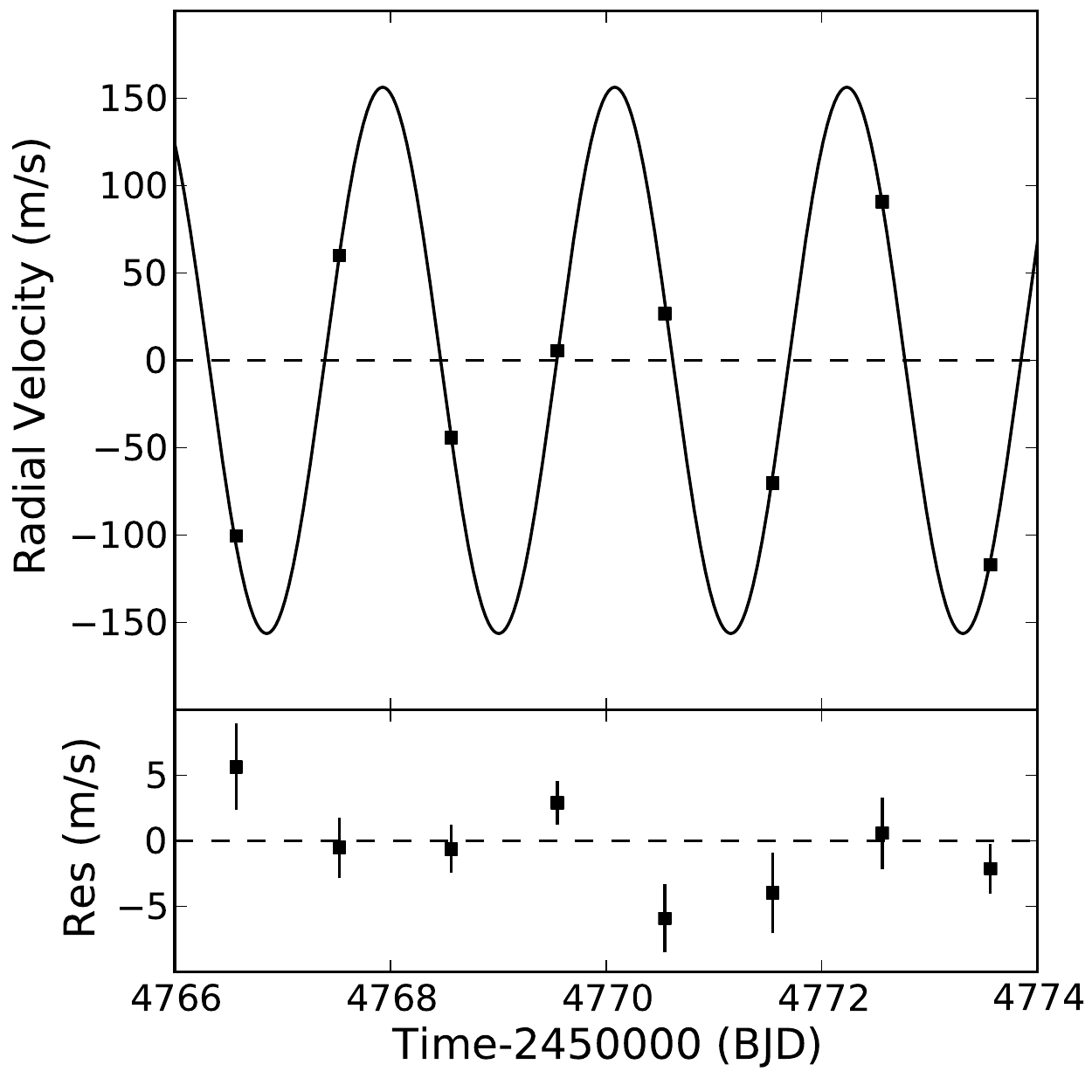}}
\resizebox{8cm}{!}{\includegraphics{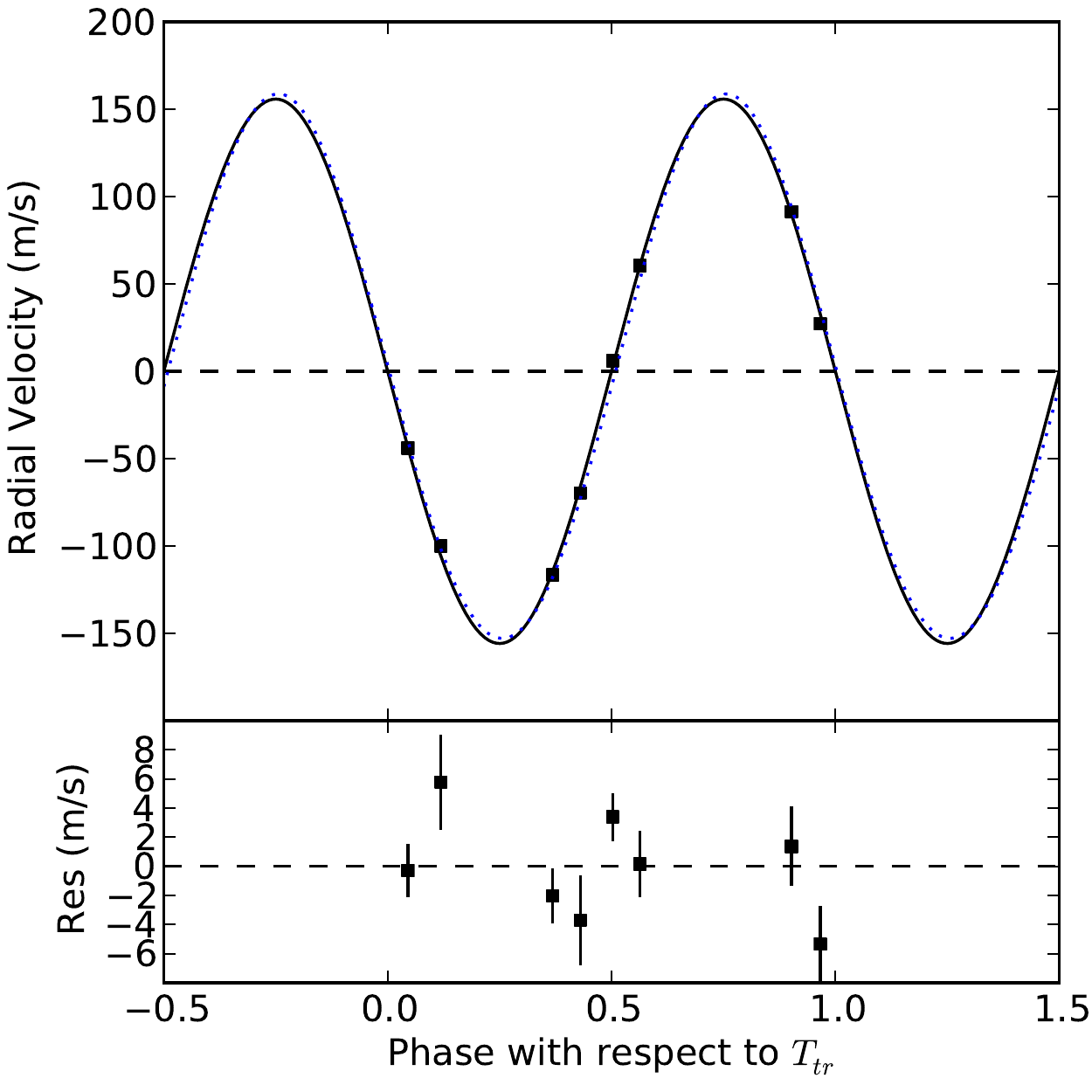}}
\caption{Plot showing our new HARPS radial velocity data for WASP-2, plotted against time (left) and orbital phase with respect to $T_{tr}$ (right). A circular orbit (solid line) and an orbit with the best-fit eccentricity (dotted line, but almost undistinguishable from the circular solution since $e=0.027$) are overplotted. The residuals relative to the circular orbit are shown in the bottom panel.}
\label{fig:wasp2ph}
\end{figure*}

We estimated the timescale of correlated noise for both the HARPS and SOPHIE data to be $\tau=1.5$ d, and we estimated $\sigma_r=10.4$ \ms\ for the SOPHIE data and $\sigma_r=6.45$ \ms\ for the HARPS data to obtain a reduced $\chi^2$ of unity for the circular orbit. We repeated this analysis with an eccentric orbit $k=5$ (3 degrees of freedom for the circular orbit, and 2 additional degrees of freedom for the eccentricity, $e\cos\omega$ and $e\sin\omega$). The orbital parameters are given in Table \ref{tab:wasp2}, and the radial velocity dataset is plotted in Figure \ref{fig:wasp2ph}, with residuals shown for a circular orbit. The Figure also shows models of a circular and an eccentric orbit (with $e=0.027$), but they are almost undistinguishable. For the circular orbit, we obtained $\chi^2=15.60$, giving a value of BIC$_c=115.08$ and for the eccentric orbit, we obtained $\chi^2=13.88$ giving a value of BIC$_e=119.02$. We repeated these calculations to obtain a reduced $\chi^2$ of unity for the eccentric orbit and estimated $\sigma_r=10.4$ \ms\ for the SOPHIE data (the SOPHIE dataset did not allow the MCMC to converge and yield a reduced $\chi^2$ of unity with an eccentric orbit) and $\sigma_r=7.05$ \ms\ for the HARPS data. For the circular orbit, we obtained $\chi^2=15.16$, and a value of BIC$_c=115.47$ and for the eccentric orbit, we obtained $\chi^2=13.49$ and a value of BIC$_e=119.47$. We therefore find that the circular orbit cannot be excluded for WASP-2. Furthermore, we exclude the possibility that $e>0.1$.

\begin{table*}
\centering
\begin{tabular}{l c  c }
\hline
Parameter	& SOPHIE, \citet{Cameron2007}					&	SOPHIE and HARPS, {\it this work}	\\ \hline
Centre-of-mass velocity $V_0$ [\ms] 	& $-$27863$\pm$7 	& $-$27862$\pm$7.4 (SOPHIE),  $-$27739.81$\pm$4.1 (HARPS),  	\\
Orbital eccentricity $e$ 				& 0 (adopted)			&  0.027$\pm0.023$ ($<$0.072)\\
Argument of periastron $\omega$ [$^o$]& 0 (unconstrained)	& 0 (unconstrained)			\\ 
$e\cos\omega$	& 				--					& $-$0.003$\pm$0.003	\\
$e\sin\omega$		&  				--				& $-$0.027$\pm$0.027	\\
Velocity semi-amplitude K [m$\,$s$^{-1}$]		& 155$\pm$7	& 156.3$\pm$2.1		\\
\hline
\end{tabular}
\caption{System parameters for WASP-2. Left: \citet{Cameron2007}. Right: Results from our HARPS radial velocity data. Median values for $V_0$ and $K$ are quoted for the circular orbits, as well as 68.3\% confidence limits obtained from the eccentric solution (see section Analysis) and 95\% limit on eccentricity.}
\label{tab:wasp2}
\end{table*}

\ \\
{\bf Other planets}\\
HD189733b and HD209458b are both on orbits that are compatible with a circular model: \citet{Laughlin2005} reported the 95\% limits on eccentricity for HD 209458b ($e<0.042$) and we estimate the upper limit for HD189733b from \citet{Triaud2009} assuming a Gaussian probability distribution, $e<0.008$). In both cases, the eccentricity is strongly constrained by the timing of the secondary eclipse. No radial velocity data was found for Kepler-5 in the literature or online, but we include the results of \citet{Kipping2011} in this study: Kepler-5b has an eccentricity of $e=0.034_{-0.018}^{+0.029}$, with a 95\% upper limit of $e<0.086 $. We therefore classify Kepler-5b as having a circular orbit. We also omitted an analysis of the two-planet system HAT-P-13, choosing to estimate the 95\% limits on the orbital eccentricity of HAT-P-13b from the literature ($e<0.022$) and classify this orbit as circular.

\subsection{Planets on eccentric orbits}
\label{sec:eccentric}
In contrast to Section~\ref{sec:noteccentric}, in this Section, we confirm the eccentricities of 10 planets. We verify the eccentricities of CoRoT-9b, GJ-436b and HAT-P-2b as a test for our procedures and we also confirm the eccentricities of HAT-P-16b and WASP-14b, with the former being the planet on a short period orbit with the smallest confirmed eccentricity, and the latter being the planet with the shortest period orbit having a confirmed eccentricity. Finally we note the confirmed orbital eccentricities of CoRoT-10b, HAT-P-15b, HD17156b, HD80606b and XO-3b.

\ \\
{\bf CoRoT-9}\\
CoRoT-9b is a $0.84$ M$_j$ planet on a $95.3$ day orbit around a G3 star (V=13.5), first reported by \cite{Deeg2010}, who found an eccentricity of $e=0.11\pm0.04$. We used the $14$ HARPS measurements from \cite{Deeg2010}, setting $\tau=1.5$ d and $\sigma_r=3.7$ \ms\ to obtain a value of reduced $\chi^2$ of unity for the circular orbit. We imposed the prior information from photometry $P=95.2738(14)$ and $T_{tr}=2454603.3447(1)$ from \cite{Deeg2010} and obtained a value of $\chi^2_{c}=14.05$ and $\chi^2_{e}=7.90$. Using $N=16$, $k_c=2$ and $k_e=4$, we obtain BIC$_c=106.17$ and BIC$_e=105.57$, which provides marginal support for an eccentric orbit at $e=0.111\pm0.046$, with the 95\% limit at $e<0.20$. We repeated the caculations, setting $\sigma_r=0$ \ms\ since this results in a reduced $\chi^2$ of less than unity for the eccentric orbit. This time, we obtained a value of $\chi^2_{c}=16.65$ and $\chi^2_{e}=9.63$. Using $N=16$, $k_c=2$ and $k_e=4$, we obtain BIC$_c=105.66$ and BIC$_e=104.18$, which supports an eccentric orbit at $e=0.111\pm0.039$.

\ \\
{\bf GJ-436}\\
GJ-436b is a $0.071$ M$_j$ planet on a $2.64$ day eccentric orbit around a M2.5 star (V=10.7), first reported by \cite{Butler2004}. \citet{Deming2007} detected the secondary eclipse using Spitzer, placing a constraint on the secondary eclipse phase $\phi_{\rm occ}=0.587\pm0.005$. This translates into $e\cos\omega=0.1367\pm0.0012$, which we apply as a Bayesian prior in the calculation of our merit function.

\begin{figure*}
\resizebox{8cm}{!}{\includegraphics{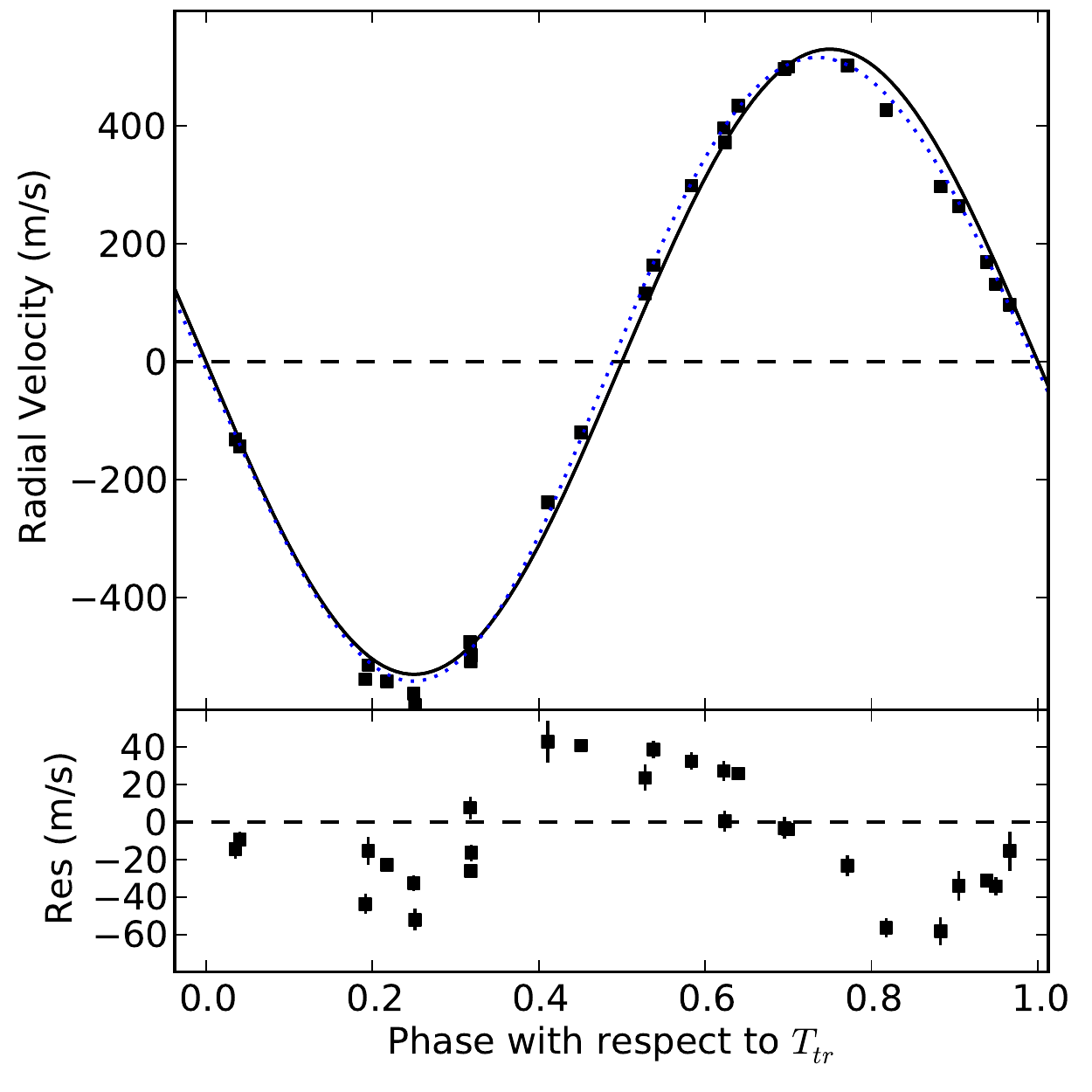}}
\resizebox{8cm}{!}{\includegraphics{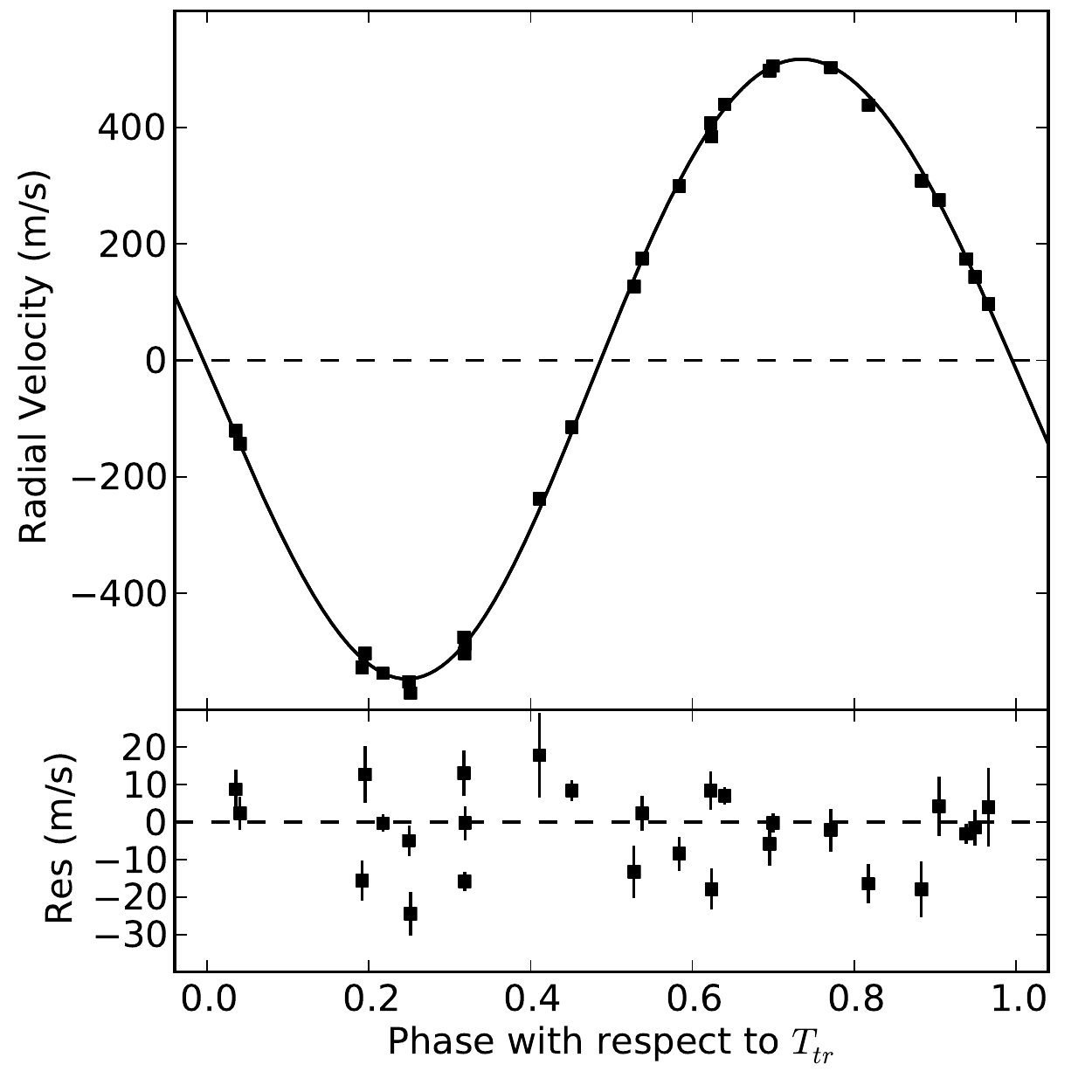}}
\caption{Plot showing radial velocity data from \citet{Buchhave2010} for HAT-P-16, plotted against orbital phase with respect to $T_{tr}$. {\bf Left:} A circular orbit is overplotted with a solid line and an eccentric orbit ($e=0.034$) is plotted with a dotted line. The bottom panel shows the residuals for a circular orbit: these show a clear periodic signal, indicating the possibility of an eccentric orbit. {\bf Right:} An eccentric orbit ($e=0.034$) is plotted with a solid line. The residuals are shown for the eccentric orbit.}
\label{fig:hatp16}
\end{figure*}

We used the $59$ HIRES measurements from \cite{Maness2007}, setting $\tau=1.5$ d and $\sigma_r=5.5$ \ms\ to obtain a value of reduced $\chi^2$ of unity for the circular orbit. We imposed the prior information from photometry $P=2.64385(9)$ from \citet{Maness2007}and $T_{tr}=2454280.78149(16)$ from \citet{Deming2007} and obtained a value of $\chi^2_{c}=59.74$ and $\chi^2_{e}=38.86$. Using $N=61$ (59 measurements and 2 priors from photometry) and $k_c=2$ for the circular orbit, we obtain BIC$_c=371.38$. Using $N=62$ (59 measurements and 3 priors from photometry) and $k_e=3$ ($V_0$, $K$, $e\sin\omega$) for the eccentric orbit, we obtain BIC$_e=354.67$, which supports an eccentric orbit at $e=0.157\pm0.024$, with the 95\% limit at $e<0.21$. We repeated the caculations, setting $\sigma_r=3.95$ \ms\ to obtain a reduced $\chi^2$ of unity for the eccentric orbit. This time, we obtained a value of $\chi^2_{c}=88.20$ and $\chi^2_{e}=59.39$. This time, we obtain BIC$_c=372.67$ and BIC$_e=348.03$, which supports an eccentric orbit at $e=0.153\pm0.017$, which is in agreement with \citet{Deming2007}, who reported $e=0.150\pm0.012$.

\ \\
{\bf HAT-P-16}\\
HAT-P-16b is a $4.19$ M$_j$ planet on a $2.78$ day orbit around a F8 star (V=10.7), first reported by \cite{Buchhave2010}. The original authors found an eccentricity of $e=0.036\pm0.004$. We re-analysed the 7 high resolution FIES measurements, 14 medium resolution FIES measurements and 6 HIRES measurements, with two priors from photometry on the period and mid-transit time. We set $\tau=1.5$ d for all instruments and set $\sigma_r=115$, 185,  and 28 \ms\ respectively for the three instruments to obtain a reduced $\chi^2$ of unity for each individually. We then analysed them together using both a circular ($\chi^2=28.83$) and an eccentric orbit ($\chi^2=3.81$). Using $N=29$, $k_c=4$ and $k_e=6$, we obtain BIC$_c=314.74$ and BIC$_e=296.45$, which supports an eccentric orbit at $e=0.034\pm0.010$. Figure~\ref{fig:hatp16} (left) shows the data from \cite{Buchhave2010}, with a circular orbit overplotted with a solid line and an eccentric orbit with the dotted line. The residuals are plotted for the circular solution and they show a clear periodic signal.

We repeated the analysis, this time setting $\sigma_r=0$ (reduced $\chi^2=0.62$, indicating over-fitting), 16 (reduced $\chi^2=33$), and 4.7 \ms\ respectively and separately for the three datasets (i.e. aiming for a reduced $\chi^2$ of unity for each dataset individually, with an eccentric orbit). We then analysed them together using both a circular ($\chi^2=347.86$) and an eccentric orbit ($\chi^2=44.62$). Using $N=29$, $k_c=4$ and $k_e=6$, we obtain BIC$_c=541.64$ and BIC$_e=245.14$, which supports an eccentric orbit at $e=0.034\pm0.003$. We thus confirm the eccentricity of HAT-P-16b, which means this is the planet with the smallest eccentricity that is reliably measured. This is in part helped by the fact that HAT-P-16b is a very massive planet, making the radial velocity signal for an eccentric orbit very clear. Figure~\ref{fig:hatp16} (right) shows the data from \cite{Buchhave2010} again, with an eccentric orbit overplotted with the dotted line. 

\ \\
{\bf WASP-14}\\
WASP-14b is a $7.3$ M$_j$ planet on a $2.24$ day orbit around a F5 star (V=9.8), first reported by \cite{Joshi2009}, who found an eccentricity of $e=0.091\pm0.003$. \citet{Husnoo2011} confirmed the eccentricity of the orbit and updated the precise value to $e=0.088\pm0.003$. This makes WASP-14b the planet that is closest to its host star but still has an eccentric orbit, taking the place from WASP-12b.

\ \\
{\bf CoRoT-10, HAT-P-2, HAT-P-15, HD17156, HD80606 and XO-3} \\
The orbits of the planets CoRoT-10b ($e=0.110\pm0.039$), HAT-P-2b ($e=0.517\pm0.003$), HAT-P-15b ($e=0.190\pm0.019$), HD17156b ($e=0.677\pm0.003$), HD80606b ($e=0.934\pm0.001$) and XO-3b ($e=0.287\pm0.005$) are clearly eccentric from existing literature \citep[See for example][ respectively]{Bonomo2010,Loeillet2008,Kovacs2010,Nutzman2010,Hebrard2010a,Hebrard2008}.

\subsection{Planets with orbits that have poorly constrained eccentricities}
\label{sec:unknown}
\begin{figure*}
\resizebox{8cm}{!}{\includegraphics{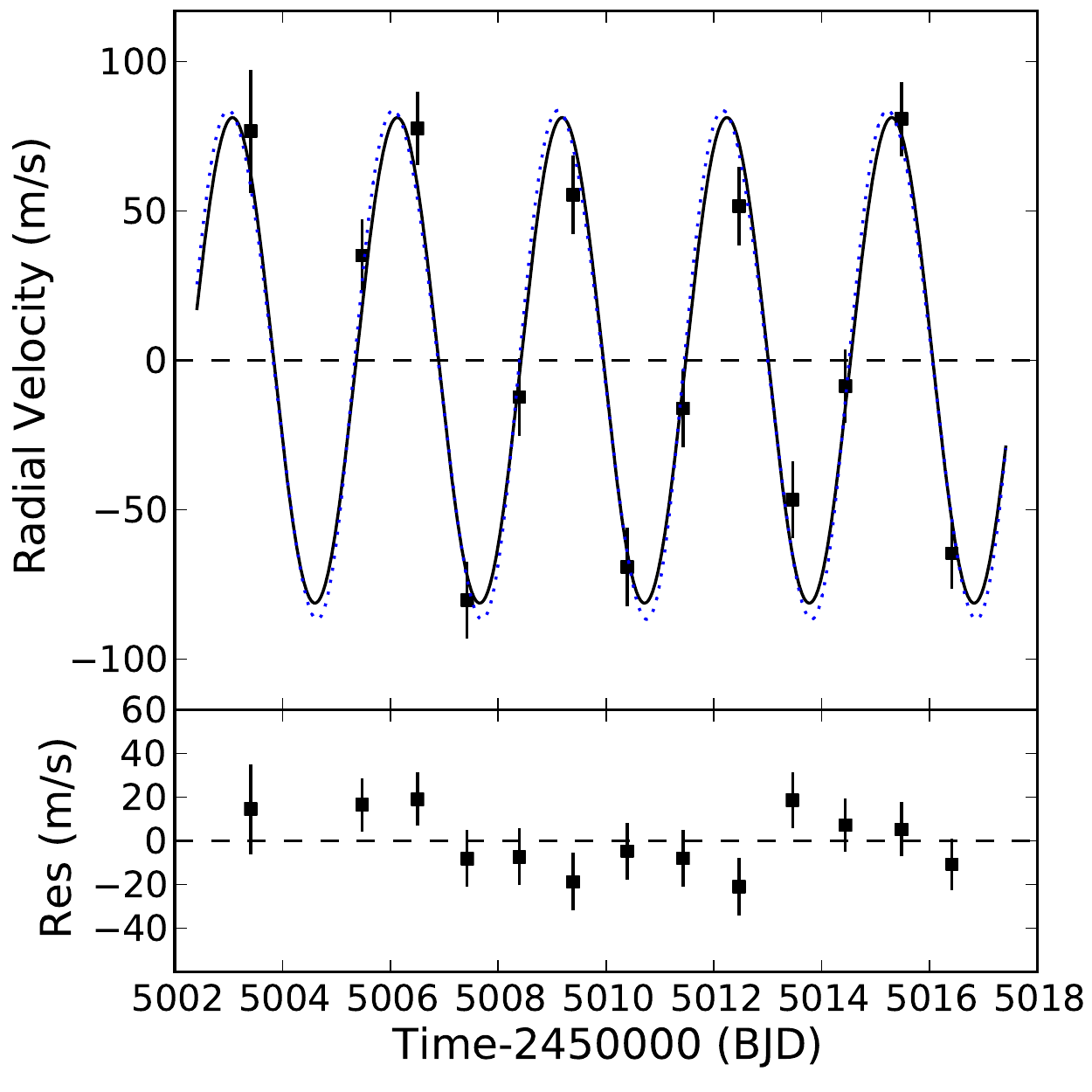}}
\resizebox{8cm}{!}{\includegraphics{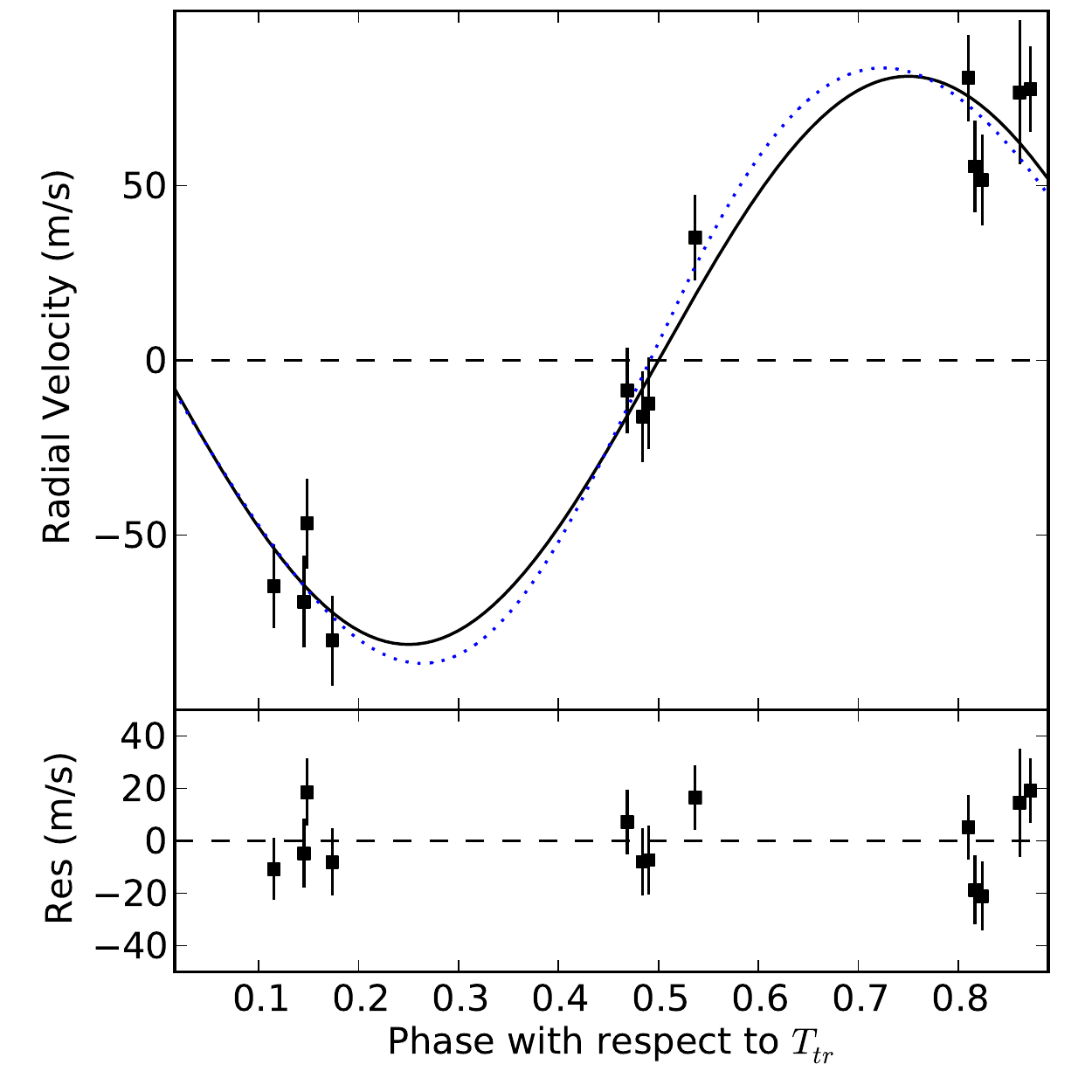}}
\caption{Plot showing our new HARPS radial velocity data for HAT-P-4, plotted against time ({\bf left}) and orbital phase with respect to $T_{tr}$ ({\bf right}). A circular orbit (solid line) and an orbit with the best-fit eccentricity (dotted line, $e=0.064$) are overplotted. The residuals relative to the circular orbit are shown in the bottom panel.}
\label{fig:hatp4sophie}
\end{figure*}

For 26 of the transiting planets that we attempted to place upper limits on their eccentricities, we obtained limits that were larger than 0.1. We considered  these eccentricities to be poorly determined. We discuss the cases of HAT-P-4b, WASP-7, XO-2b and Kepler-4b  below.

\ \\
{\bf HAT-P-4 (new SOPHIE data)}\\
HAT-P-4b is a $0.68$ M$_j$ planet on a $3.06$ day orbit around an F star (V=11.2), first reported by \cite{Kovacs2007}. We use 13 new SOPHIE radial velocity measurements and the 9 HIRES measurements in \cite{Kovacs2007} to work out the orbital parameters of HAT-P-4b. We impose the period $P=3.056536(57)$ d and mid-transit time $T_{tr}=2454248.8716(6)$ BJD as given from photometry in \cite{Kovacs2007}. We set $\tau=1.5$ d and $\sigma_r=3.35$ \ms\ for SOPHIE and $\sigma_r=3.75$\ms\ for HIRES, to obtain a reduced $\chi^2$ of unity for each dataset separately for the best-fit circular orbit. We used $22$ measurements in all, and count the two constraints from photometry as two additional datapoints ($N$=24), and used $k=3$ for the circular orbit (two $V_0$, one for each dataset, and the semi-amplitude $K$). We repeated this analysis with an eccentric orbit $k=5$ (three degrees of freedom for the circular orbit, and two additional degrees of freedom for the eccentricity, $e\cos\omega$ and $e\sin\omega$). The orbital parameters are given in Table \ref{tab:hatp4}, and the radial velocity dataset is plotted in Figure~\ref{fig:hatp4sophie}, with residuals shown for a circular orbit. The Figure also shows models of a circular and an eccentric orbit (with $e=0.064$). For the circular orbit, we obtained $\chi^2=22.05$, giving a value of BIC$_c=161.96$ and for the eccentric orbit, we obtained $\chi^2=16.77$ giving a value of BIC$_e=163.04$. We repeated these calculations by setting $\tau=1.5d$, $\sigma_r=1.81$ \ms\ for HIRES, and kept $\sigma_r=3.35$ \ms\ for SOPHIE, since we were unable to determine a value of $\sigma_r$ that would allow the MCMC chain to converge and lead to a $\chi^2$ of unity for an eccentric orbit. This time, we obtained $\chi^2=25.88$ for the circular orbit, giving a value of BIC$_c=161.96$ and for the eccentric orbit, we obtained $\chi^2=20.05$ giving a value of BIC$_e=162.49$. We find that the circular orbit cannot be excluded for HAT-P-4b, but because the eccentricity is $e=0.064\pm0.028$ with an upper limit of $e<0.11$, which is above 0.1, we classify HAT-P-4b as having a poorly constrained eccentricity.

\begin{table*}
\centering
\begin{tabular}{l c  c }
\hline
Parameter	& HIRES, \citet{Kovacs2007}					&HIRES+SOPHIE, {\it this work}	\\ \hline
Centre-of-mass velocity $V_0$ [\ms] 	& 12.1$\pm$0.9 & $20.3\pm 2.6$ (HIRES),  $-$1402.0 $\pm$4.0 (SOPHIE) \\
Orbital eccentricity $e$ 				& 0 (adopted)			&  0.064$\pm$0.028, $e<0.11$\\
Argument of periastron $\omega$ [$^o$]& 0 (unconstrained)	& 0 (unconstrained)			\\ 
$e\cos\omega$	& 				--					& $-$0.018$\pm0.012$	 \\
$e\sin\omega$		&  			--						& $-$0.061$\pm$0.027\\
Velocity semi-amplitude K [m$\,$s$^{-1}$]		& 81.1$\pm$1.9	& 81.3$\pm$2.6	\\
\hline
\end{tabular}
\caption{System parameters for HAT-P-4. Left: \citet{Kovacs2007}. Right: Results from our new SOPHIE radial velocity data and the original HIRES data. Median values for $V_0$ and $K$ are quoted for the circular orbits, as well as 68.3\% confidence limits obtained from the eccentric solution. The 95\% upper limit on eccentricity is also given.}
\label{tab:hatp4}
\end{table*}

\begin{figure*}
	\resizebox{8cm}{!}{\includegraphics{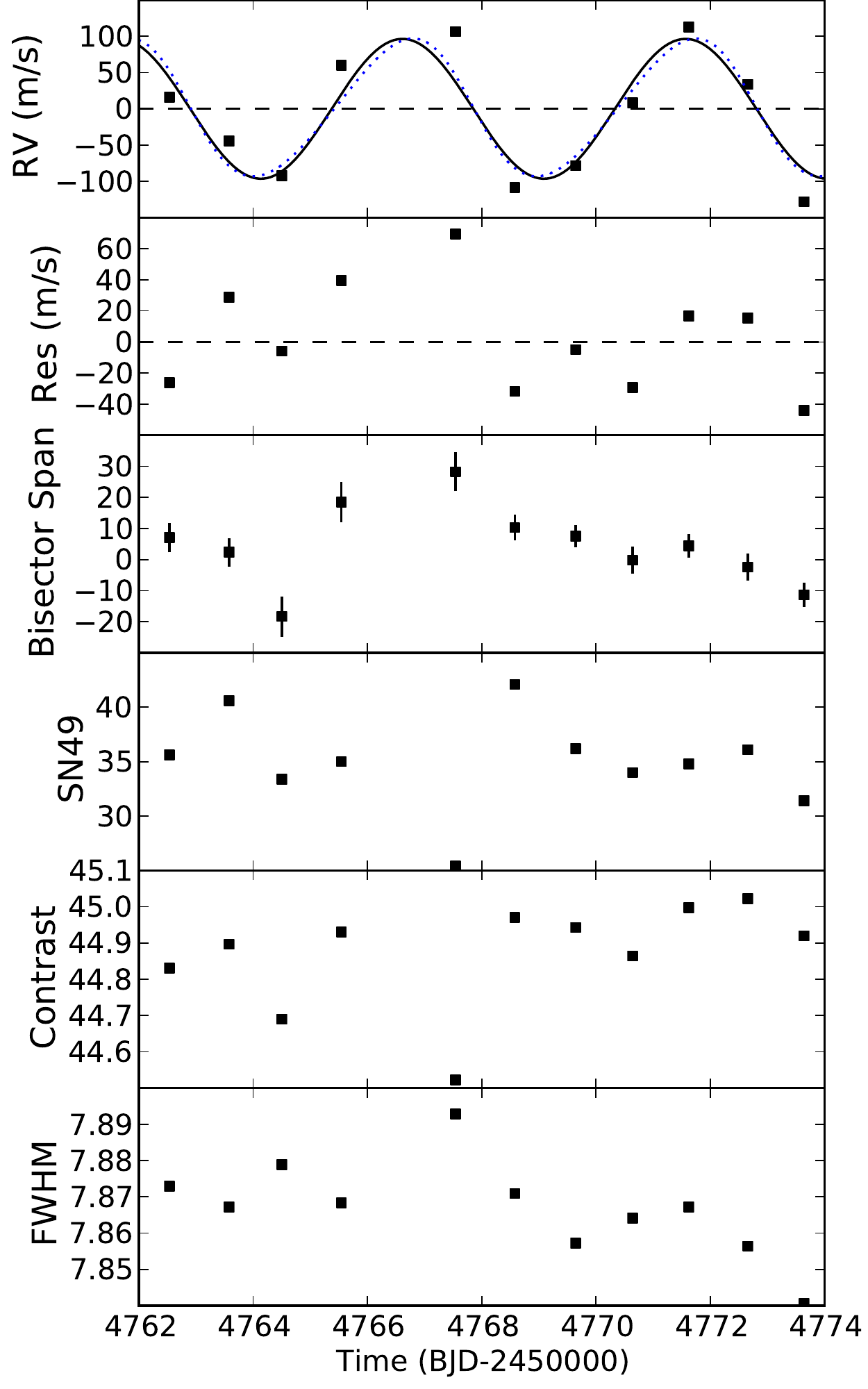}}
	\resizebox{8cm}{!}{\includegraphics{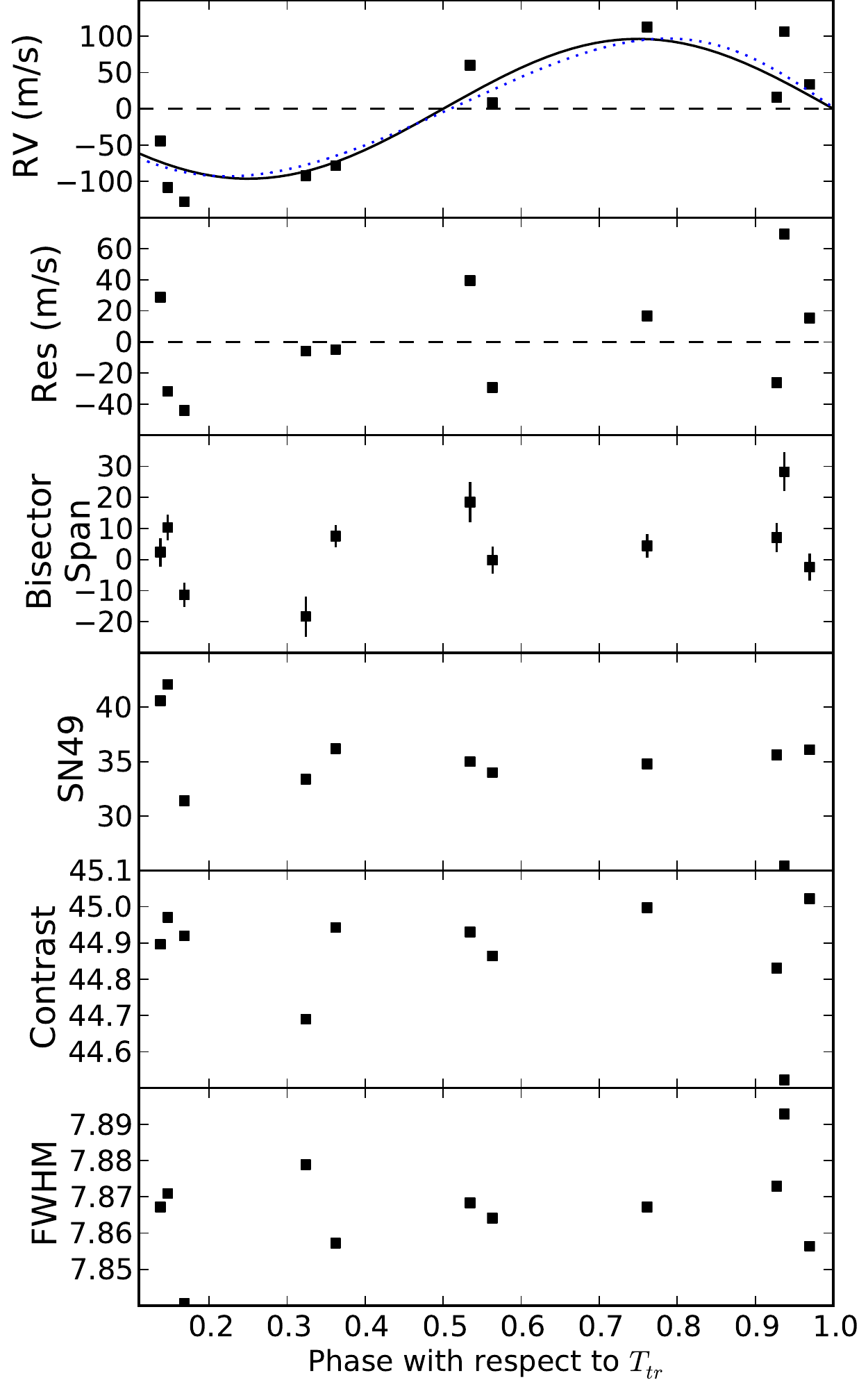}}
	\caption{HARPS measurements of WASP-7 plotted against time ({\bf left}) and phase with respect to the mid-transit time $T_{tr}$ ({\bf right}). In each case, a solid line is overplotted to represent a circular orbit and the residuals are plotted for this circular orbit. It is clear that a signal is present in the residuals (see text). An eccentric orbit with the best-fit value of $e=0.103$ is overplotted in both panels with a dotted line, but it is almost indistinguishable from the circular solution at this scale.}
	\label{fig:wasp7_circ}
\end{figure*}

\ \\
{\bf WASP-7 (new HARPS data)}\\
WASP-7b is a 1.0 M$_j$ planet on a $4.95$ day orbit around a F5 star (V=9.5), first reported by \citet{Hellier2009}. We analysed our 11 new HARPS measurements for WASP-7 as well as $11$ measurements from \citet{Hellier2009} using CORALIE, and used the photometric constraints on the orbital period $P=4.954658(55)$ and mid-transit time $T_{tr}=2453985.0149(12)$ from the same paper. For both instruments, we set $\tau=1.5$ d and for CORALIE, we set $\sigma_r=28.3$ \ms\ while for HARPS, we set $\sigma_r=210$ \ms\ in order to get a value of reduced $\chi^2$ equal to unity for the circular orbit. We performed the MCMC analysis twice: the first time fitting for the systemic velocity $v_0$ and semi-amplitude $K$, and the second time adding two parameters $e\cos\omega$ and $e\sin\omega$ to allow for an eccentric orbit. The best-fit parameters are given in Table~\ref{tab:wasp7}). We plot the radial velocity data against time (Figure~\ref{fig:wasp7_circ}, left) and phase (Figure~\ref{fig:wasp7_circ}, right). When the residuals for a circular orbit are plotted, and a scatter of about $30$ \ms\ is clearly seen, which is much larger than the median uncertainties of $\sigma=2.21$ \ms\ on the radial velocity measurements. This is similar to that found by \citet{Hellier2009} from their CORALIE data. An eccentric orbit does not reduce the scatter. The value of $\chi^2$ for the circular orbit is $22.37$ and that for an eccentric orbit is $18.11$. This leads to a value of BIC$_c=250.62$ and BIC$_e=252.72$, respectively, for 22 measurements, 2 constraints from photometry and 3 and 5 free parameters respectively (Keplerian orbits, but with two $V_0$ to account for a possible offset between the two instruments). This shows that the circular orbit is still preferred, and an eccentric orbit does not explain the scatter. We repeated this using $\sigma_r=33.8$ \ms\ for CORALIE while for HARPS, we set $\sigma_r=158.5$ \ms\ in order to get a value of reduced $\chi^2$ equal to unity for the eccentric orbit. We performed the MCMC analysis both for a circular and eccentric orbit. The value of $\chi^2$ for the circular orbit is $146.95$ and that for an eccentric orbit is $146.86$. This leads to a value of BIC$_c=367.47$ and BIC$_e=373.91$, respectively, for 22 measurements, 2 constraints from photometry and 3 and 5 free parameters respectively (Keplerian orbits, but with two $V_0$ to account for a possible offset between the two instruments). This shows that the circular orbit is still preferred, and an eccentric orbit does not explain the scatter. WASP-7 is an F5V star, with a temperature of $T_{\rm eff}=6400\pm100$ K. Despite the result of the original paper that WASP-7 is not chromospherically active above the 0.02 mag level, \citet{Lagrange2009} found evidence for other F5V stars showing radial velocity variability with a scatter at this level, for example HD 111998, HD 197692 or HD 205289, with scatters of $40$ \ms\, $30$ \ms\ and $29$ \ms\ respectively. Our derived value of eccentricity is $e=0.103\pm0.061$, with the 95\% upper limit is at $e<0.25$. We therefore classify the eccentricity of the orbit of WASP-7b as poorly constrained

In Figure~\ref{fig:wasp7_circ}, we have also plotted the bisector span, the signal to noise at order 49, the contrast and full width at half maximum for the cross-correlation function against the same time axis. The large scatter in radial velocity residuals can be seen to be correlated with both the bisector span and the full width at half maximum of the cross correlation function.

\begin{figure*}
\resizebox{8cm}{!}{\includegraphics{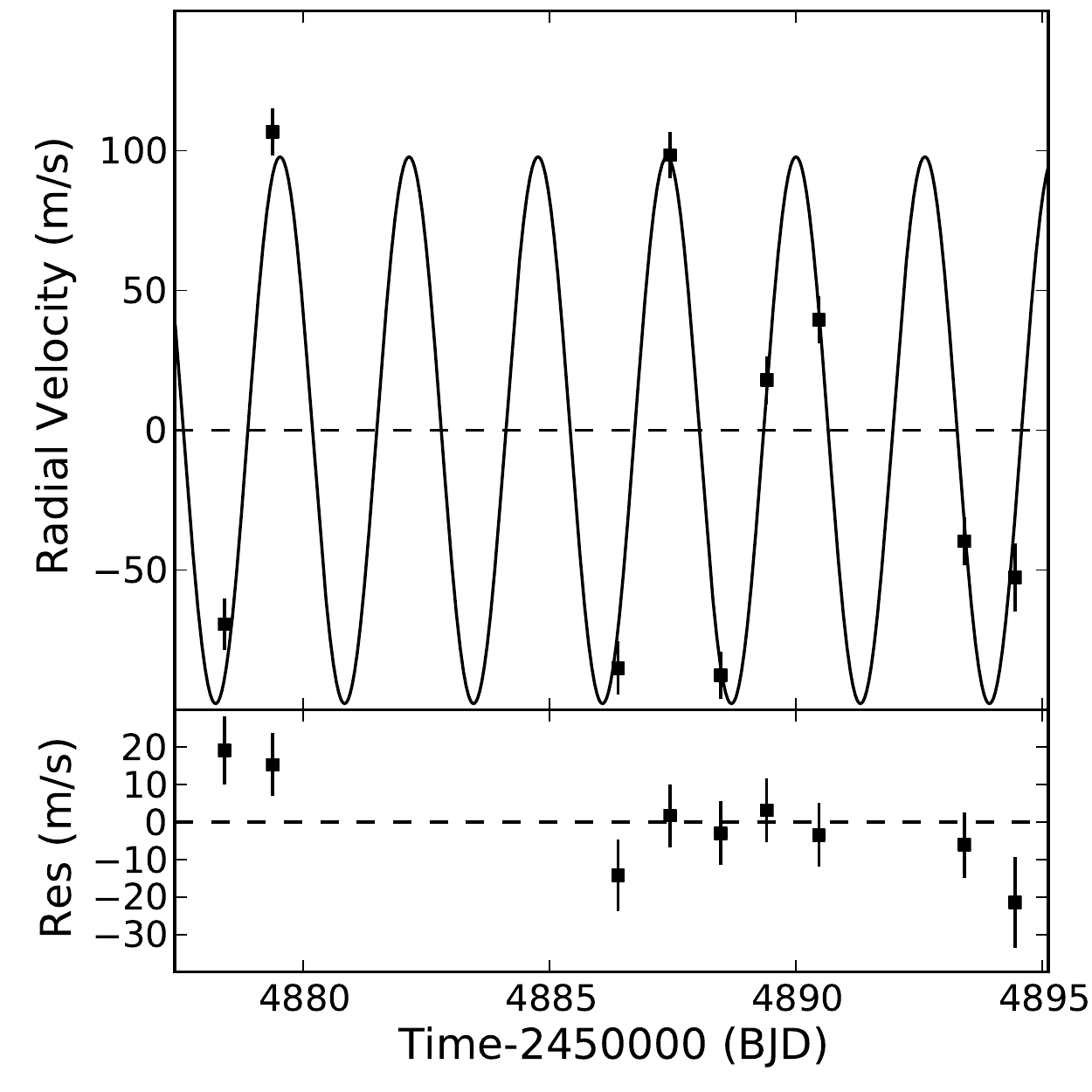}}
\resizebox{8cm}{!}{\includegraphics{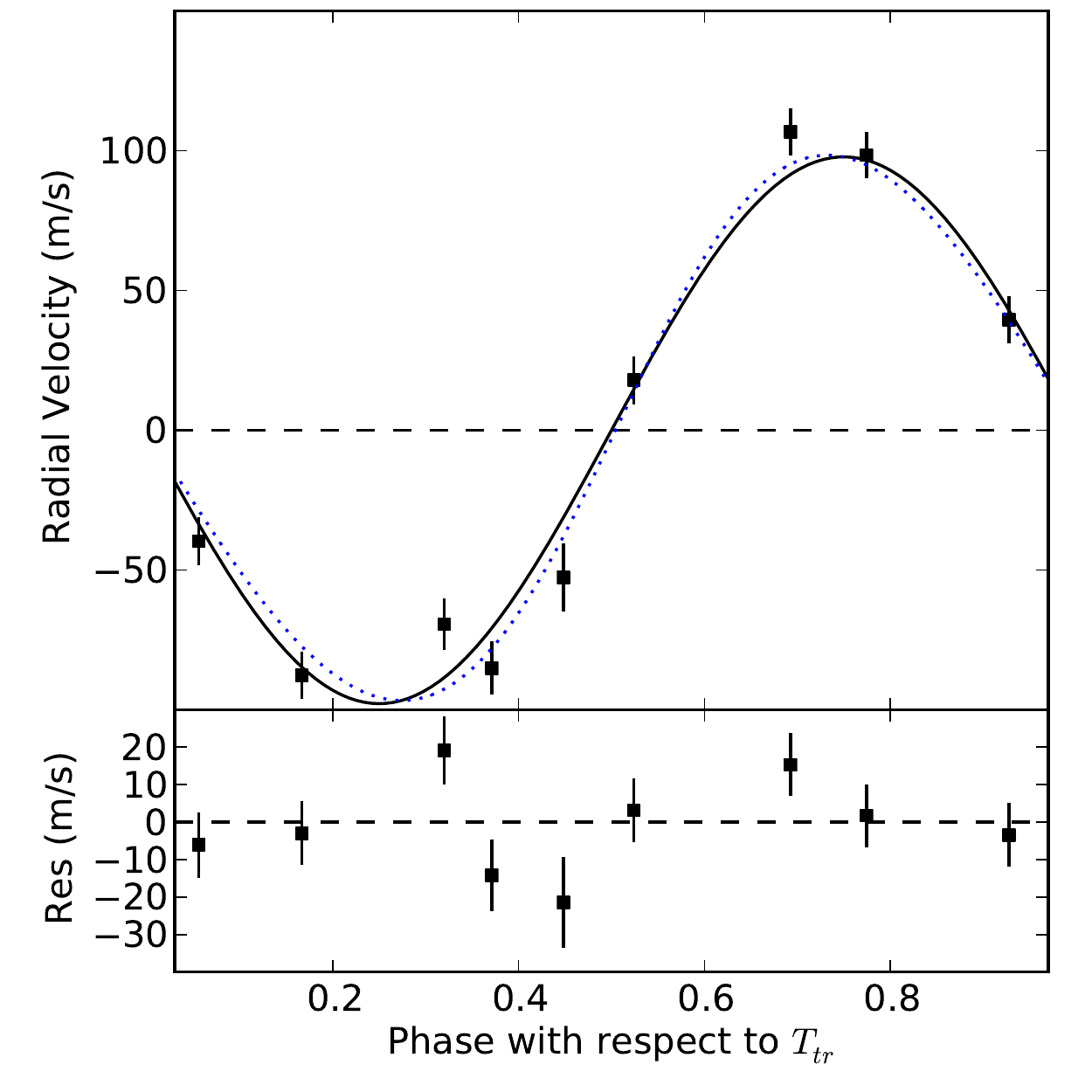}}
\caption{Plot showing our new SOPHIE radial velocity data for XO-2, plotted against orbital phase with respect to $T_{tr}$. A circular orbit (solid line) and an orbit with the best-fit eccentricity (dotted line, but almost undistinguishable from the circular solution since $e=nnnn$) are overplotted. The residuals relative to the circular orbit are shown in the bottom panel.}
\label{fig:xo2}
\end{figure*}

\begin{table*}
\centering
\begin{tabular}{l c  c }
\hline
Parameter	& \citet{Hellier2009}				&	HARPS, {\it this work}	\\ \hline
Centre-of-mass velocity $V_0$ [\ms]	& $-$29850.6$\pm$1.7& $-$29455$\pm$103  	\\
Orbital eccentricity $e$ 				& 0 (adopted)			& $0.103\pm0.061$ ($<$0.25)\\
Argument of periastron $\omega$ [$^o$]& 0 (unconstrained)	& 0 (unconstrained)			\\ 
$e\cos\omega$	& 				--					& 0.021$\pm$0.068	\\
$e\sin\omega$		&  				--					& 0.101$\pm$0.074	\\
Velocity semi-amplitude K [m$\,$s$^{-1}$]		& 97$\pm$13	& 96$\pm$14			\\
\hline
\end{tabular}
\caption{System parameters for WASP-7. Left: \citet{Hellier2009}. Right: Results from our HARPS radial velocity data. Median values for $V_0$ and $K$ are quoted for the circular orbits, as well as 68.3\% confidence limits obtained from the eccentric solution (see section Analysis).}
\label{tab:wasp7}
\end{table*}


\ \\
{\bf XO-2 (new SOPHIE data)}\\
XO-2 is a $0.6$ M$_j$ planet on a $2.62$ day orbit around a K0 star (V=11.2), first reported by \cite{Burke2007}. We use 9 new SOPHIE radial velocity measurements and the 10 HJS measurements in \cite{Burke2007} to work out the orbital parameters of XO-2. We impose the period $P=2.6158640(21)$ d and mid-transit time $T_{tr}=2454466.88467 (17)$ BJD as given from photometry in \citet{Fernandez2009}. We set $\tau=1.5$ d and $\sigma_r=5.3$ \ms\ for SOPHIE and $\sigma_r=0$ \ms\ for HJS (because the HJS data alone, with a circular orbit, yield a reduced $\chi^2$ of 0.78, indicating overfitting) to obtain a reduced $\chi^2$ of unity for the best-fit circular orbit. We used $19$ measurements in all, and count the two constraints from photometry as two additional datapoints ($N=21$), and used $k=3$ for the circular orbit (two $V_0$, one for each dataset, and the semi-amplitude $K$). We repeated this analysis with an eccentric orbit $k=5$ (two degrees of freedom for the circular orbit, and two additional degrees of freedom for the eccentricity, $e\cos\omega$ and $e\sin\omega$). The orbital parameters are given in Table \ref{tab:xo2}, and the radial velocity dataset is plotted in Figure \ref{fig:xo2}, with residuals shown for a circular orbit. The Figure also shows models of a circular and an eccentric orbit (with $e=0.064$). For the circular orbit, we obtained $\chi^2=19.65$, giving a value of BIC$_c=165.55$ and for the eccentric orbit, we obtained $\chi^2=17.57$ giving a value of BIC$_e=169.55$. We repeated the calculations using $\sigma_r=7.05$ \ms\ for SOPHIE and $\sigma_r=0$ \ms\ for HJS (because the HJS data alone, with an eccentric orbit, yield a reduced $\chi^2$ of 0.56, indicating overfitting) to obtain a reduced $\chi^2$ of unity for the best-fit eccentric orbit. For the circular orbit, we obtained $\chi^2=18.02$, giving a value of BIC$_c=165.21$ and for the eccentric orbit, we obtained $\chi^2=16.01$ giving a value of BIC$_e=169.29$. In both cases, i.e. using the optimal value of $\sigma_r$ for a circular orbit and using the optimal value of $\sigma_r$ for an eccentric orbit, a circular orbit is favoured. The 95\% upper limit is $e<0.14$, which is above 0.1, so we classify the orbital eccentricity of XO-2 as poorly constrained.

\begin{table*}
\centering
\begin{tabular}{l c  c }
\hline
Parameter	& HJS, \citet{Burke2007}						&	HJS, SOPHIE, {\it this work}	\\ \hline
Centre-of-mass velocity $V_0$ [\ms] 	& -- 					& $-1.3\pm 6.3$ (HJS), 46860.1$\pm$4.1 (SOPHIE)  	\\
Orbital eccentricity $e$ 				& 0 (adopted)			&  0.064$\pm$0.041 ($e<0.14$)\\
Argument of periastron $\omega$ [$^o$]& 0 (unconstrained)	& 0 (unconstrained) \\ 
$e\cos\omega$	& 				--					& 0.007$\pm$0.017 \\
$e\sin\omega$		&  				--				& $-0.063$$\pm$0.047 \\
Velocity semi-amplitude K [m$\,$s$^{-1}$]		& 85$\pm$8	& 98.0$\pm$4.0 \\
\hline
\end{tabular}
\caption{System parameters for XO-2. Left: \citet{Burke2007}. Right: Results from our SOPHIE radial velocity data. Median values for $V_0$ and $K$ are quoted for the circular orbits, as well as 68.3\% confidence limits obtained from the eccentric solution. The 95\% upper limit on eccentricity is also given.}
\label{tab:xo2}
\end{table*}

\ \\
{\bf Kepler-4}\\
Kepler-4b has a derived eccentricity of $e=0.25_{-0.12}^{+0.11}$, with a 95\% upper limit of $e<0.43$ \citep{Kipping2011}, so we classify it as ``poorly constrained eccentricity''.

\ \\
{\bf Other objects}\\
For the 8 objects CoRoT-6, HAT-P-1, HAT-P-3, HAT-P-6, HD149026, Kepler-6, WASP-10 and WASP-21, we found the BIC$_e$ for an eccentric orbit  was smaller than the BIC$_c$ for a circular orbit if we assume a $\sigma_r$ that yields a reduced $\chi^2$ of unity for an eccentric orbit, whereas the BIC$_c$ for a circular orbit was smaller than the BIC$_e$ for an eccentric orbit if we assume a $\sigma_r$ that yields a reduced $\chi^2$ of unity for a circular orbit. This suggests that the current RV datasets do not constrain the orbit enough for us to detect a finite eccentricity. We have already discussed the case of WASP-10b in Section~\ref{sec:noteccentric} above.

\subsection{Additional planetary systems\label{add_systems}}
In addition to the 64 planets considered so far, we now include 3 additional planets on eccentric orbits, 11 planets on orbits where $e>0.1$ is excluded at the 95\% level and two brown dwarves. The additional planets on eccentric orbits are 
HAT-P-17b \citep[][$e=0.346\pm 0.007$]{Howard2010a}, HAT-P-21b \citep[][$e=0.228\pm0.016$]{Bakos2010a} and HAT-P-31b \citep[][$e=0.245\pm0.005$]{Kipping2011b}.
The additional planets on orbits that are consistent with circular are: \\
CoRoT-18b \citep[$e<0.08$ at 3-$\sigma$,][]{Hebrard2011a},\\
HAT-P-20b \citep[$e<0.023$, estimated from][]{Bakos2010a},\\
HAT-P-22b \citep[$e<0.031$, estimated from][]{Bakos2010a}, \\
HAT-P-25b \citep[$e<0.068$, estimated from][]{Quinn2010}, \\
HAT-P-30b \citep[$e<0.074$, estimated from][]{Johnson2011}, \\
WASP-23b  \citep[$e<0.062$ at 3-$\sigma$,][]{Triaud2011}, \\
WASP-34b \citep[$e<0.058$, estimated from][]{Smalley2011}, \\
WASP-43b  \citep[$e<0.04$ at 3-$\sigma$,][]{Hellier2011}, \\
WASP-45b \citep[$e<0.095$,][]{Anderson2011}, \\
WASP-46b \citep[$e<0.065$,][]{Anderson2011} and \\
$\tau$ Bo\"otis b \citep[$e<0.045$, estimated from][]{Butler2006}. The two brown dwarves are  OGLE-TR-122b \citep[][$e=0.205\pm0.008$]{Pont2005a} and OGLE-TR-123b \citep[][$e=0$]{Pont2005}. In addition to the above, we also consider the case of WASP-38 \citep{Barros2011}, which has an eccentricity of $e=0.031\pm0.005$, indicating it is in the process of circularisation, just like WASP-14 and HAT-P-16.

\begin{table*}
\centering
\begin{tabular}{ l          r         r@{$\pm$}l        r@{}l           c        r@{$\pm$}l        }
\hline
Name	&   Eccentricity		&  \multicolumn{2}{r|}{Eccentricity}  &  \multicolumn{2}{r|}{95\% limit}  &  E      	& \multicolumn{2}{r|}{$M_p$(M$_j$)}\\
		&   (literature)			& \multicolumn{2}{r|}{\it (this work)}  & \multicolumn{2}{r|}{}			&		& \multicolumn{2}{r|}{\it (this work)}	\\
\hline
CoRoT-1b                             &  --                                   &  0.006 & 0.012                        &  $(<$ & $0.042)$                      &  C                                    &  1.06 & 0.14                  \\
CoRoT-2b                             &  --                                   &  0.036 & 0.033                        &  $(<$ & $0.10)$                       &  P                                    &  3.14 & 0.17                  \\
CoRoT-3b                             &  $0.008^{+0.015}_{-0.005}$             &  0.012 & 0.01                         &  $(<$ & $0.039)$                      &  C                                    &  21.61 & 1.2                 \\
CoRoT-4b                             &  $0\pm0.1$                            &  0.27 & 0.15                          &  $(<$ & $0.48)$                       &  P                                    &  0.659 & 0.079                  \\
CoRoT-5b                             &  $0.09^{+0.09}_{-0.04}$               &  0.086 & 0.07                         &  $(<$ & $0.26)$                       &  P                                    &  0.488 & 0.032                  \\
CoRoT-6b                             &  $<0.1$                               &  0.18 & 0.12                          &  $(<$ & $0.41)$                       &  P                                    &  2.92 & 0.30                  \\
CoRoT-9b                             &  0.11$\pm$0.04                        &  0.11 & 0.039                         &  $(<$ & $0.20)$                       &  E                                    &  0.839 & 0.070                  \\
CoRoT-10b                            &  0.53$\pm$0.04                        &  0.53 & 0.04                          &  --  &                                 &  E                                   &  2.75 & 0.16                  \\
GJ-436b                              &  0.150$\pm$0.012                      &  0.153 & 0.017                        &  -- &                                   &  E                                  &  0.069 & 0.006                  \\
GJ-1214b                             &  $<0.27$~(95\%)                       &  0.12 & 0.09                          &  $(<$ & $0.34)$                       &  P                                    &  0.020 & 0.003                  \\
HAT-P-1b                             &  $<0.067$~(99\%)                      &  0.048 & 0.021                        &  $(<$ & $0.087)$                      &  P                                    &  0.514 & 0.038                  \\
HAT-P-2b                             &  0.517$\pm$0.003                      &  0.517 & 0.003                        &  --  &                                 &  E                                   &  8.76 & 0.45                  \\
HAT-P-3b                             &  --                                   &  0.1 & 0.05                           &  $(<$ & $0.20)$                       &  P                                    &  0.58 & 0.17                  \\
HAT-P-4b                             &  --                                   &  0.063 & 0.028                        &  $(<$ & $0.107)$                      &  P                                    &  0.677 & 0.049                  \\
HAT-P-5b                             &  --                                   &  0.053 & 0.061                        &  $(<$ & $0.24)$                       &  P                                    &  1.09 & 0.11                  \\
HAT-P-6b                             &  --                                   &  0.047 & 0.017                        &  $(<$ & $0.078)$                      &  P                                    &  1.031 & 0.053                  \\
HAT-P-7b                             &  $<0.039$~(99\%)                      &  0.014 & 0.01                         &  $(<$ & $0.037)$                      &  C                                    &  1.775 & 0.070                  \\
HAT-P-8b                             &  --                                   &  0.011 & 0.019                        &  $(<$ & $0.064)$                      &  C                                    &  1.340 & 0.051                  \\
HAT-P-9b                             &  --                                   &  0.157 & 0.099                        &  $(<$ & $0.40)$                       &  P                                    &  0.767 & 0.10                  \\
HAT-P-11b                            &  0.198$\pm$0.046                      &  0.28 & 0.32                          &  $(<$ & $0.80)$                       &  P                                    &  0.055 & 0.022                  \\
HAT-P-12b                            &  --                                   &  0.071 & 0.053                        &  $(<$ & $0.22)$                       &  P                                    &  0.187 & 0.033                  \\
HAT-P-13b                            &  0.014$^{+0.005}_{-0.004}$            &  0.014 & 0.005                        &  $(<$ & $0.022)$                      &  C                                    &  0.855 & 0.046                  \\
HAT-P-14b                            &  0.107$\pm$0.013                      &  0.11 & 0.04                          &  $(<$ & $0.18)$                       &  P                                    &  2.23 & 0.12                  \\
HAT-P-15b                            &  0.190$\pm$0.019                      &  0.19 & 0.019                         &  --   &                                &  E                                   &  1.949 & 0.077                  \\
HAT-P-16b                            &  0.036$\pm$0.004                      &  0.034 & 0.003                        &  $(<$ & $0.039)$                      &  ES                                   &  4.20 & 0.11                  \\
HD17156b                             &  0.677$\pm$0.003                      &  0.675 & 0.004                        &  --    &                               &  E                                   &  3.223 & 0.087                  \\
HD80606b                             &  0.934$\pm$0.001                      &  0.933 & 0.001                        &  --      &                             &  E                                   &  3.99 & 0.33                  \\
HD149026b                            &  --                                   &  0.121 & 0.053                        &  $(<$ & $0.21)$                       &  P                                    &  0.354 & 0.031                  \\
HD189733b                            &  0.004$^{+0.003}_{-0.002}$            &  0.004 & 0.003                        &  $(<$ & $0.0080)$                     &  C                                    &  1.139 & 0.035                  \\
HD209458b                            &  0.014$\pm$0.009                      &  0.014 & 0.009                        &  $(<$ & $0.042)$                      &  C                                    &  0.677 & 0.033                  \\
Kepler-4b                            &  $0.25_{-0.12}^{+0.11}$~($<0.43$)     &  0.25 & 0.12                          &  $(<$ & $0.43)$                       &  P                                    &  0.077 & 0.028                  \\
Kepler-5b                            &  $0.034_{-0.018}^{+0.029}$~($<0.086$) &  0.034 & 0.029                        &  $(<$ & $0.086)$                      &  C                                    &  2.120 & 0.079                  \\
Kepler-6b                            &  $0.056_{-0.028}^{+0.044}$~($<0.13$)  &  0.057 & 0.026                        &  $(<$ & $0.12)$                       &  P                                    &  0.659 & 0.038                  \\
Kepler-7b                            &  $0.102_{-0.047}^{+0.104}$~($<0.31$)  &  0.065 & 0.045                        &  $(<$ & $0.19)$                       &  P                                    &  0.439 & 0.044                  \\
Kepler-8b                            &  $0.35_{-0.11}^{+0.15}$~($<0.59$)     &  0.011 & 0.24                         &  $(<$ & $0.39)$                       &  P                                    &  0.57 & 0.11                  \\
TrES-1b                              &  --                                   &  0.019 & 0.054                        &  $(<$ & $0.21)$                       &  P                                    &  0.757 & 0.061                  \\
TrES-2b                              &  --                                   &  0.023 & 0.014                        &  $(<$ & $0.051)$                      &  C                                    &  1.195 & 0.063                  \\
TrES-3b                              &  --                                   &  0.066 & 0.048                        &  $(<$ & $0.16)$                       &  P                                    &  1.86 & 0.12                  \\
TrES-4b                              &  --                                   &  0.21 & 0.21                          &  $(<$ & $0.66)$                       &  P                                    &  0.93 & 0.17                  \\
WASP-1b                              &  --                                   &  0.19 & 0.22                          &  $(<$ & $0.65)$                       &  P                                    &  0.89 & 0.15                  \\
WASP-2b                              &  --                                   &  0.027 & 0.023                        &  $(<$ & $0.072)$                      &  C                                    &  0.852 & 0.080                  \\
WASP-3b                              &  --                                   &  0.009 & 0.013                        &  $(<$ & $0.048)$                      &  C                                    &  1.99 & 0.13                  \\
WASP-4b                              &  --                                   &  0.005 & 0.003                        &  $(<$ & $0.011)$                      &  C                                    &  1.205 & 0.044                  \\
WASP-5b                              &  0.038$^{+0.026}_{-0.018}$            &  0.012 & 0.007                        &  $(<$ & $0.026)$                      &  C                                    &  1.571 & 0.063                  \\
WASP-6b                              &  0.054$^{+0.018}_{-0.015}$            &  0.041 & 0.019                        &  $(<$ & $0.075)$                      &  C                                    &  0.480 & 0.038                  \\
WASP-7b                              &  --                                   &  0.074 & 0.063                        &  $(<$ & $0.23)$                       &  P                                    &  1.07 & 0.16                  \\
WASP-10b                             &  0.057$^{+0.014}_{-0.004}$            &  0.052 & 0.031                        &  $(<$ & $0.11)$                       &  P                                    &  3.15 & 0.12                  \\
WASP-11b                             &  --                                   &  0.091 & 0.054                        &  $(<$ & $0.21)$                       &  P                                    &  0.470 & 0.035                  \\
WASP-12b                             &  $0.049\pm0.015$                      &  0.018 & 0.018                        &  $(<$ & $0.05)$                       &  C                                    &  1.48 & 0.14                  \\
WASP-13b                             &  --                                   &  0.14 & 0.1                           &  $(<$ & $0.32)$                       &  P                                    &  0.458 & 0.064                  \\
WASP-14b                             &  0.091$\pm$0.004                      &  0.088 & 0.003                        &  $(<$ & $0.090)$                      &  ES                                   &  7.26 & 0.59                  \\
WASP-15b                             &  --                                   &  0.056 & 0.048                        &  $(<$ & $0.17)$                       &  P                                    &  0.548 & 0.059                  \\
WASP-16b                             &  --                                   &  0.009 & 0.012                        &  $(<$ & $0.047)$                      &  C                                    &  0.846 & 0.072                  \\
WASP-17b                             &  $0.129^{+0.106}_{-0.068}$            &  0.121 & 0.093                        &  $(<$ & $0.32)$                       &  P                                    &  0.487 & 0.062                  \\
WASP-18b                             &  $0.009\pm0.001$                      &  0.007 & 0.005                        &  $(<$ & $0.018)$                      &  C                                    &  10.16 & 0.87                 \\
WASP-19b                             &  0.02$\pm$0.01                        &  0.011 & 0.013                        &  $(<$ & $0.047)$                      &  C                                    &  1.15 & 0.10                  \\
WASP-21b                             &  --                                   &  0.048 & 0.024                        &  $(<$ & $0.11)$                       &  P                                    &  0.308 & 0.018                  \\
WASP-22b                             &  $0.023\pm0.012$                      &  0.022 & 0.016                        &  $(<$ & $0.057)$                      &  C                                    &  0.56 & 0.13                  \\
WASP-26b                             &  --                                   &  0.033 & 0.025                        &  $(<$ & $0.086)$                      &  C                                    &  1.018 & 0.034                  \\
XO-1b                                &  --                                   &  0.042 & 0.088                        &  $(<$ & $0.30)$                       &  P                                    &  0.911 & 0.088                  \\
XO-2b                                &  --                                   &  0.064 & 0.041                        &  $(<$ & $0.14)$                       &  P                                    &  0.652 & 0.032                  \\
XO-3b                                &  0.287$\pm$0.005                      &  0.287 & 0.005                        &  --   &                                &  E                                   &  11.81 & 0.53                 \\
XO-4b                                &  --                                   &  0.28 & 0.15                          &  $(<$ & $0.50)$                       &  P                                    &  1.56 & 0.30                  \\
XO-5b                                &  --                                   &  0.01 & 0.01                          &  $(<$ & $0.036)$                      &  C                                    &  1.065 & 0.036                  \\
\hline
\end{tabular}
\caption{\tiny Table showing the objects which we considered in this study. We have included the fifth column to show if the object is on a circular orbit (``C'', ie circular according to the BIC test and 95\% limit on $e$ is less that 0.1), ``E'', for objects that are on eccentric orbits (either determined to be eccentric using the BIC test, or the orbit is clearly eccentric from the radial velocity plot), or ``P'', for objects which we fail to place any useful constraints on the eccentricity (ie the 95\% limit on $e$ is larger than 0.1), or it is unclear from model selection whether the orbit is circular or eccentric.}
\label{tab:results}
\end{table*}

\clearpage

\begin{figure*}
	\resizebox{16cm}{!}{\includegraphics{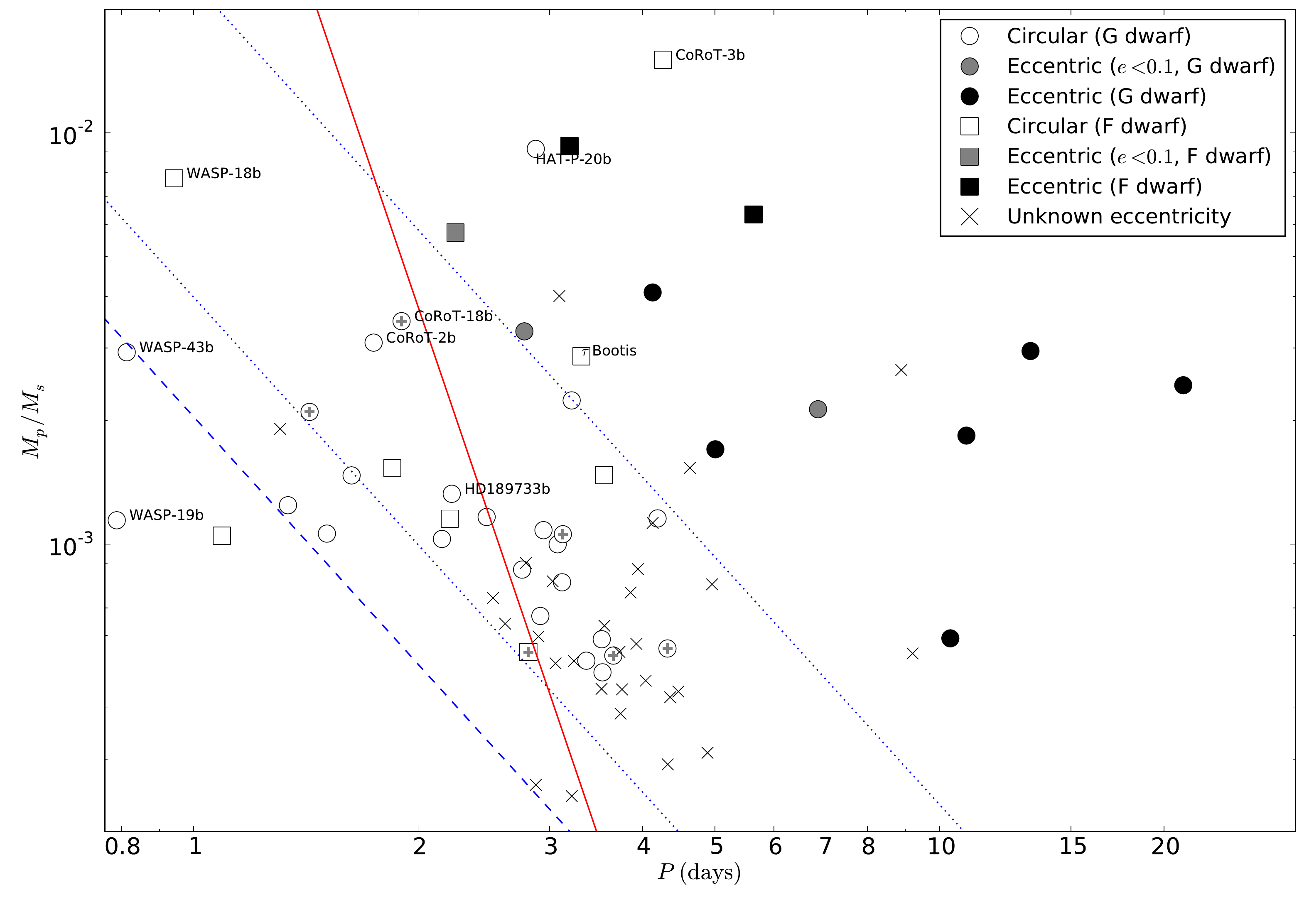}}
	\caption{Plot showing the mass ratio v/s period plane, for transiting planets with orbital period $P<20$ days. The low mass hot Jupiters on orbits that are consistent with circular around G dwarfs migrate inwards until they stop at a minimum period for a given mass, conglomerating on the mass-period relation of \citet{Mazeh2005}. 
The heavier planets can move in towards the star, and synchronise their rotations, as CoRoT-3b and $\tau$ Bo\"otis b did, or if they lack the angular momentum to synchronise the star, they can continue migrating inwards towards their destruction, as WASP-18b appears to be doing.
The labelled symbols (except for WASP-18b) represent objects on orbits that are consistent with circular where the host star rotation is significantly faster than the expected rotation from the isochrones of Strassmeier \& Hall (1988). Five objects have been marked with a $+$ symbol to mark objects with upper limits greater than $e<0.05$ that are described in Section~\ref{add_systems}. The dashed line represents $\alpha=2$ for $R_p=1.2 R_j$, while the dotted lines represent a value in the range $\alpha=2.5$--$4.5$ in the equation $a=\alpha a_R$ (see text). The solid line represents a circularisation isochrone at 1 Gyr for tides in the planet alone.}.
	\label{fig:mass_period}
\end{figure*}

\section{Discussion}
\label{sec:discussion}

\ \\
{\bf The Mass-Period plane}\\
We now discuss the results of the previous sections in the context of tidal evolution in hot Jupiters. Figure~\ref{fig:mass_period} shows a plot of the mass ratio $M_p/M_s$ against orbital period for transiting planets with orbital period $P<20$ days. The empty symbols represent orbits that are consistent with circular, and the black symbols represent eccentric orbits, whereas grey symbols represent objects with small ($e<0.1$), but significant eccentricities. The circles represent the G dwarfs and the squares represent F dwarfs. It appears that the low mass hot Jupiters on orbits that are consistent with circular around G dwarfs migrate inwards  until they stop at a minimum period for a given mass, conglomerating on the mass-period relation of \citet{Mazeh2005}. In this case, the heavier planets can move in further before they are stopped. Planets heavier than about 1.2 $M_j$ can migrate inwards and raise tides on the star, leading to a spin-up of the host star, and even synchronisation in some cases where enough angular momentum can be transferred from the orbital motion into the stellar rotation. In cases where the planetary angular momentum is insufficient, the process can lead to a run-away migration until the planet is destroyed inside the star.

The Roche limit for a planet is defined by $R_p=0.462a_R(M_p/M_s)^{-3}$. If we write the stopping distance $a=\alpha a_R$, \cite{Ford2006a} argued that slow migration on quasi-circular orbits would result in a value of $\alpha=1$, with the only surviving planets being those that stop at their Roche limit. On the other hand, if the planets were brought in on an eccentric orbit (eg: dynamical interactions within a system or capture from interstellar space), and then circularised by tidal interaction, the value of $\alpha$ should be two. In Figure~\ref{fig:mass_period}, the dashed line shows this case, with $\alpha=2$. This does not appear to be a very good fit for the hot Jupiters that are on orbits consistent with circular. The dotted lines show the range $\alpha=2.5$--$4.5$. As mentionned in \citet{Pont2011}, this larger value of $\alpha$ could indicate the planets had larger radii at the time their orbits were circularised. Subsequent thermal evolution of the planets would have shrunk them \citep[eg:][]{Baraffe2004}, leaving them further out from their current Roche limits.

\ \\
{\bf Circularisation Timescales}\\
The process of tidal circularisation, spin-orbit alignment and synchronisation are expected to occur roughly in this order, and over a similar timescale. For close-in systems, this timescale is expected to be small compared to the lifetime of the system.  \citet{Hut1981} derived equations for the tidal evolution due to the equilibrium tide using the assumption of weak friction, and constant time-lag $\Delta t$. \citet{Leconte2010} re-visited this model and showed that the orbital eccentricity evolves according to

\begin{eqnarray}
\frac{1}{e}\frac{{\rm d} e}{{\rm d} t} =  11 \frac{a}{GM_sM_p} & \{K_p\left[\Omega_e(e)x_p\frac{\omega_p}{n} - \frac{18}{11}N_e(e)\right] \\
   & +K_s\left[\Omega_e(e)x_s\frac{\omega_s}{n} - \frac{18}{11}N_e(e)\right] \} \nonumber,
\end{eqnarray}

\noindent where $\Omega_e(e)$ and $N_e(e)$ are functions of $e$ and approximately equal to unity for small $e$; $x_p$ and $x_s$ are the cosines of the angle between the orbital plane and the planet and stellar equators respectively. $\omega_p$ and $\omega_s$ are the angular frequencies of rotation of the planet and star, and the two terms

\begin{equation}
K_p = \frac{3}{2}k_{2,p}\Delta t_p \left(\frac{GM_p^2}{R_p}\right)\left(\frac{M_s}{M_p}\right)^2\left(\frac{R_p}{a}\right)^6n^2
\end{equation}

\noindent and 
\begin{equation}
K_s = \frac{3}{2}k_{2,s}\Delta t_s \left(\frac{GM_s^2}{R_s}\right)\left(\frac{M_p}{M_s}\right)^2\left(\frac{R_s}{a}\right)^6n^2
\end{equation}

\noindent describe the effect of tides on the planet by the star, and vice-versa, respectively. $n$ is the mean orbital motion and the semi-major axis is denoted $a$. Under the assumption of a constant-time delay between the exciting tidal potential and the response of the equilibrium tide in the relevant body, $k_{2,p} \Delta t_p$ and $k_{2,s} \Delta t_s$ are constants where $k_{2}$ are the potential Love numbers of degree 2 and $\Delta t$ are the constant time lags in each of the two bodies.

We now consider two limits, firstly the case where only the tides in the planet dominate, and then the case where only tides in the star dominate. When tides in the planet dominate, $K_s\sim 0$ so that we obtain a timescale 

\begin{equation}
\tau_p = -\left(\frac{1}{e}\frac{{\rm d}e}{{\rm d}t}\right)^{-1} = \frac{2}{21G}\frac{1}{k_{2,p}\Delta t_{p}}\frac{M_p}{M_s^2}\frac{a^8}{R_p^5}\label{eqn:planet_isochrones}
\end{equation}

where we have assumed that $\Omega_e=N_e\approx 1$, i.e. the equation is valid to lowest order in $e$; $\omega_p/n \sim 1$, i.e. synchronisation of the planetary rotation with the orbit and $x_p\sim 1$, i.e. the planet's equator coincides with the orbital plane. 
A similar equation can be written for tides in the star, even though $\omega_s/n$ is not typically unity. As long as $\omega_s/n<18/11$, for small $e$, the effect of tides in the star will lead to a decrease in orbital eccentricity. We can therefore write,

\begin{equation}
\tau_s = -\left(\frac{1}{e}\frac{{\rm d}e}{{\rm d}t}\right)^{-1} = \frac{2}{21G}\frac{1}{k_{2,s}\Delta t_{s}}\frac{M_s}{M_p^2}\frac{a^8}{R_s^5}\label{eqn:star_isochrones}
\end{equation}

We take some typical values of $k_{2,p}\Delta t_{p}\sim 0.01$ s and $k_{2,s}\Delta t_{s}\sim 1$ s, which would correspond to tidal quality factors \citep{Goldreich1966} of about $10^6$ and $10^4$ respectively, in the constant-$Q$ model (in contrast to the constant $\Delta t$ model that we consider here) for an orbital period of about 5 d.

\begin{figure*}
	\resizebox{16cm}{!}{\includegraphics{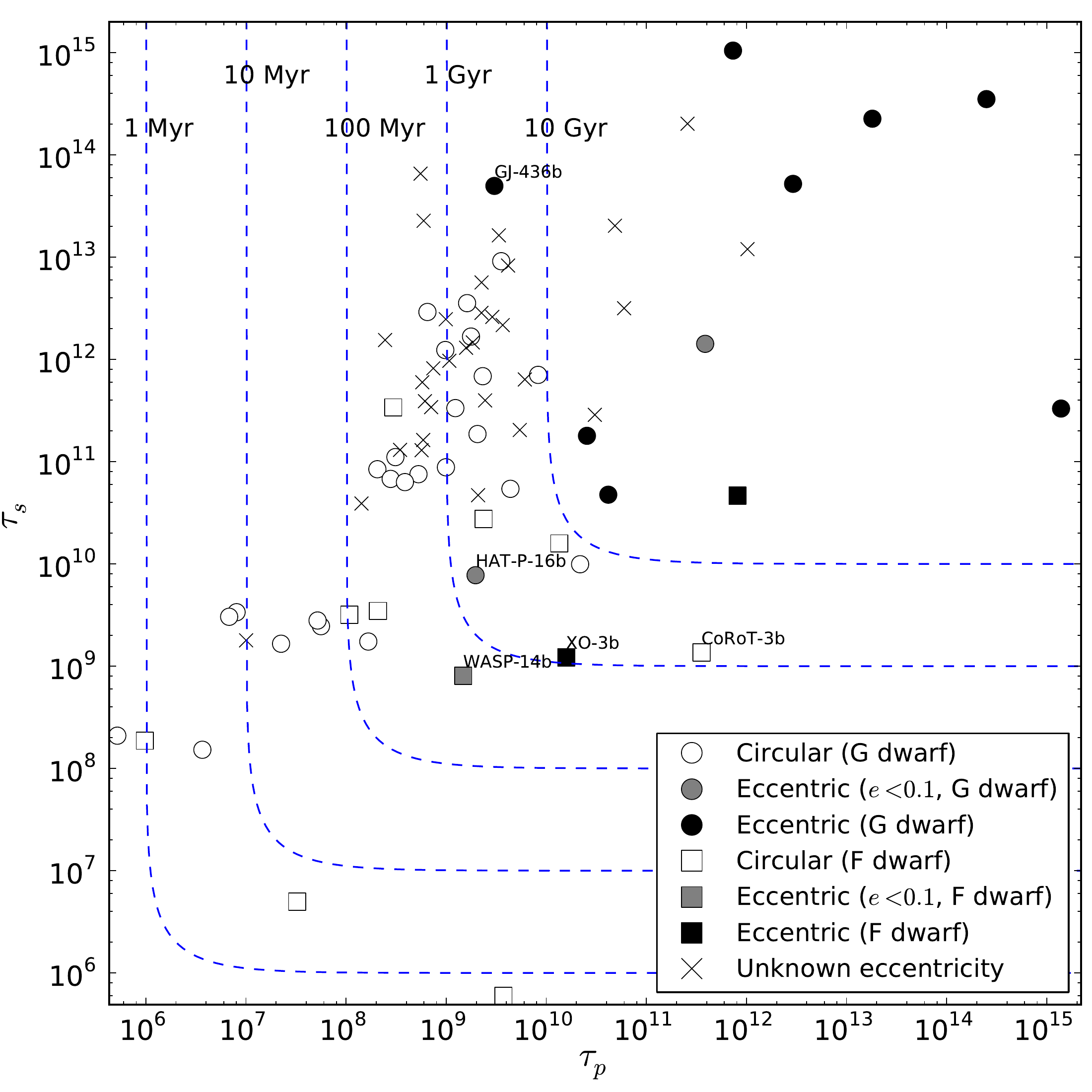}}
	\caption{Plot showing the timescale of circularisation assuming tides inside the star alone (vertical axis) against the timescale of circularisation assuming tides in the planet alone (horizontal axis). The dotted lines represent lines of constant circularisation timescale. For the G dwarfs (circles), orbits that are consistent with circular and eccentric orbits are cleanly segregated by the 10 Gyr isochrone, with HAT-P-16b ($e=0.034\pm0.003$) caught in the process of circularisation. For the F dwarfs (squares), WASP-14b has a small eccentricity $e=0.008\pm0.003$ and XO-3b has an eccentricity of $0.287\pm0.005$, whereas CoRoT-3b is on an orbit that is consistent with circular. The short timescale for tides in the star, coupled with the relatively small eccentricities of WASP-14 and XO-3, suggest that tidal effects in the star are still operating.}
	\label{fig:timescale_circ}
\end{figure*}

We expect planets that are further out to be only weak affected by tides, whereas close-in planets will experience strong tides. Some of these close-in planets will be heavy enough and close enough to exert their own influence on the star by raising stellar tides. This can be seen in Figure~\ref{fig:timescale_circ}, where we have plotted the timescale of circularisation assuming tides inside the star alone against the timescale of circularisation assuming tides in the planet alone. The open symbols represent orbits that are consistent with circular, and the black symbols represent eccentric orbits, whereas the grey symbols represent objects with small ($e<0.1$), but significant eccentricities. The dashed lines represent lines of constant circularisation timescale, at 1 Myr, 10 Myr, 100 Myr, 1 Gyr and 10 Gyr. For the G dwarfs, orbits that are consistent with circular and eccentric orbits are cleanly segregated by the 10 Gyr isochrone, with HAT-P-16b ($e=0.034\pm0.003$) caught in the process of circularisation. For the F dwarfs (open symbols), WASP-14b ($T_{\mathrm{eff}}=6475 \pm 100$ K) has a small eccentricity $e=0.008\pm0.003$ and XO-3b ($T_{\mathrm{eff}}=6429 \pm 100$ K) has an eccentricity of $0.287\pm0.005$, whereas CoRoT-3b ($T_{\mathrm{eff}}=6740\pm 140$ K) is on an orbit that is consistent with circular. This suggests that in the dissipation factor in hotter stars may vary in an unknown fashion, although the small eccentricity of WASP-14b and the moderately small eccentricity of XO-3, together with the short timescale for stellar tides indicate that tides in the star are clearly important even in these cases.

\ \\
{\bf Hot Neptunes}\\
GJ-436b is a planet on an eccentric orbit ($e=0.153\pm0.017$) in a region of the mass-scale plane where tidal effects on the planet are expected to be significant. The planet is a hot Neptune so it is possible that the structure is different enough that the tidal quality factor $Q$ is very much higher, leading to a longer circularisation timescale. In this case, GJ-436b would simply not have had enough time to circularise its orbit. Another possibility that was initially suggested by \citet{Maness2007}, is that a second companion may be present in the system and is pumping up the eccentricity of GJ-436b by secular interactions. Further measurements with radial velocity \citep{Ribas2009} and photometry \citep{Ballard2010} appear to rule this possibility out.

\ \\
{\bf Synchronisation}\\

\begin{figure*}
	\resizebox{8.5cm}{!}{\includegraphics{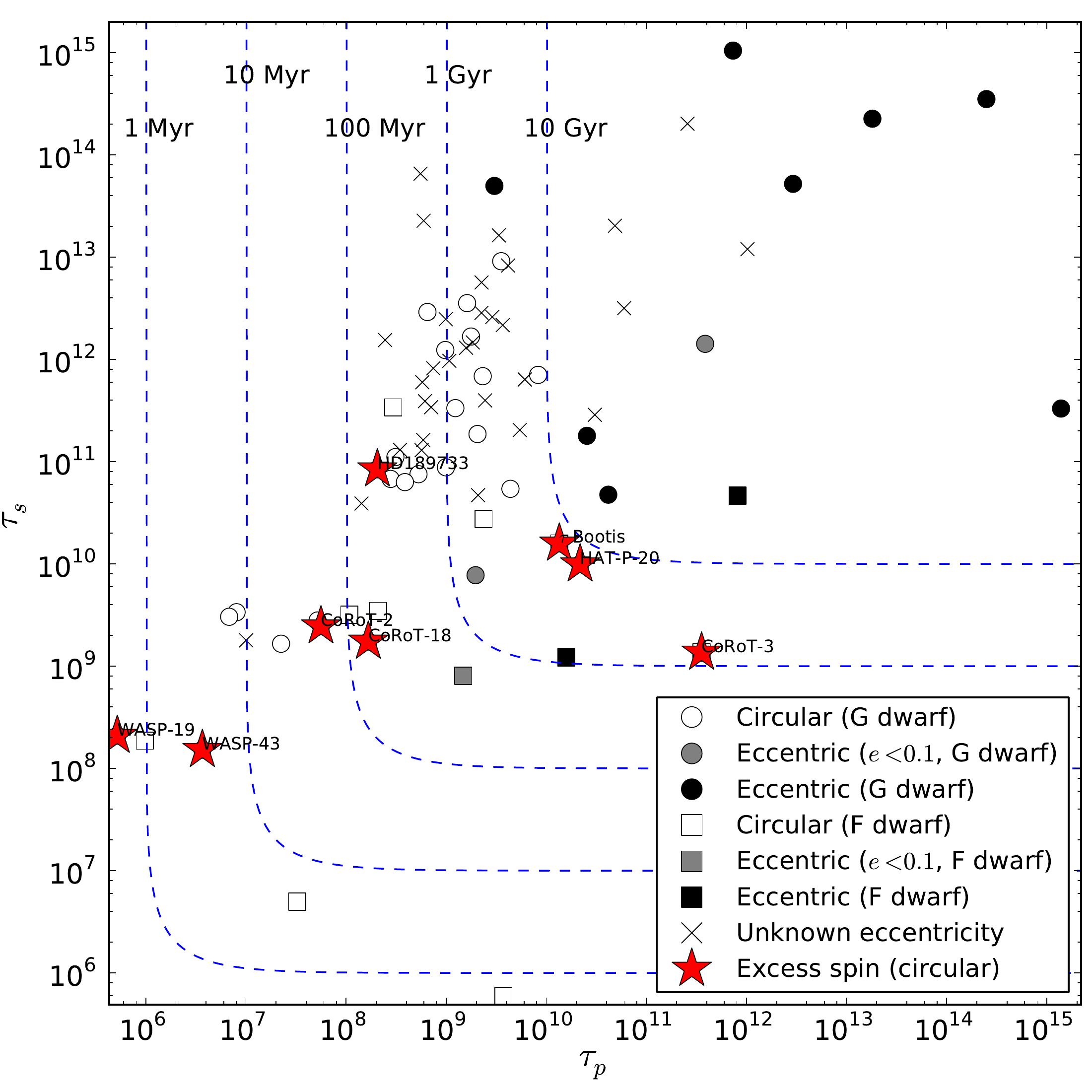}}
	\resizebox{8.5cm}{!}{\includegraphics{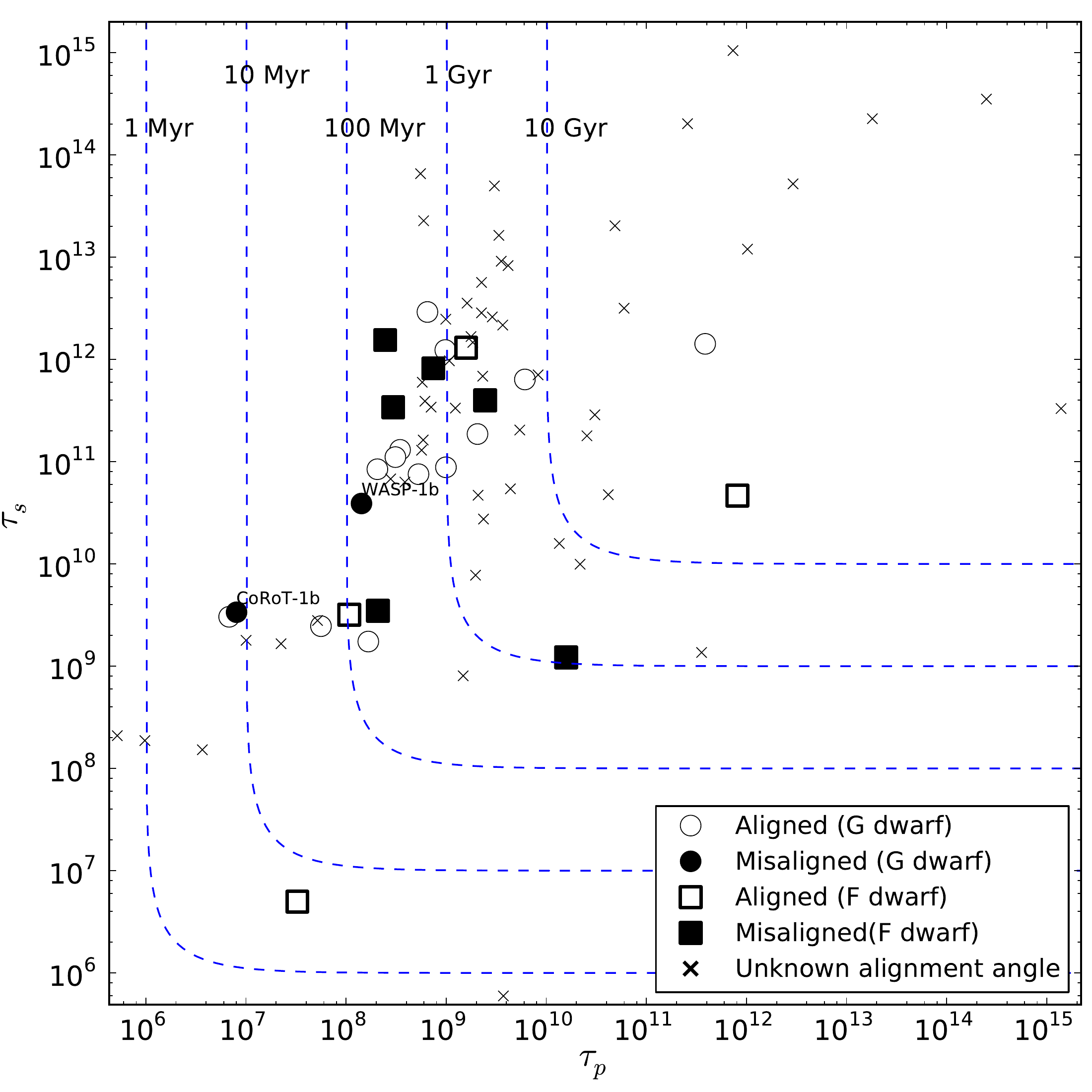}}
	\caption{The axes are the same as Figure~\ref{fig:timescale_circ}. {\bf Left:} The star symbols represent objects where there is evidence of spin-up. These are stars that rotate faster than predicted by the isochrones of Strassmeier \& Hall (1988). In the case of the two hot stars CoRoT-3 and $\tau$ Bo\"otis b, the stellar rotation have even become synchronised with the orbital period. No objects with a  stellar tidal dissipation timescale larger than about $\tau_s>10^{11}$ years show any evidence of excess rotation, supporting the case for tidal involvement in the objects with excess rotation. \newline{\bf Right:} The circles indicate aligned systems ($\lambda<30^{\circ}$), whereas the star symbols represent misaligned systems ($\lambda>30^{\circ}$). In this case, the G dwarfs are aligned (CoRoT-1 and WASP-1 are actually hot stars, and WASP-8 is outside the region of strong tides in the star). The F dwarfs, on the other hand, display a spread in terms of aligned and misaligned, even in cases of strong tides, in agreement with Winn et al. (2010).}
	
	\label{fig:timescale2}
\end{figure*}

Tidal dissipation leading to orbital circularisation can occur in either the planet, the star, or both, according to the timescale for each case. On the other hand, synchronisation of the host star rotation with the orbital motion would depend on tidal effects inside the star alone. This would occur on a similar timescale as circularisation in the case of dissipation in the star alone. Figure~\ref{fig:timescale2} shows the same axes as Figure~\ref{fig:timescale_circ}, but on the left panel, the red star symbols represent objects with excess stellar rotation. In the case of CoRoT-3b and $\tau$ Bo\"otis b, the rotation of the host star has been synchronised with the orbital period. Pont (2009) also pointed out that HD 189733 and CoRoT-2b were rotating faster than expected from the isochrones of Strassmeier \& Hall (1988), even if the stellar rotations were not synchronised. We can now confirm that four more objects are clearly in this regime: CoRoT-18, HAT-P-20, WASP-19 and WASP-43. The rotation periods of these stars and the expected rotation periods are shown in Table~\ref{tab:excess_rot}. From Figure~\ref{fig:timescale2}, we note that the estimated timescale for orbital circularisation due to tidal effects in the star alone is less than 5 Gyr for the objects WASP-19, WASP-43, CoRoT-2, CoRoT-18 and CoRoT-3. This means that tidal dissipation in the star could lead to the excess rotation well within the lifetime of these stars. On the other hand, the two objects $\tau$ Bo\"otis b and HAT-P-20 have timescales $\tau_s\sim10$ Gyr, while HD 189733b has $\tau_s\sim80$ Gyr. Even in this case, it should be noted that the tidal dissipation strength would have to be stronger by a single order of magnitude for these objects to have been spun up by tidal dissipation inside the star. Given that the tidal time lag is uncertain by up to about two orders of magnitude, this does not sound implausible. In contrast, orbital circularisation in many of these cases may well have occured due to dissipation in the planet instead. Planets that are unable to spin-up their parent stars to synchronisation may be doomed to destruction. \citet{Hellier2009} pointed out that the existence of WASP-18 at its current position in the mass-period plane suggests that either the tidal dissipation in the system is several orders of magnitude smaller than expected, or that the system is caught at a very special time while it is in the last $10^{-4}$ of the estimated lifetime of the system. The latter possibility sounds more plausible, considering the striking paucity  of heavy planets at short period.

\begin{table}
\centering
\begin{tabular}{ l           r@{$\pm$}l         c      }
\hline
Name			&   \multicolumn{2}{r|}{$P_{\rm rot}$ (d)} 	& Expected $P_{\rm rot}$ (d) \\
\hline
CoRoT-2			& 4.52 & 0.02 		&   36 \\
CoRoT-18		& 6.3 & 0.9 		&   49 \\
HAT-P-20			& 11.3 & 2.2 		&  57  \\
HD 189733		& 12.95 & 0.01 	&  57  \\
WASP-19			& 10.5 & 0.2 		&   42 \\
WASP-43			& 7.6 & 0.7 		&   57\\
\hline
\end{tabular}
\caption{Table showing the systems with excess rotation in the left panel of Figure~\ref{fig:timescale2}.  $P_{\textrm{rot}}$ is the stellar rotation period today, and `Expected $P_{\textrm{rot}}$' is the expected rotation period of the star as estimated from the rotation isochrones of Strassmeier \& Hall (1988).}
\label{tab:excess_rot}
\end{table}

\ \\
{\bf Spin-orbit alignment}\\
The right panel of Figure~\ref{fig:timescale2} shows the same axes (timescales), but now the circles represent G stars and the squares represent F stars. The empty symbols represent aligned systems ($\lambda<30^{\circ}$), and the filled symbols represent misaligned systems ($\lambda>30^{\circ}$). In this case, the G dwarfs are aligned, except for CoRoT-1($T_{\mathrm{eff}}=5950\pm150$ K) and WASP-1 ($T_{\mathrm{eff}}=6110\pm245$ K) are actually hot stars, and WASP-8 is outside the region of strong tides in the star. CoRoT-1, WASP-1 and the F dwarfs, display a spread in terms of aligned and misaligned, even in cases of strong tides. \citet{Winn2010b} found a link between the presence of a convective core and spin-orbit alignment by tidal effects. Thus, exoplanets could migrate inwards by planet-planet scattering, giving rise to orbits with a range of eccentricities and spin-orbit angles. Planets in orbit around cooler stars ($T_{\rm eff}<6250$ K, where the stellar convective region is significant), can have their orbital angular momentum aligned with the stellar rotation, while planets in orbit around hot stars ($T_{\rm eff}>6250$ K, where the extent of the convective region is negligible) manage to keep their initial misalignment.

\section{Conclusion}
We have recalculated estimates of orbital eccentricity for a population of known transiting planets and included a noise treatment to account for systematic effects in the data. As \citet{Laughlin2005} showed using synthetic data, analysis of radial velocity data can result in a derived eccentricity at a few $\sigma$ level even in cases where the orbit is in fact consistent with circular. In a similar way, correlated noise in the instrument or atmosphere, stellar activity, or additional companions to the host star can cause a spurious eccentricity detection, the cases of WASP-12 and WASP-10 being two examples highlighted in this paper. 

Once these confusing effects are accounted for, a much clearer picture emerges, highlighting the importance of tidal interactions in close-in exoplanet systems. The present observations support a scenario where low mass hot Jupiters migrate inwards and circularise their orbits until they stop at a minimum period for a given mass, conglomerating on the mass-period relation of \citet{Mazeh2005}. The heavier planets are able to move further inwards before they stop. Planets heavier than about 1.2 $M_j$ can raise tides on the star as they migrate inwards, leading to a spin-up of the host star \citep{Pont2009a}, and even spin-orbit synchronisation in some cases where enough angular momentum can be transferred from the orbital motion into the stellar rotation. This appears to be the case for CoRoT-3b, $\tau$ Bo\"otis b, HD 189733, CoRoT-2b, CoRoT-18, HAT-P-20, WASP-19 and WASP-43, where the first two are synchronised, and the rest show clear evidence of excess rotational angular momentum in the star. If the planetary angular momentum is insufficient, the process can lead to a run-away migration and the planet is destroyed, as appears to be the case for WASP-18b \citep{Hellier2009a}. This is also supported by the lack of such heavy planets at short period. As suggested by \cite{Winn2010b}, tidal effects in G dwarfs are also responsible for aligning the spin of the star with the orbit of the planet, whereas the same effect is much less effective in the case of the hotter F stars. Overall, therefore, the present data on close-in exoplanets support the case for a prominent role for tidal interactions between the planet and the host star in the orbital evolution of hot Jupiters.

\subsection*{Acknowledgements}
We are grateful to the anonymous referee for the encouraging remarks and the enormous amount of detailed feedback and sound advice, which helped us to make this paper better. We thank the editor for his encouraging comments. 
FP is grateful for the STFC grant and Halliday fellowship ST/F011083/1. We also thank the SOPHIE Exoplanet Consortium for arranging a flexible observation schedule for our programme, and the whole OHP/SOPHIE team for support. NH thanks Gilles Chabrier for his clear explanations of the equilibrium tides model.

\clearpage\clearpage

\bibliography{husnoo}{}

\end{document}